\documentclass{CSML}
\pdfoutput=1

\usepackage{lastpage}
\newcommand{\dOi}{14(1:6)2018}
\lmcsheading{}{1--\pageref{LastPage}}{}{}%
{Nov.~23, 2016}{Jan.~16, 2018}{}


\ACMCCS{[{\bf Theory of computation}]: AAA---BBB; [{\bf Mathematics of
  computing}]: CCC---DDD}
\usepackage{hyperref}
\hypersetup{hidelinks}
\usepackage{enumerate}

\usepackage{multirow}
\usepackage{amssymb}
\usepackage{algorithmic}
\usepackage{amsmath}
\usepackage{etoolbox}
\usepackage{mathrsfs}
\usepackage{lineno}
\usepackage{color}
\usepackage{array}
\usepackage{framed,multicol}

\usepackage{graphicx}

\usepackage{subfig}
\usepackage{hhline}

\usepackage{wrapfig}
\usepackage{float}
\usepackage{footmisc}

\theoremstyle{plain}
\title{Fencing Programs with Self-Invalidations and Self-Downgrades}

\usepackage[utf8]{inputenc}
\usepackage[ruled,nofillcomment,noend,linesnumbered]{algorithm2e}
\usepackage{stmaryrd, dsfont}
\usepackage{cleveref}
\usepackage{amssymb}
\usepackage{color}
\usepackage{textcomp}
\usepackage{wrapfig}
\usepackage{tikz}
\usepackage{url}
\usepackage{subfig}
\usetikzlibrary{calc}
\usetikzlibrary{shapes.geometric} 
\usetikzlibrary{backgrounds}
\usetikzlibrary{fit}
\usetikzlibrary{decorations.pathmorphing}
\usetikzlibrary{patterns}

\pgfdeclarelayer{mybackground}
\pgfdeclarelayer{background}
\pgfdeclarelayer{foreground}
\pgfsetlayers{background,mybackground,foreground,main}

\tikzstyle{tracenode}=[font=\scriptsize,inner sep=0mm,anchor=west,fill=yellow]
\tikzstyle{traceedge}=[line width=0.5pt,->,>=latex,color=blue!50!black]

\tikzstyle{runnode}=[fill=black,circle,minimum size=3pt,inner sep=0mm]
\tikzstyle{confrunnode}=[font=\scriptsize,inner sep=0mm]
\tikzstyle{lblnode}=[font={\tt\tiny},above=-1pt]
\tikzstyle{runedge}=[line width=0.5pt,->,>=latex,color=black]
\tikzstyle{segmentnode}=[font=\scriptsize,inner sep=0.5pt,draw=black,fill=yellow,line width=0.5pt, anchor=north,circle]

\tikzstyle{llcnode}=[fill=yellow!10,rounded corners,line width=0.5pt,draw,align=center,font={\tt\scriptsize},inner sep=2pt,anchor=west,name=llc]

\tikzstyle{procnollnode}=[fill=red!10,rounded corners,line width=0.5pt,draw,align=center,font={\tt\scriptsize},inner sep=2pt,anchor=west,name=p0]

\tikzstyle{proconenode}=[fill=blue!10,rounded corners,line width=0.5pt,draw,align=center,font={\tt\scriptsize},inner sep=2pt,anchor=west,name=p1]

\tikzstyle{pgmtextnode}=[font={\tt\scriptsize},inner sep=0pt]

\tikzstyle{fitnode}=[draw=black, rounded corners,line width=1pt]

\newcommand{\confpic}[4]{
\begin{tikzpicture}[background rectangle/.style={rounded corners,bottom color=green!5,top color=green!5,line width=1pt,draw},show background rectangle,inner sep=-3pt]

\node [name=dummy] {};
\node [llcnode] at (dummy) {#2};
\node [procnollnode] at ($(llc.east)+(1pt,0pt)$) {#3};
\node [proconenode] at ($(p0.east)+(1pt,0pt)$) {#4};

\node[font=\scriptsize] at  ($(p0.south)+(0pt,-5pt)$) {#1};
\end{tikzpicture}
}

\newcommand{\confpicarrow}[4]{
\begin{tikzpicture}[background rectangle/.style={rounded corners,bottom color=green!5,top color=green!5,line width=1pt,draw},show background rectangle,inner sep=-3pt]

\node [name=dummy] {};
\node [llcnode] at (dummy) {#2};
\node [procnollnode] at ($(llc.east)+(1pt,0pt)$) {#3};
\node [proconenode] at ($(p0.east)+(1pt,0pt)$) {#4};

\node[font=\scriptsize] at  ($(p0.south)+(0pt,-5pt)$) {#1};

\node[font=\tiny,name=llclbl] at  ($(llc.north)+(0pt,15pt)$) {LLC};
\node[font=\scriptsize,name=p0lbl] at  ($(p0.north)+(0pt,15pt)$) {{\tt P0}};
\node[font=\scriptsize,name=p1lbl] at  ($(p1.north)+(0pt,15pt)$) {{\tt P1}};

\draw[->,decorate,decoration={snake,post length=2pt},color=red] ($(llclbl.south)+(0,-3pt)$) -- (llc);
\draw[->,decorate,decoration={snake,post length=2pt},color=red] ($(p0lbl.south)+(0,-3pt)$)  -- (p0);
\draw[->,decorate,decoration={snake,post length=2pt},color=red] ($(p1lbl.south)+(0,-3pt)$)  -- (p1);

\end{tikzpicture}
}

\renewcommand{\emptyset}{\varnothing}
\newcommand{\dirty}{{\tt dirty}}
\newcommand{\clean}{{\tt clean}}

\newcommand{\run}{\pi}
\newcommand{\rrun}{\rho}
\newcommand{\bad}{{\it Bad}}

\newcommand{\memlocset}{\mathcal{M}}

\newcommand{\fence}{{\tt fence}}
\newcommand{\ssfence}{{\tt ssfence}}
\newcommand{\llfence}{{\tt llfence}}
\newcommand{\assign}[2]{#1\hspace{2pt}{\tt := }\hspace{2pt}#2}
\newcommand{\plainsyncwr}{{\tt syncwr}}
\newcommand{\syncwr}[2]{\plainsyncwr: \assign{#1}{#2}}

\newcommand{\cas}{{\tt cas}}

\newcommand{\ifetch}{{\tt fetch}}
\newcommand{\ievict}{{\tt evict}}
\newcommand{\iwrllc}{{\tt wrllc}}
\newcommand{\fetch}[2]{\ifetch({#1,#2})}
\newcommand{\evict}[2]{\ievict({#1,#2})}
\newcommand{\wrllc}[2]{\iwrllc({#1,#2})}

\newcommand{\tuple}[1]{\left(#1\right)}

\newcommand{\bnfvar}[1]{\langle{\it {#1}}\rangle}
\newcommand{\bnfkey}[1]{\tt {#1}}
\newcommand{\bnfstr}[1]{\textrm{\textquotesingle}{#1}\textrm{\textquotesingle}}

\newcommand{\valset}{{\mathcal V}}

\newcommand{\cbranch}{{\tt cbranch}}
\newcommand{\proc}{p}

\newcommand{\regval}{{\sf RVal}}
\newcommand{\loneconf}{{\sf L1}}
\newcommand{\valid}{{\sf Valid}}
\newcommand{\lstatus}{{\sf LStatus}}
\newcommand{\lval}{{\sf LVal}}
\newcommand{\memloc}{x}
\newcommand{\localconf}{{\sf LConf}}
\newcommand{\llc}{{\sf LLC}}
\newcommand{\conf}{c}
\newcommand{\movesto}[1]{\stackrel{#1}{\longrightarrow}}
\newcommand{\xvar}{x}
\newcommand{\yvar}{y}
\newcommand{\zvar}{z}
\newcommand{\reg}{\$r}
\newcommand{\update}[2]{\left[{#1}\leftarrow{#2}\right]}
\newcommand{\stmt}{\sigma}
\newcommand{\procof}[1]{{\sf Proc}\left(#1\right)}
\newcommand{\stmtof}[1]{{\sf Stmt}\left(#1\right)}
\newcommand{\lbl}{\lambda}

\newcommand{\nextof}[1]{{\sf Next}\left(#1\right)}
\newcommand{\evt}{\omega}
\newcommand{\prog}{{\mathcal P}}
\newcommand{\transition}{t}

\tikzstyle{background rectangle}=
[rounded corners,bottom color=green!20,top color=green!20,inner frame sep=0pt]

\tikzstyle{pgmnode}=
[rectangle,line width=1.5pt,color=blue!50!black,fill=blue!10,draw,inner sep=2mm, 
rounded corners, minimum width=2mm,
minimum height=2mm,text=blue!50!black,align=center,font=\scriptsize]

\tikzstyle{crmnode}=
[circle,line width=1.5pt,color=blue!50!black,fill=blue!10,draw,inner sep=1mm, 
rounded corners, minimum width=2mm,
minimum height=2mm,text=blue!50!black,align=center,font={\scriptsize}]

\tikzstyle{storenode}=
[cylinder,line width=1.5pt,color=red!50!black,fill=red!10,draw,inner sep=2mm, 
text=red!50!black,align=center,font={\scriptsize}]

\tikzstyle{textnode}=
[font={\scriptsize}]

\tikzstyle{cndnode}=
[line width=1.5pt,color=green!50!black,fill=green!10]

\tikzstyle{blockedge}=[line width=1pt,->,>=stealth]

\tikzstyle{graphnode}=[inner sep=1pt,fill=yellow!10,draw,circle,font=\scriptsize]

\newcommand{\finish}{{\tt end}}

\newlength{\algoinddepth}
\newlength{\algostdind}
\newenvironment{algo1}[2]{
    \newcommand{\thename}{#1}
    \newcommand{\linctr}{#2}
    \newcommand{\commentcolor}{gray}
    \newcounter{\linctr}
    \setcounter{\linctr}{0}
    \setlength{\algoinddepth}{0pt}
    \setlength{\algostdind}{10pt}
    \newcommand{\icmt}[1]{{\color{\commentcolor}// ##1}} %% Inline comment
    \newcommand{\comment}[1]{\hphantom{MM: }\rule{\algoinddepth}{0pt}\icmt{##1}\\}
    \newcommand{\linunterm}[1]{\refstepcounter{\linctr}\hphantom{MM}\llap{\arabic{\linctr}}: \rule{\algoinddepth}{0pt}##1}
    \newcommand{\lnlabel}[1]{\addtocounter{\linctr}{-1}\refstepcounter{\linctr}\label{##1}}
    \newcommand{\lin}[1]{\linunterm{##1}\\}
    \newcommand{\incind}{\global\addtolength{\algoinddepth}{\algostdind}}
    \newcommand{\decind}{\global\addtolength{\algoinddepth}{-\algostdind}}
    \newcommand{\key}[1]{\textrm{\textbf{##1}}}
\newcommand{\while}[2]{%
\lin{\key{while}(##1)\{\incind}%
##2%
\decind%
\lin{\}}%
}
\newcommand{\ifunterm}[3][]{%
\lin{\key{if}(##2)\{\ifstrempty{##1}{}{ \icmt{##1}}\incind}%
##3%
\decind%
\linunterm{\}}%
}
\newcommand{\ifterm}[3][]{\ifunterm[##1]{##2}{##3}\\}

\newcommand{\elseterm}[2][]{%
\key{else}\{\ifstrempty{##1}{}{ \icmt{##1}}\incind\\%
##2%
\decind%
\lin{\}}%
}
  \tt
  \begin{tabular}{@{}l@{}}
    {\bf \thename}\\
    \hline
}{
  \end{tabular}
  \normalfont
}

\begin{document}

\title[Mending Fences with Self-Invalidation and Self-Downgrade]{Mending Fences with Self-Invalidation and Self-Downgrade\rsuper*}%\Thanks
\titlecomment{\rsuper* A preliminary version of this paper has appeared at FORTE 2016,
  held as Part of the 11th International Federated Conference on Distributed Computing Techniques,
  DisCoTec 2016}

\author[P.A.~Abdulla]{Parosh Aziz Abdulla\rsuper{a}}
\address{\lsuper{a}Uppsala University, Sweden}	%required
\email{parosh.abdulla@it.uu.se}
\email{mohamed\_faouzi.atig@it.uu.se}
\email{stefanos.kaxiras@it.uu.se}
\email{carl.leonardsson@it.uu.se}
\email{yunyun.zhu@it.uu.se}
\author[M.F.~Atig]{Mohamed Faouzi Atig\rsuper{a}}
\address{\vskip-6pt}
\author[S.~Kaxiras]{Stefanos Kaxiras\rsuper{a}}
\address{\vskip-6pt}
\author[C.~Leonardsson]{Carl Leonardsson\rsuper{a}}
\address{\vskip-6pt}
\author[A.~Ros]{Alberto Ros\rsuper{b}}
\address{\lsuper{b}Universidad de Murcia, Spain}
\email{aros@ditec.um.es}
\author[Y.~Zhu]{Yunyun Zhu\rsuper{a}}
\address{\vskip-6pt}

\begin{abstract}
  \noindent Cache coherence protocols based on self-invalidation and self-downgrade
  have recently seen increased popularity due to their simplicity, potential
  performance efficiency,
  and low energy consumption. However, such protocols result in
   memory
  instruction reordering,  thus causing extra program behaviors that are often
  not intended by the programmers.
We propose a novel formal  model that captures the semantics of programs
running under such protocols,
and features a set of fences that interact with the coherence layer.
Using the model, we design an algorithm to analyze the reachability
and 
check whether a program satisfies a given safety property
with the current set of fences.
We describe
a method for insertion
 of {\it optimal} sets of fences that ensure correctness of the
 program under such protocols. The
 method relies on a {\em counter-example} guided fence insertion procedure.
 One feature of our  method is that it can
 handle a variety  of fences (with different costs).
 This diversity makes optimization more difficult
 since one has to optimize the total cost of the inserted fences,
 rather than just their number.
 To demonstrate the strength of our approach,
 we have implemented a prototype and run it on a wide range
 of examples and benchmarks.
 We have also, using simulation, evaluated the performance of the resulting fenced programs.

\end{abstract}

\maketitle

\section{Introduction}

\subsection*{Background.}

Traditional cache coherence protocols, either directory-based or snooping-based,
 are
{\it transparent} to the programmer
in the sense that
they respect the memory consistency model of the system, and hence
there is  no effect on memory ordering due to the coherence protocol.
On the other hand, there is an ever larger demand on hardware designers
to increase {\it efficiency}
both in performance and power consumption.
The quest to increase performance while maintaining transparency has led
to complex coherence protocols
with many states and relying on directories,
invalidations, broadcasts, etc, often at the price of
high verification cost, area (hardware cost) and increased energy consumption.
Therefore, many researchers have recently proposed
ways to simplify coherence without
compromising performance but at the price
of relaxing the memory consistency model~\cite{lebeck95,choi11,kaxiras11,aros-pact12,sung13,skaxiras-isca13,hower14,aros-hpca15,sung15,aros-isca15,mdavari-pact15,kkoukos-taco16,aros-micro16}. %
Principal techniques among these proposals are Self-Invalidation ({\sc Si}) and Self-Downgrade ({\sc Sd}).

A protocol with Self-Invalidation ({\sc Si}) allows
old copies of the data to be kept,
without invalidation on each store operation by another core.
This eliminates the need for tracking readers~\cite{lebeck95}.
In an {\sc Si} protocol, invalidation of data from a cache is caused by
synchronization instructions executed by the core local to the cache.

%\begin{figure}
%  \includegraphics[width=200pt]{TSO.eps}
%  \caption{TSO}
%  \label{fig:TSO}
%\end{figure}

Correspondingly,
in  a protocol with Self-Downgrade ({\sc Sd}),
downgrades are not
caused by read operations in other cores, but again by synchronization instructions.
{\sc Sd}
eliminates the need to track the last writer
of a cache line \cite{aros-pact12}.

A protocol
with both self-invalidation and self-downgrade ({\sc SiSd}) does not
need a directory, thus removing a main source of complexity and
scalability constraints in traditional
cache coherence protocols \cite{aros-pact12}.
But this comes at a price:
{\sc SiSd} protocols induce
{\it weak memory semantics} that allow reordering or memory
instructions.
The behavior of a program may now
deviate from its behavior
under the standard {\it Sequentially Consistent  (SC)}  semantics \cite{lamport79},
sometimes leading to subtle errors that are hard
to detect and correct.

In the context of weak memory, hardware designers provide
memory {\em fence} instructions to
help the programmer to eliminate the undesired  behaviors.
A fence instruction, executed by a thread,
limits the allowed reorderings between
instructions issued before and after the fence instruction.
To enforce consistency under {\sc SiSd},
fences should also be made visible to caches,
such that necessary invalidations or downgrades may be performed.
In this paper,
we consider different types of fences.
Each type eliminates a different kind of non-SC behavior, and
may have different impact on the program performance.
In fact, unnecessary fences may significantly jeopardize program performance.
This is particularly true for the fences considered in this work,
since they both incur latency, and
affect the performance
of the cache coherence subsystem as a whole.
These fences cause the invalidation of
the contents of the cache. Hence the more fences the less caching and the
higher traffic we have.
Thus, it is desirable to find the \emph{optimal} set of fences, which
guarantee correctness at minimal performance cost.
There are multiple ways of defining optimality. The one we adopt is 
to calculate the number of occurrences of fences in the source program,
with the costs of different fences taken into account.

\subsection*{Challenge.}
One possibility to make {\sc SiSd} transparent to the program
is to require the programmer to ensure that the program
does not contain any data races.
In fact, data race freedom is often required by designers
of {\sc SiSd} protocols in order to guarantee
correct program behavior \cite{choi11,kaxiras11}.
However, this approach would unnecessarily disqualify large sets
of programs, since many data races are in reality not harmful.
Examples of correct programs with races include
lock-free data structures
(e.g.,  the Chase-Lev Work-stealing queue algorithm
\cite{DBLP:conf/spaa/ChaseL05}),
transactional memories (e.g., the TL2 algorithm
\cite{DBLP:conf/wdag/DiceSS06}),
and synchronization library primitives
(e.g. {\tt pthread\_spin\_lock}
in {\tt glibc}).
In this paper, we consider a different approach
where fences are inserted to restore correctness.
This means that we may insert sufficiently many fences
to achieve program correctness
without needing to eliminate all its races or non-SC behaviors.
The challenge then is to
find sets of fences that guarantee program correctness
without compromising efficiency.
Manual fence placement is time-consuming and error-prone
due to the complex behaviors
of multithreaded programs
\cite{Herlihy:2008:AMP:1734069}.
Thus, we would like to provide the programmer
with a tool for {\it automatic} fence placement.
There are several
requirements to be met in the design of fence insertion
algorithms.
First, a set of fences should be {\it sound}, i.e., it should have
enough fences to enforce a sufficiently ordered behavior
for the program to be correct.
Second, the set should be {\it optimal},
in the sense that it 
has a lowest total cost among all sound sets
of fences.
We define the cost to be the number of occurrences of
fences in the source program, with the costs of different
fences taken into account.
In general, there may exist several different optimal sets
of fences for the same program.
Our experiments (Section~\ref{sec:results}) show that different
choices of sound fence sets may impact performance and
network traffic.

To carry out fence insertion
we need to be able to perform {\it program verification},
i.e., to check correctness of the program with a given set of fences.
This is necessary in order to be able to decide whether the  set
of fences is sound, or whether additional fences are
needed to ensure correctness.
A critical task in the design of
formal verification algorithms is to define
the  program semantics under
the given memory model.

\subsection*{Our Approach.}
We present a method for automatic fence
insertion in programs running in the presence
of {\sc SiSd}.
The method is applicable to a large class of self-invalidation and
self-downgrade protocols such as the ones in
\cite{lebeck95,choi11,kaxiras11,aros-pact12,sung13,skaxiras-isca13,hower14,aros-hpca15,sung15,aros-isca15,mdavari-pact15,kkoukos-taco16}.
Our goal is to eliminate incorrect behaviors that occur due to
the memory model induced by {\sc SiSd}.
We will not concern ourselves with other sources of consistency
relaxation, such as compiler optimizations.
We formulate the correctness of programs as {\it safety properties}.
A safety property is an assertion that some specified ``erroneous'', or
``bad'', program states can never occur during execution.
Such bad states may include e.g., states where a programmer-specified
assert statement fails, or where uninitialized data is read.
To check a safety property, we check the reachability of the set of ``bad'' states.

We provide an algorithm for checking the reachability of a
set of  bad states for a given  program running under {\sc SiSd}.
In the case that such states are reachable, our algorithm provides
a counter-example (i.e., an execution of the program that leads to one of
the bad states).
This counter-example is used by our fence insertion procedure to add
fences in order to remove the counter-examples introduced by {\sc
  SiSd} semantics.
Thus, we get a counter-example guided procedure for inferring the optimal
sets of fences.
The termination of the obtained procedure is guaranteed
under the assumption that  each call to the reachability algorithm terminates.
As a  special case, our tool   detects when
a program behaves incorrectly
already under SC.
Notice that in such a case,
the program cannot be corrected by inserting any set of fences.

\subsection*{Contributions.}
We make the following main contributions:
\begin{itemize}
\item We define a novel formal model that captures the semantics of programs
running under {\sc SiSd},
and employs a set of fences that interact with the coherence layer.
The semantics support the essential features of typical assembly code.
\item We develop a tool, \textsc{Memorax}, available at
\url{https://github.com/memorax/memorax}, that we have run
successfully on a wide range of examples
under {\sc SiSd} and under {\sc Si}.
Notably, our tool detects for the first time four
bugs in programs in the \mbox{Splash-2}
benchmark suite~\cite{splash95}, which have been fixed in a recent Splash-3 
release~\cite{sakalis16}.
Two of these are present even under  SC, while the other two
arise under  {\sc SiSd}.
We employ the tool to infer fences
of different kinds  and evaluate the relative performance of the fence-augmented
programs by simulation in GEMS.
\end{itemize}

We augment the semantics with a
reachability analysis algorithm that can check whether
a program satisfies a given safety property
with the current set of fences.
Inspired by an algorithm in \cite{LiuNPVY12}
(which uses dynamic analysis instead of verification as backend), we describe
a counter-example guided fence insertion
procedure that automatically
infers the optimal sets of fences necessary for the correctness of
the program.
The procedure relies on the counter-examples
provided by the reachability algorithm in order to
refine the set of fences.
 One feature of our  method is that it can
 handle different types of fences with different costs.
 This diversity makes optimization more difficult
 since one has to optimize the total cost of the inserted fences,
 rather than just their number.
Upon termination, the procedure will return all optimal
sets of fences.

\subsection*{Related Work.}
Adve and Hill proposed SC-for-DRF as a contract between software and hardware: If the software is data race free,
the hardware behaves as sequentially consistent~\cite{adve90}.
Dynamic self-invalidation (for DRF programs) was first proposed by Lebeck and Wood \cite{lebeck95}. Several recent works employ
self-invalidation to simplify coherence,
including SARC coherence~\cite{kaxiras11}, DeNovo~\cite{choi11,sung13,sung15}, and VIPS-M~\cite{aros-pact12,skaxiras-isca13,aros-hpca15,aros-isca15,kkoukos-taco16}.

A number of techniques for automatic fence insertion have been
proposed, for different memory models and with different
approaches. However, to our knowledge, we propose the first counter-example guided fence insertion procedure in the presence of a variety of fences (with different costs).
In our previous work~\cite{abdulla2012counter}, we propose counter-example guided fence insertion for programs under TSO with respect to
safety properties (also implemented in \textsc{Memorax}). Considering  the  {\sc SiSd} model makes  the problem significantly  more difficult. TSO offers only one fence, whereas the {\sc SiSd} model offers a variety of fences with different costs. This diversity makes the optimization more difficult since one has to minimize the total cost of the fences rather than just their number.

The work  presented in \cite{kuperstein2010automatic} proposes an insertion procedure for  different memory models w.r.t.
safety properties. This procedure computes the set of needed fences in order to not reach each   state in the transition graph. Furthermore, this procedure assigns a unique  cost for all fences. The procedure is not counter-example based, and requires some modification to the reachability procedure.

In~\cite{bouajjani2013checking}, the tool \textsc{Trencher} is
introduced, which inserts fences under TSO to enforce robustness  (formalised by  Shasha and Snir in \cite{shasha1988efficient}), also
using an exact, model-checking based technique.
\textsc{Musketeer}~\cite{AlglaveKNP14} uses static analysis
to efficiently overapproximate the fences necessary to enforce
robustness under several different memory models.
In contrast to our work,
the fence insertion procedures in \cite{bouajjani2013checking} and \cite{AlglaveKNP14} first enumerate  all solutions and then use linear programming to find the optimal set of fences.

In~\cite{guerraouiHS11} and~\cite{guerraouiHS09}, the authors discuss fence insertion and verification of
satisfiability modulo theory (SMT) of relaxed memory models under transactional memories.
Our work, in comparison, addresses the reachability problem.

The program semantics under
{\sc SiSd} is different from those under other
weak memory models (e.g. TSO and POWER). Hence existing techniques
cannot be directly applied. To our knowledge, ours is the first work 
that defines the {\sc SiSd} model, proposes a 
reachability analysis and describes a fence insertion procedure under {\sc SiSd}.

There exist works on the verification
of cache coherence protocols.
This paper is orthogonal to these works since
we are concerned with verification of
{\it programs} running on such architectures
and not the protocols themselves.

%%% Local Variables:
%%% mode: latex
%%% TeX-master: "main.tex"
%%% End:

\section{Self-Invalidation, Self-Downgrade, and their Fences}
\label{sec:sisd}

In this section, we recall the notions of self-invalidation and
self-downgrade, and describe the main features of the system architecture and
the protocol we consider. We also introduce two fences that are defined under 
the protocol.

\subsection{Self-Invalidation and Self-Downgrade}

Self-invalidation eliminates the need to track sharers of a cache line in a
directory structure \cite{lebeck95}. We consider that invalidation of shared data in caches
is caused by fences inserted in the programs and not as a result
of writes from other cores. 

Correspondingly, self-downgrade
eliminates the need to track the last writer (i.e., the owner, in a
{\sc Moesi}-like protocol) of a cache line \cite{aros-pact12}. 
This is because downgrades are also not
performed as a consequence of read operations, but by means of
fence instructions inserted in the programs. 

A protocol
that implements self-invalidation together with self-downgrade does not
need a directory, thus removing one of the main sources of complexity and 
scalability constraints in traditional cache coherence protocols \cite{aros-pact12}.

We first set the stage for the architecture and the coherence protocol
we study in this work, by discussing some of their details: i.e., 
how memory accesses are resolved, and how the
self-invalidation and the self-downgrade are performed when a fence is
encountered.

\paragraph{System architecture}

We assume a standard multicore architecture with a number of cores, each with a private 
L1 cache.
The proposals and algorithms described in this paper
are more widely applicable to systems with several levels of private
caches. The last level cache (LLC) of the system is logically shared
among all the cores.

\subsection{Cache coherence protocol}

We also assume a very simple version of a self-invalidation/self-downgrade protocol 
with only three stable states in the L1 cache
(invalid --I--, clean --C--, and dirty --D--) and only two stable states in
the LLC (invalid --I-- and valid --V--).
There are no invalidations or downgrades, which means that there are
no transient states to account for the arrival of such coherence actions.
There are no requests other than from the L1s to the LLC 
(and from LLC to memory). There is no distinction of data into private 
or shared as in \cite{aros-pact12}, as this would distract from our 
discussion. Such optimizations are straightforward extensions in our
approach.

\paragraph{Basic actions:}

To connect with the formal specification of the system behavior that
follows in Section~\ref{sec:model} we present here some necessary
---if somewhat mundane--- details of the basic actions in our assumed
system. %You can refer to these descriptions as explanations to help in
%the understanding of the formal specifications:

\begin{itemize}
\item
A \textit{read} request that misses in the L1 cache issues a request to
the LLC. If it hits in the LLC, a reply containing the data is
sent. In case of a miss, main memory is accessed to get the data
block. When the data arrives to the L1 cache, the miss is resolved and
the data can be accessed. The block is stored in an L1 cache line in clean
state.
\item
A write request is always resolved immediately, even if the block
is not present in the L1 (in this case, the miss status handling register --MSHR-- can temporarily hold the new data). 
This is because writes are assumed to be data-race-free, i.e,
they are always ordered with respect to conflicting reads \cite{adve90}. 
In this case, writes do not require ``write permission''. 
\item
After writing in an L1 cache line (e.g., one word), if the data block is
missing it is fetched from the LLC. The block is merged with the written
word. Before merging the data, the cache line is in a transient state
and once merged transitions to dirty.
\item
An atomic read-modify-write (RMW) request (e.g., test-and-set --TAS--) 
needs to reach the LLC, get the data block and
send it back to the L1 cache. During this operation, the corresponding 
cache line in the LLC is blocked, so no other RMW request can
proceed. 
When the data arrives at the L1 cache it is read and possibly modified.
If modified, the data is written-back in the LLC,
unblocking the LLC line at the same time. Otherwise the LLC line will
be simply unblocked. This blocking operation
---common in other protocols for directory operations that 
generate new messages (indirection, invalidations, etc.)--- is only
necessary in this protocol for RMW requests. Once the transaction
finishes, the data block remains in the L1 in clean state.
\item
Evictions of clean cache lines only require a change of state to
invalid. However, evictions of dirty cache lines need to write back to the
LLC the data that have been modified locally. 
This is necessary to avoid
overwriting unrelated data in the LLC cache line 
(a different part of the LLC line may have been modified 
independently without a conflict).
When modified data are written-back an acknowledgement message 
is sent to the L1 to signal the completion of the corresponding writes. 
\item
To keep track of the locally modified data in an L1 cache line, 
it is necessary to keep information in the form of a dirty bit per word (byte), 
either with the L1 cache lines \cite{choi11}, or 
in the write-buffer or MSHRs \cite{aros-pact12}.
\end{itemize}

\subsection{Self-Invalidation and Self-Downgrade fences}

Since the described protocol has neither invalidation on writes nor
downgrades on reads, we need to ensure that a read operation sees the
latest value written, when this is intended by the program.
Typically, the program contains synchronization 
to enforce an order between conflicting writes and reads. 

In this context, to ensure that a read gets the latest value of a corresponding write, two
things need to happen: first, the data in the writer's cache must be self-downgraded 
and put back in the LLC sometime after the write but before the read; 
second, if there is a (stale) copy in the reader's cache, it 
must be self-invalidated sometime before the read. The self-downgrade and self-invalidation 
also need to be ordered the same as the write and read are ordered by synchronization.

Prior proposals \cite{lebeck95,kaxiras11,choi11,sung13,ashby11,aros-pact12,skaxiras-isca13}  
invariably offer SC for data-race-free (DRF) programs \cite{adve90}.
In general, such proposals can be thought of as employing a single fence causing
the self-invalidation and self-downgrade of cached data,
\emph{on every synchronization in the DRF program} (e.g., \cite{aros-pact12}.) 

Our approach is fundamentally different. We make no
assumption as to what constitutes synchronization (perhaps ordinary accesses relying on SC semantics,
or algorithms involving atomic RMW operations). 
We insert \emph{fences} in a program to cause self invalidation and
self downgrade in such a way as to produce the desired behavior.

With only a single fence, ensuring that a read sees the latest value of a write causes
the self-invalidation \emph{and} self-downgrade of \emph{both} the reader's and the writer's
cache. In many cases, this is unnecessary.
 
One of the contributions of our work is to propose two separate  
fences, which we call ``load-load fence'' (\llfence{}) and ``store-store fence'' (\ssfence{}) to address the above problem. 
An \llfence{} self-invalidates only \emph{clean} data in the cache (at word level),
while an \ssfence{} writes back only \emph{dirty} data to the LLC (again at word level),
so all the data in the L1 cache of the process are clean.

The separation of self-invalidation and self-downgrade into two fences
affects performance in two ways: first, we reduce the fence latency when 
we do not have to self-downgrade; second, we eliminate extraneous 
misses (that cost in performance) when we only need to self-downgrade.

\subsection{Improving self-invalidation of partially dirty cache lines: the DoI state}\label{sec:doi}

The \llfence{} defined above operates efficiently for cache lines that
are entirely clean, or entirely dirty. In particular, for clean data
they take a single cycle, while for dirty data they do not perform any
action. However, cache lines that contain both clean and dirty words
are not self-invalidated efficiently. Consider, for example, a cache
line with one clean word and one dirty word (its dirty bit is set).
The \llfence{} must invalidate the clean word (if it were to be
accessed afterwards), without affecting the dirty word. If we
invalidate the whole cache line we also have to write the dirty data
to the LLC. This would have the same impact (for this cache line) as a
single full fence, and would offer no advantage from using the
\llfence{} instead.

In order to improve the efficiency of \llfence{} operations we propose that they operate at word granularity, being able to self-invalidate the clean words in a single cycle and leaving untouched the dirty words. Thus, we introduce a new state for L1 cache lines, 
called \textit{DoI} (dirty or invalid), for exactly this purpose. 
A cache line in this state contains words that are either dirty (with the dirty bit set) 
or invalid (with the dirt bit unset).
An \llfence{} transitions any partially dirty cache line to the \textit{DoI} state.
No write back is performed for its dirty words. 
This allows an efficient, one-cycle implementation of \llfence{}, since now the only necessary action for a \llfence{} is to change the cache-line state from dirty to \textit{DoI}.

%%% Local Variables:
%%% mode: latex
%%% TeX-master: "main.tex"
%%% End:

\section{Overview}
\label{overview:section}
In this section, we give an informal overview of the main concepts in
our framework. 
We describe
the semantics (configurations and runs) of programs running under {\sc SiSd},
the notion of safety properties,
the weak memory model induced by {\sc SiSd},
the roles of fences, and
optimal sets of fences.
This will be formalized in later sections.

\begin{figure} \begin{center}

\begin{tikzpicture}
[background rectangle/.style={rounded corners,fill=green!5,line width=1pt,draw},show background rectangle]

\node[name=dummy]{};

\node[pgmtextnode,name=data,scale=1.1] at (dummy)
{\tt data x=0 y=0 z=0};

%%%%%%%%%%%%%%% Proc 0 %%%%%%%%%%%%%%%

\node[pgmtextnode,name=p0,anchor=north west,scale=1.1] at ($(data.south west)+(0pt,-14pt)$)
{\tt process P0};

\node[pgmtextnode,name=r0,anchor=north west,scale=1.1] at ($(p0.south west)+(0pt,-2pt)$)
{\tt registers \$r0};

\node[pgmtextnode,name=b0,anchor=north west,scale=1.1] at ($(r0.south west)+(0pt,-2pt)$)
{\tt begin};

\node[pgmtextnode,name=l1,anchor=north west,scale=1.1] at ($(b0.south west)+(2pt,-2pt)$)
{\tt L1:  \vphantom{y\$}x := 1;};

\node[pgmtextnode,name=l2,anchor=north west,scale=1.1] at ($(l1.south west)+(0pt,-2pt)$)
{\tt L2:  \vphantom{y\$}y := 1;};

\node[pgmtextnode,name=l3,anchor=north west,scale=1.1] at ($(l2.south west)+(0pt,-2pt)$)
{\tt L3:  \vphantom{y\$}\$r0 := z;};

\node[pgmtextnode,name=e0,anchor=north west,scale=1.1] at ($(l3.south west)+(-2pt,-2pt)$)
{\tt end};

\begin{pgfonlayer}{foreground}
\node[fitnode,dotted,fill=red!5,fit= (b0) (p0) (r0) (l1) (l2)  (l3) (e0)]{};
\end{pgfonlayer}

%%%%%%%%%%%%%%% Proc 1 %%%%%%%%%%%%%%%

\node[pgmtextnode,name=p1,anchor=west,scale=1.1] at ($(p0.east)+(32pt,0pt)$)
{\tt process P1};

\node[pgmtextnode,name=r1a,anchor=north west,scale=1.1] at ($(p1.south west)+(0pt,-2pt)$)
{\tt registers \$r1};
\node[pgmtextnode,name=r1b,anchor=north west,scale=1.1] at ($(r1a.south west)+(0pt,-2pt)$)
{\tt \$r2 \$r3};

\node[pgmtextnode,name=b1,anchor=north west,scale=1.1] at ($(r1b.south west)+(0pt,-2pt)$)
{\tt begin};

\node[pgmtextnode,name=l4,anchor=north west,scale=1.1] at ($(b1.south west)+(2pt,-2pt)$)
{\tt L4:  z := 1;};

\node[pgmtextnode,name=l5,anchor=north west,scale=1.1] at ($(l4.south west)+(0pt,-2pt)$)
{L5:  \$r1 := x;};

\node[pgmtextnode,name=l6,anchor=north west,scale=1.1] at ($(l5.south west)+(0pt,-2pt)$)
{\tt L6:  \$r2 := y;};

\node[pgmtextnode,name=l7,anchor=north west,scale=1.1] at ($(l6.south west)+(0pt,-2pt)$)
{\tt L7:  \$r3 := x;};

\node[pgmtextnode,name=e1,anchor=north west,scale=1.1] at ($(l7.south west)+(-2pt,-2pt)$)
{\tt end};

\begin{pgfonlayer}{foreground}
\node[fitnode,dotted,fill=blue!5,fit= (b1) (p1) (r1a) (r1b) (l4) (l5)  (l6) (l7) (e1)]{};
\end{pgfonlayer}
\end{tikzpicture}
\caption{A simple program $\prog$.}
\label{running:fig}
\end{center}\end{figure}

\paragraph{Example.}

We will use the toy program $\prog$ in Figure~\ref{running:fig}
as a running example.
The program is written in a simple assembly-like
programming language.
The syntax and semantics of the language are formally  defined
in Section~\ref{sec:model}.
$\prog$ consists of two processes
{\tt P0} and {\tt P1} that share three variables
{\tt x}, {\tt y}, and {\tt z}.
Process {\tt P0} has one register {\tt \$r0}, and
process {\tt P1} has three registers {\tt \$r1},
{\tt \$r2}, and {\tt \$r3}.
Process {\tt P0} has three instructions labeled with
{\tt L1}, {\tt L2}, {\tt L3}, and
process {\tt P1} has four instructions labeled with
{\tt L4}, {\tt L5}, {\tt L6}, {\tt L7}.

To simplify the presentation, we assume that each cache line
holds only one variable.
We also assume that the underlying protocol
contains both {\sc Si} and {\sc Sd}.
It is straightforward to extend our framework
to the case where a cache line may hold several variables, and to the case
where the protocol only contains one of {\sc Si} and {\sc Sd}.
In $\prog$, all the instructions have unique labels.
Therefore, to simplify the presentation, we identify
each label with the corresponding instruction, e.g.,
{\tt L1} and {\tt x := 1} in {\tt P0}.

\paragraph{Configurations.}

%\begin{figure} \begin{center}
%
%\confpicarrow{$\conf_0$}{x=0\\[-3pt]y=0\\[-3pt]z=0}
%{L1\\[-3pt] \$r0=0}{L4 \$r1=0\\[-3pt]\$r2=0 \$r3=0}
%
%\confpic{$\conf_1$}{x=0\\[-3pt]y=0\\[-3pt]z=0}
%{L3\\[-3pt]\$r0=0\\[-3pt]\underline{x=1} \underline{y=1}}
%{L5 \$r1=0\\[-3pt]\$r2=0 \$r3=0\\[-3pt] \underline{z=1}}
%
%\confpic{$\conf_2$}{x=1\\[-3pt]y=1\\[-3pt]z=0}
%{L3\\[-3pt]\$r0=0}
%{L6 \$r1=0\\[-3pt]\$r2=0 \$r3=0\\[-3pt] x=0 \underline{z=1}}
%
%\confpic{$\conf_3$}{x=1\\[-3pt]y=1\\[-3pt]z=0}
%{L3\\[-3pt]\$r0=0}
%{{\tt end} \$r1=0\\[-3pt]\$r2=1 \$r3=0\\[-3pt] x=0 y=1 \underline{z=1} }
%
%\confpic{$\conf_4$}{x=0\\[-3pt]y=1\\[-3pt]z=0}
%{L3\\[-3pt]\$r0=0\\[-3pt]\underline{x=1}}
%{{\tt end} \$r1=0\\[-3pt]\$r2=1 \$r3=0\\[-3pt] x=0 \underline{z=1} }
%
%\confpic{$\conf_5$}{x=1\\[-3pt]y=0\\[-3pt]z=0}
%{end \\[-3pt]  \$r0=0 \\[-3pt] \underline{y=1} z=0}
%{end \$r1=0\\[-3pt]\$r2=0 \$r3=0\\[-3pt] x=0 \underline{z=1} }
%
%\caption{Configurations.}
%
%\label{confs:fig}
%\end{center}\end{figure}

\begin{figure*}[h]
\centering

\begin{minipage}[]{0.51\linewidth}
  \centering
\confpicarrow{$\conf_0$}{x=0\\[-3pt]y=0\\[-3pt]z=0}
{L1\\[-3pt] \$r0=0}{L4 \$r1=0\\[-3pt]\$r2=0 \$r3=0}

\confpic{$\conf_1$}{x=0\\[-3pt]y=0\\[-3pt]z=0}
{L3\\[-3pt]\$r0=0\\[-3pt]\underline{x=1} \underline{y=1}}
{L5 \$r1=0\\[-3pt]\$r2=0 \$r3=0\\[-3pt] \underline{z=1}}

\confpic{$\conf_2$}{x=1\\[-3pt]y=1\\[-3pt]z=0}
{L3\\[-3pt]\$r0=0}
{L6 \$r1=0\\[-3pt]\$r2=0 \$r3=0\\[-3pt] x=0 \underline{z=1}}

\confpic{$\conf_3$}{x=1\\[-3pt]y=1\\[-3pt]z=0}
{L3\\[-3pt]\$r0=0}
{{\tt end} \$r1=0\\[-3pt]\$r2=1 \$r3=0\\[-3pt] x=0 y=1 \underline{z=1} }
\end{minipage}
\begin{minipage}[]{0.47\linewidth}
  \centering
\confpicarrow{$\conf_4$}{x=0\\[-3pt]y=1\\[-3pt]z=0}
{L3\\[-3pt]\$r0=0\\[-3pt]\underline{x=1}}
{{\tt end} \$r1=0\\[-3pt]\$r2=1 \$r3=0\\[-3pt] x=0 \underline{z=1} }

\confpic{$\conf_5$}{x=1\\[-3pt]y=0\\[-3pt]z=0}
{end \\[-3pt]  \$r0=0 \\[-3pt] \underline{y=1} z=0}
{end \$r1=0\\[-3pt]\$r2=0 \$r3=0\\[-3pt] x=0 \underline{z=1} }
\end{minipage}
\caption{Configurations}
\label{confs:fig}
\end{figure*}

A {\it configuration}  is a snapshot of the
global state of the system,  and consists
of two parts, namely the {\it local} and {\it shared} parts.
The local part defines the local states of the processes, i.e.,
it defines for each process:
(i) its next instruction to be executed,
(ii) the values stored in its registers, and
(iii) the variables (memory locations)
 that are currently cached in its L1,
together with their status:
{\it invalid},
{\it clean}, or {\it dirty};
and the current value of the variable in case it is valid.
The shared part  defines, for each
variable, its value in the LLC.
Figure~\ref{confs:fig} shows different configurations
of $\prog$.
Each configuration is depicted as three fields,
representing the LLC, {\tt P0}, and {\tt P1} respectively.
$\prog$ starts its execution from the {\it initial configuration}
$\conf_0$, where the values  of all variables are  $0$ in the LLC.
{\tt P0} and {\tt P1} are about to execute the instructions labeled
{\tt L1} and {\tt L4}, respectively.
The values of all registers are $0$.
None of the variables is valid in the L1 of the processes.
In contrast, in $\conf_4$, the value of {\tt y}
is  $1$ in the LLC.
{\tt P0} is about to execute the instruction
{\tt L3}, while {\tt P1} has ended its execution.
The value of the register {\tt \$r2} is $1$.
The variables {\tt x},  and {\tt z} are valid in the L1
of {\tt P1}, with values $0$ and $1$ respectively.
The variable {\tt z} is dirty in the L1 of
 {\tt P1} (marked by underlining {\tt z=1}),
while {\tt x} is clean (not underlined).
Finally, there is a dirty copy of {\tt x} with value $1$ in {\tt P0}.

\paragraph{Safety Properties.}
Suppose that, together with the program $\prog$,
we are given a {\it safety property} $\phi$
which states that a certain
set $\bad$ of configurations will not occur
during any execution of $\prog$.
For the example, we assume that $\bad$ is the set of configurations
where (i) {\tt P1} has ended its execution, and (ii) the registers
{\tt \$r2} and {\tt \$r3} have values $1$ and $0$ respectively.
For instance, $\conf_3$ and $\conf_4$ are members of $\bad$.
We are interested in checking whether $\prog$ satisfies $\phi$
under the {\sc SiSd} semantics.
Note that the set $\bad$ is not reachable in $\prog$ under SC semantics,
which also means that $\prog$ satisfies $\phi$ under SC.

\paragraph{Runs.}
The semantics of a program boils down to defining a
{\it transition relation} on the set of configurations.
The execution of the program can be viewed as a {\it run},
consisting
of a sequence of {\it transitions}, i.e.,
events that take the program from one configuration
to another by changing the local states of the processes
and the shared parts.
Such a transition will either be performed by a given process
when it executes an instruction,
or it occurs due to a system event.
We consider three kinds of system events: $\ifetch$, $\ievict$, and
$\iwrllc$.
They model respectively fetching
a value from LLC to L1, invalidating an L1 entry, and writing a dirty
L1 entry through to the LLC. The system events are decoupled from
program instructions and execute independently. 

\begin{figure} [h] \begin{center}
\begin{tikzpicture}
[background rectangle/.style={rounded corners,fill=green!5,line width=1pt,draw},show background rectangle]

\node[confrunnode,name=c0,scale=1.5]{$\conf_0$};
\node[runnode,name=c01,anchor=west] at ($(c0.east)+(20pt,0pt)$) {};
\node[confrunnode,name=rrun1,anchor=north,scale=1.2, text=red]
at ($(c01.south)+(5.5pt,-2pt)$) {$\rrun_1$};
\node[runnode,name=c02,anchor=west] at ($(c01.east)+(20pt,0pt)$) {};
\node[confrunnode,name=c1,anchor=west,scale=1.5] at ($(c02.east)+(20pt,0pt)$) {$\conf_1$};

\draw[runedge] (c0) to node[above,lblnode,name=l4,scale=1.2]{L4*} (c01);
\draw[runedge] (c01) to node[above,lblnode,name=l1,scale=1.2]{L1*} (c02);
\draw[runedge] (c02) to node[above,lblnode,name=l2,scale=1.2]{L2*} (c1);

\begin{pgfonlayer}{foreground}
\node[fitnode,dotted,fill=blue!5,fit= (l4) (l1) (l2) (rrun1),inner xsep=-3pt,inner ysep=1pt]{};
\end{pgfonlayer}

\node[runnode,name=c11,anchor=west] at ($(c1.east)+(20pt,0pt)$) {};
\node[runnode,name=c12,anchor=west] at ($(c11.east)+(50pt,0pt)$) {};
\node[confrunnode,name=rrun2,anchor=north,scale=1.2,text=red]
at ($(c12.south)+(0pt,-2pt)$) {$\rrun_2$};
\node[confrunnode,name=c2,anchor=west,scale=1.5] at ($(c12.east)+(50pt,0pt)$) {$\conf_2$};

\draw[runedge] (c1) to node[above,lblnode,name=l5,scale=1.2]{L5*} (c11);
\draw[runedge] (c11) to node[scale=0.9,above,lblnode,name=pevictx,scale=1.2]{evict*(P0,x) } (c12);
\draw[runedge] (c12) to node[scale=0.9,above,lblnode,name=pevicty,scale=1.2]{evict*(P0,y) } (c2);

\begin{pgfonlayer}{foreground}
\node[fitnode,dotted,fill=blue!5,fit= (l5) (pevictx) (pevicty) (rrun2),inner xsep=-3pt,inner ysep=1pt]{};
\end{pgfonlayer}

\node[runnode,name=c21,anchor=west] at ($(c2.east)+(20pt,0pt)$) {};
\node[confrunnode,name=rrun3,anchor=north,scale=1.2,text=red]
at ($(c21.south)+(0pt,-2pt)$) {$\rrun_3$};

\node[confrunnode,name=c3,anchor=west,scale=1.5] at ($(c21.east)+(15pt,0pt)$) {$\conf_3$};

\draw[runedge] (c2) to node[above,lblnode,name=l6,scale=1.2]{L6*} (c21);
\draw[runedge] (c21) to node[above,lblnode,name=l7,scale=1.2]{L7} (c3);

\begin{pgfonlayer}{foreground}
\node[fitnode,dotted,fill=blue!5,fit= (l6) (l7) (rrun3),inner xsep=-3pt,inner ysep=1pt]{};
\end{pgfonlayer}

\end{tikzpicture}
\caption{The run $\run_1$.}\label{run1:fig}
\end{center}\end{figure} 

In Figure~\ref{run1:fig} we show one example run $\run_1$
 of $\prog$.
It consists of three sequences $\rrun_1$, $\rrun_2$, $\rrun_3$
of transitions,
and takes us from $\conf_0$ through
$\conf_1$  and $\conf_2$ to $\conf_3$.
$\rrun_1$:
Starting from $\conf_0$, {\tt P1} executes {\tt L4}.
Since {\tt z} is invalid in the L1 of {\tt P1}, it is fetched
from the LLC.
In Figure~\ref{run1:fig}, the star in {\tt L4*} is to simplify
the notation, and it indicates that
the instruction {\tt L4} is preceded by {\tt fetch}
event\footnote{In the examples of this section, {\tt fetch}
events always precede read or write events.
In general, {\tt fetch} events may occur anywhere along the run.}
of
the process (here {\tt P1}) on the relevant variable (here {\tt z}).
Consequently,
a dirty copy of {\tt z} with value $1$ is stored in
the L1 of {\tt P1}.
Next, {\tt P0} executes {\tt L1} and {\tt L2},
putting dirty copies (with values $1$) of {\tt x} and {\tt y}
in its L1, reaching the configuration $\conf_1$.
$\rrun_2$:
{\tt P1} executes {\tt L5}, fetching {\tt x} from the LLC,
and storing a clean copy with value $0$.
{\tt P0} evicts the variable {\tt x}.
An {\tt evict} event may only be performed
on clean variables.
To simplify the notation, we augment
the {\tt evict} event in Figure~\ref{run1:fig}
by a star.
This indicates that it is preceded by
an {\tt wrllc} event on {\tt x}.
The latter updates the value of {\tt x} to $1$
in the LLC, and makes {\tt x} clean in the L1
of {\tt P0}.
Next, {\tt P0} evicts {\tt y} in a similar manner,
thus reaching $\conf_2$.
$\rrun_3$:
{\tt P1} executes {\tt L6}.
Since {\tt y} is invalid in the L1 of {\tt P1}, it is fetched from
the LLC, and stored with value $1$ as clean.
The register {\tt \$r2} will be assigned the value $1$.
Finally, {\tt P1} executes {\tt L7}.
Since {\tt x} is valid, it need not be fetched from the memory ({\tt
  L7} is therefore not starred in Figure~\ref{run1:fig}), and hence
{\tt \$r3} is assigned the value $0$. Thus we reach $\conf_3$, which is
in the set $\bad$.
And so $\prog$ violates the safety property $\phi$
under the {\sc SiSd} semantics.

\paragraph{Weak Memory Model.}
Although the configuration $\conf_3$ is not reachable from $\conf_0$
under SC semantics, we demonstrated above that it is reachable
under {\sc SiSd} semantics.
The reason is that {\sc SiSd} introduces a weak memory semantics
in the form of {\it reorderings} of events.
In the case of $\run_1$, we have a {\it read-read} reordering.
More precisely, the read event {\tt L7} overtakes
the read event {\tt L6}, in the sense that
{\tt L6} is issued before {\tt L7}, but the value
assigned to {\tt\$r2} in {\tt L6} (coming from the write on {\tt y} in
{\tt L2})
is more recent than the value assigned
to {\tt\$r3} in {\tt L7} (which is the initial value of {\tt x}).
To prevent event reorderings, we use {\it fences}.
In this paper, we use four types of fences, namely
$\llfence$, $\ssfence$, $\fence$, and $\plainsyncwr$.
In this section, we only describe the first three types.

\begin{figure} [h] \begin{center}

\begin{tikzpicture}
[background rectangle/.style={rounded corners,fill=green!5,line width=1pt,draw},show background rectangle]

\node[name=dummy]{};

\node[pgmtextnode,name=data,scale=1.1] at (dummy)
{\tt data x=0 y=0 z=0};

%%%%%%%%%%%%%%% Proc 0 %%%%%%%%%%%%%%%

\node[pgmtextnode,name=p0,anchor=north west,scale=1.1] at ($(data.south west)+(0pt,-14pt)$)
{\tt process P0};

\node[pgmtextnode,name=r0,anchor=north west,scale=1.1] at ($(p0.south west)+(0pt,-2pt)$)
{\tt registers \$r0};

\node[pgmtextnode,name=b0,anchor=north west,scale=1.1] at ($(r0.south west)+(0pt,-2pt)$)
{\tt begin};

\node[pgmtextnode,name=l1,anchor=north west,scale=1.1] at ($(b0.south west)+(2pt,-2pt)$)
{\tt L1:  \vphantom{y\$}x := 1;};

\node[pgmtextnode,name=l2,anchor=north west,scale=1.1] at ($(l1.south west)+(0pt,-2pt)$)
{\tt L2:  \vphantom{y\$}y := 1;};

\node[pgmtextnode,name=l3,anchor=north west,scale=1.1] at ($(l2.south west)+(0pt,-2pt)$)
{\tt L3:  \vphantom{y\$}\$r0 := z;};

\node[pgmtextnode,name=e0,anchor=north west,scale=1.1] at ($(l3.south west)+(-2pt,-2pt)$)
{\tt end};

\begin{pgfonlayer}{foreground}
\node[fitnode,dotted,fill=red!5,fit= (b0) (p0) (r0) (l1) (l2)  (l3) (e0)]{};
\end{pgfonlayer}

%%%%%%%%%%%%%%% Proc 1 %%%%%%%%%%%%%%%

\node[pgmtextnode,name=p1,anchor=west,scale=1.1] at ($(p0.east)+(32pt,0pt)$)
{\tt process P1};

\node[pgmtextnode,name=r1a,anchor=north west,scale=1.1] at ($(p1.south west)+(0pt,-2pt)$)
{\tt registers \$r1};
\node[pgmtextnode,name=r1b,anchor=north west,scale=1.1] at ($(r1a.south west)+(0pt,-2pt)$)
{\tt \$r2 \$r3};

\node[pgmtextnode,name=b1,anchor=north west,scale=1.1] at ($(r1b.south west)+(0pt,-2pt)$)
{\tt begin};

\node[pgmtextnode,name=l4,anchor=north west,scale=1.1] at ($(b1.south west)+(2pt,-2pt)$)
{\tt L4:  z := 1;};

\node[pgmtextnode,name=l5,anchor=north west,scale=1.1] at ($(l4.south west)+(0pt,-2pt)$)
{L5:  \$r1 := x;};

\node[pgmtextnode,name=l6,anchor=north west,scale=1.1] at ($(l5.south west)+(0pt,-2pt)$)
{\tt L6:  \$r2 := y;};

\node[pgmtextnode,name=l8,anchor=north west,fill=red!20,scale=1.1] at ($(l6.south west)+(0pt,-2pt)$)
{\tt L8: llfence;};

\node[pgmtextnode,name=l7,anchor=north west,scale=1.1] at ($(l8.south west)+(0pt,-2pt)$)
{\tt L7:  \$r3 := x;};

\node[pgmtextnode,name=e1,anchor=north west,scale=1.1] at ($(l7.south west)+(-2pt,-2pt)$)
{\tt end};

\begin{pgfonlayer}{foreground}
\node[fitnode,dotted,fill=blue!5,fit= (b1) (p1) (r1a) (r1b) (l4) (l5)  (l6) (l7) (l8) (e1)]{};
\end{pgfonlayer}
\end{tikzpicture}
\caption{The program $\prog_1$.}
\label{p1:fig}
\end{center}\end{figure}

%\begin{figure} \begin{center}
%\begin{tikzpicture}
%[background rectangle/.style={rounded corners,fill=green!5,line width=1pt,draw},show background rectangle]
%\node[name=dummy]{};
%\node[pgmtextnode,name=p1,anchor=west,scale=1.1] at (dummy)
%{\tt process P1};
%\node[pgmtextnode,name=dots1,anchor=north west,scale=1.1] at ($(p1.south west)+(0pt,-4pt)$)
%{$\;\;\;\;\;\;\;\;\;\;\;\;\cdot$};
%\node[pgmtextnode,name=l6,anchor=north west,scale=1.1] at ($(dots1.south west)+(0pt,-4pt)$)
%{\tt L6:  \$r2 := y;};
%\node[pgmtextnode,name=l8,anchor=north west,fill=red!20,scale=1.1] at ($(l6.south west)+(0pt,-2pt)$)
%{\tt L8: llfence;};
%\node[pgmtextnode,name=l7,anchor=north west,scale=1.1] at ($(l8.south west)+(0pt,-2pt)$)
%{\tt L7:  \$r3 := x;};
%\node[pgmtextnode,name=dots2,anchor=north west,scale=1.1] at ($(l7.south west)+(0pt,-4pt)$)
%{$\;\;\;\;\;\;\;\;\;\;\;\;\cdot$};
%
%\begin{pgfonlayer}{foreground}
%\node[fitnode,dotted,fill=blue!5,fit= (p1) (dots1)  (l6) (l7) (dots2)]{};
%\end{pgfonlayer}
%\end{tikzpicture}
%\caption{{\tt P1} in $\prog_1$ and $\prog_2$.}
%\label{p1:fig}
%\end{center}\end{figure} 

\begin{figure} [h] \begin{center}
\begin{tikzpicture}
[background rectangle/.style={rounded corners,fill=green!5,line width=1pt,draw},show background rectangle]

\node[confrunnode,name=c0,scale=1.5]{$\conf_0$};
\node[runnode,name=c01,anchor=west] at ($(c0.east)+(20pt,0pt)$) {};
\draw[runedge] (c0) to node[above,lblnode,name=l1,scale=1.2]{L1*} (c01);
\node[runnode,name=c02,anchor=west] at ($(c01.east)+(20pt,0pt)$) {};
\draw[runedge] (c01) to node[above,lblnode,name=l2,scale=1.2]{L2*} (c02);
\node[runnode,name=c03,anchor=west] at ($(c02.east)+(50pt,0pt)$) {};
\draw[runedge] (c02) to node[scale=0.9,above,lblnode,name=p0evicty,scale=1.2]{evict*(P0,y)} (c03);
\node[runnode,name=c04,anchor=west] at ($(c03.east)+(20pt,0pt)$) {};
\draw[runedge] (c03) to node[above,lblnode,name=l4,scale=1.2]{L4*} (c04);
\node[runnode,name=c05,anchor=west] at ($(c04.east)+(20pt,0pt)$) {};
\draw[runedge] (c04) to node[lblnode,name=l5,scale=1.2]{L5*} (c05);
\node[runnode,name=c06,anchor=north] at ($(c05.south)+(0pt,-20pt)$) {};
\draw[runedge] (c05) to node[,sloped,right,lblnode,name=l6,scale=1.2]{L6*} (c06);
\node[runnode,name=c07,anchor=east] at ($(c06.west)+(-43pt,0pt)$) {};
\draw[runedge] (c06) to node[scale=0.9,above,lblnode,name=p0evictx,scale=1.2]{evict(P1,x)} (c07);
\node[runnode,name=c08,anchor=east] at ($(c07.west)+(-50pt,0pt)$) {};
\draw[runedge] (c07) to node[scale=0.9,above,lblnode,name=p0evicty,scale=1.2]{evict(P1,y)} (c08);
\node[runnode,name=c09,anchor=east] at ($(c08.west)+(-20pt,0pt)$) {};
\draw[runedge] (c08) to node[above,lblnode,name=l8,scale=1.2]{L8} (c09);
\node[confrunnode,name=c4,anchor=east,scale=1.5] at ($(c09.west)+(-20pt,0pt)$) {$\conf_4$};
\draw[runedge] (c09) to node[above,lblnode,name=l7,scale=1.2]{L7} (c4);
\end{tikzpicture}
\caption{The run $\run_2$.}
\label{run2:fig}
\end{center}\end{figure}

\paragraph{LL Fences.}
To forbid the run $\run_1$, we insert an
$\llfence$ between {\tt L6} and {\tt L7},
obtaining a new program $\prog_1$
(Figure~\ref{p1:fig}).
Intuitively, an $\llfence$ (load-load-fence)
blocks when there are clean
entries in the L1 of the process, and hence forbids
the reordering of two read (load) operations.
For instance, in the above example, {\tt L8} cannot be executed before
{\tt x} has become invalid in the {\tt L1} of
 {\tt P1}, and hence the new value $1$ of
{\tt x} will be assigned to {\tt \$r3} in {\tt L7}.
Therefore, $\prog_1$ does not contain the run $\run_1$ any more.

\paragraph{SS Fences.}
Despite the fact that the insertion of {\tt L8} eliminates the run $\run_1$,
the program $\prog_1$ still does not satisfy the safety
property $\phi$:
the set $\bad$ is still reachable from $\conf_0$ in $\prog_1$,
this time through a run $\run_2$ (Figure~\ref{run2:fig})
that leads to the configuration $\conf_4$ (which is also
a member of $\bad$).
In $\run_2$, {\tt P0} performs {\tt L1} and {\tt L2}
and then evicts only {\tt y}, which means that the values
of {\tt x} and {\tt y} in the LLC will be $0$ resp.\ $1$.
Now, {\tt P1}
will perform the instructions {\tt L4}, {\tt L5}, {\tt L6}.
Next {\tt P1}
evicts {\tt x}  and then {\tt y} which means that now {\tt P1}
does not contain any clean variables, and hence {\tt L8} is enabled.
Notice that these {\tt evict} events are not followed by stars
(since they concern clean copies of the variables).
Finally, {\tt P1} executes {\tt L7}.
Since {\tt x} is invalid in the {\tt L1} of
{\tt P0}, it is fetched from
the LLC (where its value is $0$ since it was never evicted by
{\tt P0}), and hence {\tt \$r3} will be assigned the value $0$.
Thus, we are now in $\conf_4$.

\begin{figure} \begin{center}

\begin{tikzpicture}
[background rectangle/.style={rounded corners,fill=green!5,line width=1pt,draw},show background rectangle]

\node[name=dummy]{};

\node[pgmtextnode,name=data,scale=1.1] at (dummy)
{\tt data x=0 y=0 z=0};

%%%%%%%%%%%%%%% Proc 0 %%%%%%%%%%%%%%%

\node[pgmtextnode,name=p0,anchor=north west,scale=1.1] at ($(data.south west)+(0pt,-14pt)$)
{\tt process P0};

\node[pgmtextnode,name=r0,anchor=north west,scale=1.1] at ($(p0.south west)+(0pt,-2pt)$)
{\tt registers \$r0};

\node[pgmtextnode,name=b0,anchor=north west,scale=1.1] at ($(r0.south west)+(0pt,-2pt)$)
{\tt begin};

\node[pgmtextnode,name=l1,anchor=north west,scale=1.1] at ($(b0.south west)+(2pt,-2pt)$)
{\tt L1:  \vphantom{y\$}x := 1;};

\node[pgmtextnode,name=l8,anchor=north west,fill=blue!20,scale=1.1] at ($(l1.south west)+(0pt,-2pt)$)
{\tt L9: ssfence;};

\node[pgmtextnode,name=l2,anchor=north west,scale=1.1] at ($(l8.south west)+(0pt,-2pt)$)
{\tt L2:  \vphantom{y\$}y := 1;};

\node[pgmtextnode,name=l3,anchor=north west,scale=1.1] at ($(l2.south west)+(0pt,-2pt)$)
{\tt L3:  \vphantom{y\$}\$r0 := z;};

\node[pgmtextnode,name=e0,anchor=north west,scale=1.1] at ($(l3.south west)+(-2pt,-2pt)$)
{\tt end};

\begin{pgfonlayer}{foreground}
\node[fitnode,dotted,fill=red!5,fit= (b0) (p0) (r0) (l1) (l2)  (l3) (l8) (e0)]{};
\end{pgfonlayer}

%%%%%%%%%%%%%%% Proc 1 %%%%%%%%%%%%%%%

\node[pgmtextnode,name=p1,anchor=west,scale=1.1] at ($(p0.east)+(32pt,0pt)$)
{\tt process P1};

\node[pgmtextnode,name=r1a,anchor=north west,scale=1.1] at ($(p1.south west)+(0pt,-2pt)$)
{\tt registers \$r1};
\node[pgmtextnode,name=r1b,anchor=north west,scale=1.1] at ($(r1a.south west)+(0pt,-2pt)$)
{\tt \$r2 \$r3};

\node[pgmtextnode,name=b1,anchor=north west,scale=1.1] at ($(r1b.south west)+(0pt,-2pt)$)
{\tt begin};

\node[pgmtextnode,name=l4,anchor=north west,scale=1.1] at ($(b1.south west)+(2pt,-2pt)$)
{\tt L4:  z := 1;};

\node[pgmtextnode,name=l5,anchor=north west,scale=1.1] at ($(l4.south west)+(0pt,-2pt)$)
{L5:  \$r1 := x;};

\node[pgmtextnode,name=l6,anchor=north west,scale=1.1] at ($(l5.south west)+(0pt,-2pt)$)
{\tt L6:  \$r2 := y;};

\node[pgmtextnode,name=l7,anchor=north west,scale=1.1] at ($(l6.south west)+(0pt,-2pt)$)
{\tt L7:  \$r3 := x;};

\node[pgmtextnode,name=e1,anchor=north west,scale=1.1] at ($(l7.south west)+(-2pt,-2pt)$)
{\tt end};

\begin{pgfonlayer}{foreground}
\node[fitnode,dotted,fill=blue!5,fit= (b1) (p1) (r1a) (r1b) (l4) (l5)  (l6) (l7) (e1)]{};
\end{pgfonlayer}
\end{tikzpicture}
\caption{The program $\prog_2$.}
\label{p2:fig}
\end{center}\end{figure}

%\begin{figure} \begin{center}
%\centering
%\begin{tikzpicture}
%[background rectangle/.style={rounded corners,fill=green!5,line width=1pt,draw},show background rectangle]
%
%\node[name=dummy]{};
%
%
%\node[pgmtextnode,name=p1,anchor=west,scale=1.1] at (dummy)
%{\tt process P0};
%
%\node[pgmtextnode,name=dots1,anchor=north west,scale=1.1] at ($(p1.south west)+(0pt,-4pt)$)
%{$\;\;\;\;\;\;\;\;\;\;\;\;\cdot$};
%
%\node[pgmtextnode,name=l6,anchor=north west,scale=1.1] at ($(dots1.south west)+(0pt,-4pt)$)
%{\tt L1:  x := 1;};
%
%\node[pgmtextnode,name=l8,anchor=north west,fill=blue!20,scale=1.1] at ($(l6.south west)+(0pt,-2pt)$)
%{\tt L9: ssfence;};
%
%\node[pgmtextnode,name=l7,anchor=north west,scale=1.1] at ($(l8.south west)+(0pt,-2pt)$)
%{\tt L2:  y := 1;};
%
%\node[pgmtextnode,name=dots2,anchor=north west,scale=1.1] at ($(l7.south west)+(0pt,-4pt)$)
%{$\;\;\;\;\;\;\;\;\;\;\;\;\cdot$};
%
%
%\begin{pgfonlayer}{foreground}
%\node[fitnode,dotted,fill=red!5,fit= (p1) (dots1)  (l6) (l8) (l7) (dots2)]{};
%\end{pgfonlayer}
%\end{tikzpicture}
%\caption{{\tt P0} in $\prog_2$.}
%\label{p2:fig}
%\end{center}\end{figure} 

%
%
Notice again that $\run_2$ is not possible under SC
(in the SC semantics, fences have no effect, so they are equivalent
to empty statements).
The reason why $\run_2$ is possible under {\sc SiSd} is due to a
{\it write-write} reordering.
More precisely, the write event {\tt L1}
is issued before the write event {\tt L2}, but
{\tt L2} takes effect (updates the LLC) before {\tt L1}.
To forbid the run $\run_2$, we insert an
$\ssfence$ between {\tt L1} and {\tt L2},
obtaining the program $\prog_2$
(Figure~\ref{p2:fig}).
An $\ssfence$ (store-store-fence) is only enabled when there are no dirty
entries in the L1  of the process. Hence it forbids
the reordering of two write operations.
For instance, in the above example, {\tt L9} cannot be executed before
{\tt x} has been evicted by  {\tt P0}, and hence the value of
{\tt x} in the LLC will be updated to $1$.

\begin{figure} \begin{center}

\begin{tikzpicture}
[background rectangle/.style={rounded corners,fill=green!5,line width=1pt,draw},show background rectangle]

\node[confrunnode,name=c0,scale=1.5]{$\conf_0$};
\node[runnode,name=c01,anchor=west] at ($(c0.east)+(20pt,0pt)$) {};
\draw[runedge] (c0) to node[above,lblnode,scale=1.2]{L4*} (c01);

\node[runnode,name=c02,anchor=west] at ($(c01.east)+(20pt,0pt)$) {};
\draw[runedge] (c01) to node[above,lblnode,scale=1.2]{L5*} (c02);

\node[runnode,name=c03,anchor=west] at ($(c02.east)+(20pt,0pt)$) {};
\draw[runedge] (c02) to node[scale=0.9,above,lblnode,scale=1.2]{L6*} (c03);

\node[runnode,name=c04,anchor=west] at ($(c03.east)+(50pt,0pt)$) {};
\draw[runedge] (c03) to node[above,lblnode,scale=1.2]{evict(P1,x)} (c04);

\node[runnode,name=c05,anchor=west] at ($(c04.east)+(50pt,0pt)$) {};
\draw[runedge] (c04) to node[scale=0.9,above,lblnode,scale=1.2]{evict(P1,y)} (c05);

\node[runnode,name=c06,anchor=north] at ($(c05.south)+(0pt,-20pt)$) {};
\draw[runedge] (c05) to node[right,sloped,lblnode,scale=1.2]{L8} (c06);

\node[runnode,name=c07,anchor=east] at ($(c06.west)+(-23pt,0pt)$) {};
\draw[runedge] (c06) to node[above,lblnode,scale=1.2]{L7*} (c07);

\node[runnode,name=c08,anchor=east] at ($(c07.west)+(-23pt,0pt)$) {};
\draw[runedge] (c07) to node[above,lblnode,scale=1.2]{L1*} (c08);

\node[runnode,name=c09,anchor=east] at ($(c08.west)+(-50pt,0pt)$) {};
\draw[runedge] (c08) to node[above,lblnode,scale=1.2]{evict*(P0,x)} (c09);

\node[runnode,name=c10,anchor=east] at ($(c09.west)+(-18pt,0pt)$) {};
\draw[runedge] (c09) to node[above,lblnode,scale=1.2]{L9} (c10);

\node[runnode,name=c11,anchor=east] at ($(c10.west)+(-22pt,0pt)$) {};
\draw[runedge] (c10) to node[above,lblnode,scale=1.2]{L2*} (c11);

\node[confrunnode,name=c5,anchor=east,scale=1.5] at ($(c11.west)+(-22pt,0pt)$) {$\conf_5$};
 \draw[runedge] (c11) to node[above,lblnode,scale=1.2]{L3*} (c5);

\end{tikzpicture}

\caption{The run $\run_3$.}
\label{run3:fig}

\end{center}\end{figure} 

In fact, no configuration in $\bad$ is reachable from $\conf_0$ in $\prog_2$,
which means that $\prog_2$ indeed satisfies the property $\phi$.
Thus, we have found a sound set of fences for $\prog$ w.r.t.\ $\phi$.
It is interesting to  observe that, although $\prog_2$ is correct w.r.t.\
$\phi$, the program still contains runs
that are impossible under SC, e.g.,
the run $\run_3$ given in Figure~\ref{run3:fig}.

\paragraph{Full Fences.}
Consider a safety property $\phi'$ defined by
(unreachability of) a new set of configurations $\bad'$.
The set $\bad'$ contains
all configurations in  $\bad$,  and also
all configurations where
(i) the processes {\tt P0} and {\tt P1}
have both terminated, and
(ii) both {\tt \$r0} and {\tt \$r3} have values $0$.
We show that $\prog_2$ violates $\phi'$,
i.e., the set $\bad'$ is reachable from $\conf_0$ in $\prog_2$.
To that end, we construct the run $\run_3$, depicted in
Figure~\ref{run3:fig}.
(The run can be explained similarly to $\run_1$ and
$\run_2$.)
At the end of $\run_3$,
we reach the configuration $\conf_5$
which is in $\bad'$.

Notice that the run $\run_3$ is not possible under SC,
while it is feasible under the {\sc SiSd} semantics even in the presence of
the two fences at {\tt L8} and {\tt L9}.
The reason $\run_3$ is possible under {\sc SiSd} is {\it write-read} reordering.
More precisely, read events may overtake write events
(although not the other way round).
In $\run_3$, the write event {\tt L4}
is issued before the read events {\tt L5}, {\tt L6}, and {\tt L7}, but
{\tt L4} does not take effect (does not update the LLC) before
the read events.
There are several ways to prevent the reachability of the set $\bad'$.
One is to replace the $\llfence$ at {\tt L8} and
the $\ssfence$ at {\tt L9} by the full fence
$\fence$, thus obtaining the program $\prog_3$
(Figure~\ref{p3:fig}).
A full fence $\fence$ is only enabled when the L1 of the process is
empty, and hence it forbids
all reorderings of  events of the process.
In $\prog_3$, no configuration in $\bad'$ is reachable from $\conf_0$.
Thus we have inserted a sound set of fences in $\prog$ w.r.t.\ the set $\bad'$.

\begin{figure} [h] \begin{center}

\begin{tikzpicture}
[background rectangle/.style={rounded corners,fill=green!5,line width=1pt,draw},show background rectangle]

\node[name=dummy]{};

\node[pgmtextnode,name=data,scale=1.1] at (dummy)
{\tt data x=0 y=0 z=0};

%%%%%%%%%%%%%%% Proc 0 %%%%%%%%%%%%%%%

\node[pgmtextnode,name=p0,anchor=north west,scale=1.1] at ($(data.south west)+(0pt,-14pt)$)
{\tt process P0};

\node[pgmtextnode,name=r0,anchor=north west,scale=1.1] at ($(p0.south west)+(0pt,-2pt)$)
{\tt registers \$r0};

\node[pgmtextnode,name=b0,anchor=north west,scale=1.1] at ($(r0.south west)+(0pt,-2pt)$)
{\tt begin};

\node[pgmtextnode,name=l1,anchor=north west,scale=1.1] at ($(b0.south west)+(2pt,-2pt)$)
{\tt L1:  \vphantom{y\$}x := 1;};

\node[pgmtextnode,name=l8,anchor=north west,fill=blue!20,scale=1.1] at ($(l1.south west)+(0pt,-2pt)$)
{\tt L9: fence;};

\node[pgmtextnode,name=l2,anchor=north west,scale=1.1] at ($(l8.south west)+(0pt,-2pt)$)
{\tt L2:  \vphantom{y\$}y := 1;};

\node[pgmtextnode,name=l3,anchor=north west,scale=1.1] at ($(l2.south west)+(0pt,-2pt)$)
{\tt L3:  \vphantom{y\$}\$r0 := z;};

\node[pgmtextnode,name=e0,anchor=north west,scale=1.1] at ($(l3.south west)+(-2pt,-2pt)$)
{\tt end};

\begin{pgfonlayer}{foreground}
\node[fitnode,dotted,fill=red!5,fit= (b0) (p0) (r0) (l1) (l2)  (l3) (l8) (e0)]{};
\end{pgfonlayer}

%%%%%%%%%%%%%%% Proc 1 %%%%%%%%%%%%%%%

\node[pgmtextnode,name=p1,anchor=west,scale=1.1] at ($(p0.east)+(32pt,0pt)$)
{\tt process P1};

\node[pgmtextnode,name=r1a,anchor=north west,scale=1.1] at ($(p1.south west)+(0pt,-2pt)$)
{\tt registers \$r1};
\node[pgmtextnode,name=r1b,anchor=north west,scale=1.1] at ($(r1a.south west)+(0pt,-2pt)$)
{\tt \$r2 \$r3};

\node[pgmtextnode,name=b1,anchor=north west,scale=1.1] at ($(r1b.south west)+(0pt,-2pt)$)
{\tt begin};

\node[pgmtextnode,name=l4,anchor=north west,scale=1.1] at ($(b1.south west)+(2pt,-2pt)$)
{\tt L4:  z := 1;};

\node[pgmtextnode,name=l5,anchor=north west,scale=1.1] at ($(l4.south west)+(0pt,-2pt)$)
{L5:  \$r1 := x;};

\node[pgmtextnode,name=l6,anchor=north west,scale=1.1] at ($(l5.south west)+(0pt,-2pt)$)
{\tt L6:  \$r2 := y;};

\node[pgmtextnode,name=l9,anchor=north west,fill=red!20,scale=1.1] at ($(l6.south west)+(0pt,-2pt)$)
{\tt L8: fence;};

\node[pgmtextnode,name=l7,anchor=north west,scale=1.1] at ($(l9.south west)+(0pt,-2pt)$)
{\tt L7:  \$r3 := x;};

\node[pgmtextnode,name=e1,anchor=north west,scale=1.1] at ($(l7.south west)+(-2pt,-2pt)$)
{\tt end};

\begin{pgfonlayer}{foreground}
\node[fitnode,dotted,fill=blue!5,fit= (b1) (p1) (r1a) (r1b) (l4) (l5)  (l6) (l7) (l9) (e1)]{};
\end{pgfonlayer}
\end{tikzpicture}
\caption{The program $\prog_3$.}
\label{p3:fig}
\end{center}\end{figure}

\paragraph{Optimal Sets of Fences.}
We will describe some optimal sets of fences for the
program $\prog$.
As we will notice, this task is not trivial even for
$\prog$.
Our framework allows to make use of
different kinds of fences.
We saw above three examples of fences (and we introduce another
one in Section~\ref{sec:model}).
The motivation is that different kinds of fences may have
different costs.
Using a more ``light-weight'' fence may both increase
program performance and reduce network traffic
(see Section~\ref{sec:results}).
In that respect,
the cost of a full fence is higher than
that  of an $\llfence$ or an $\ssfence$.
The cost assignment is provided by the user of our tool.
Let us assume that an $\llfence$ or an $\ssfence$ costs 1 unit, and
that a full fence costs 2 units.
Let $F_1$ be the set of fences where there is an $\ssfence$ after
{\tt L1}, and an $\llfence$ after {\tt L6}.
Then, $F_1$ is optimal for the program $\prog$ w.r.t.\
the property $\phi$.
First, $F_1$ is sound since
$\prog_2$ (which is the result of inserting the two fences in $\prog$)
satisfies $\phi$, i.e., it does not reach $\bad$ from $\conf_0$.
Second, $F_1$ has
the minimal cost that guarantees unreachability of $\bad$.
The set $F_2$ which we get by replacing both the
$\llfence$ and $\ssfence$ by full fences is also sound.
It is also minimal w.r.t.\ the number of fences (which is $2$).
However, it is not optimal w.r.t.\ $\phi$ since it has a larger cost
than  $F_1$.
On the other hand, $F_2$ is optimal w.r.t.\ the set $\phi'$.
In fact, there are several optimal sets of fences w.r.t.\
$\phi'$ (12 sets to be exact, as reported by our tool).
One such a set is $F_3$ which we get by inserting an $\ssfence$ after
{\tt L1}, an $\llfence$ after {\tt L2}, and an
$\ssfence$ followed by an $\llfence$ after {\tt L6}.
The set $F_3$ is not minimal w.r.t.\ the number of fences, but optimal w.r.t. the property
$\phi'$.
Notice that there are at least $2^{15}$ ways to insert three types of fences
in the simple program of Figure~\ref{running:fig}.
(Each type may or may not be inserted in any particular position.)

%%% Local Variables:
%%% mode: latex
%%% TeX-master: "main.tex"
%%% End:

\section{Programs -- Syntax and Semantics}\label{sec:model}

In this section, we formalize {\sc SiSd} and {\sc Si} protocols, by
introducing a simple assembly-like programming language, and defining
its syntax and semantics.

\subsection{Syntax} 

{
  \newcounter{tracelinectr}
  \newcounter{voffctr}
  \setcounter{voffctr}{0}
  \setcounter{tracelinectr}{0}
  \crefalias{tracelinectr}{line}
  \newcommand{\ind}{\rule{10pt}{0pt}}

  \newcommand{\traceline}[2]{%
  \footnotesize
    \node (#1) at ($(0,-0.4*\value{tracelinectr})+(0,-0.4*\value{voffctr})$) [anchor=west] {\tt\refstepcounter{tracelinectr}\hphantom{12}\llap{\arabic{tracelinectr}}:#2};
  }

}

The syntax of programs is given by the grammar in
Figure~\ref{cprogram:fig}.
A program has a finite set of processes which share a
number of variables (memory locations) $\memlocset$. 
A variable $\xvar\in\memlocset$ should be interpreted as one
machine word at a particular memory address.
For simplicity,
we assume that all the variables and process registers
take their values from a common finite domain $\valset$ of values.
Each process contains a
sequence of instructions, each consisting of a program
label and a statement.
\noindent
To simplify the presentation, we assume that all instructions
(in all processes)
have unique labels.
For a label $\lbl$, we apply three functions:
$\procof\lbl$ returns the process
$\proc$ in which the label occurs.
$\stmtof\lbl$ returns the statement whose label id is
$\lbl$. $\nextof\lbl$ returns the label of
the next statement in the process code, or $\finish$ if there is no next
statement.

\begin{figure}[h]%[10]{r}{.63\textwidth}

\begin{framed}

  \footnotesize{
\[
\begin{array}{@{}r@{\;}c@{\;}l@{}}
\bnfvar{pgm} &::=&
\bnfkey{data} \; {\bnfvar{vdecl}}^+
{\bnfvar{proc}}^+
\\\\
\bnfvar{vdecl}&::=&
\bnfvar{var}\;\bnfstr{=}\; \left(\bnfstr{*}\;|\;\bnfvar{val}\right)
\\\\
\bnfvar{proc}&::=&
\bnfkey{process}\;\bnfvar{pid}\;

\bnfkey{registers}\;{\bnfvar{reg}}^*\;\bnfvar{stmts}
\\\\
\bnfvar{stmts}&::=&
\bnfkey{begin}\;{(\bnfvar{label}\;\bnfstr{:}\;\bnfvar{stmt}\;\bnfstr{;})}^+\;\bnfkey{end}
\\\\
\bnfvar{stmt}&::=&
\bnfvar{var}\;\bnfstr{:=}\;\bnfvar{expr}
\;\;\;|\;\;\;
\bnfvar{reg}\;\bnfstr{:=}\;\bnfvar{var}
\;\;\;|\;\\
&&\bnfvar{reg}\;\bnfstr{:=}\;\bnfvar{expr}
\;\;\;\;|\;\;\;
 \bnfkey{llfence} \;\;\;|\;\;\; \bnfkey{fence}
\;\;\;|\\
&&\bnfkey{cas}\;\bnfstr{(}\;\bnfvar{var}\;\bnfstr{,}\;\bnfvar{expr}\;\bnfstr{,}\;\bnfvar{expr}\;\bnfstr{)}
 \;\;\;|\\
&&\bnfkey{syncwr}\;\bnfstr{:}\;\bnfvar{var}\;\bnfstr{:=}\;\bnfvar{expr}
\;\;\;|\;\;\;\bnfkey{ssfence}\;\;\;|\;
\\
&&
\bnfkey{cbranch}\;\bnfstr{(}\;\bnfvar{bexpr}\;\bnfstr{)}\;\bnfvar{label}
\\
\end{array}
\]}

\caption{The grammar of concurrent programs.}
\label{cprogram:fig}
\end{framed}

\end{figure}

\begin{figure}[ht]%[10]{r}{.63\textwidth}

\begin{framed}

  \small
      {
\[
\begin{array}%{lcr}
  {@{}r@{\;}c@{\;}l@{}}
  $\conf$ &: &\mbox{($\localconf,\llc)$} 
  \\\\
  $\localconf$ & : &\mbox{for each process $\proc$, returns the local configuration of $\proc$} 
  \\\\
  \mbox{$(\lbl,\regval,\loneconf)$} & : & \mbox{Local configuration of $\proc$}
  \\\\
%\end{array}
%\begin{array}{@{}r@{\;}c@{\;}l@{}}%{lcr}
  \mbox{$\lbl$} & : & \mbox{the label of the next statement to execute in $\proc$}
  \\\\
  \mbox{$\regval$} & : & \mbox{the values of the local registers in $\proc$}
  \\\\
  \mbox{$\loneconf$} & : & \mbox{$(\valid,\lstatus,\lval)$, the state of the L1 cache of $\proc$}
  \\\\
  \mbox{$\valid\subseteq\memlocset$} & : &\mbox{the set of shared variables currently in the valid state} 
  \\\\
  \mbox{$\lstatus:\valid\to\{\dirty, \clean\}$} & : &\mbox{for each $\memloc\in\valid$, whether $x$ is dirty or clean} 
  \\\\
  \mbox{$\lval:\valid\to\valset$} & : &\mbox{for each $\memloc\in\valid$, its current value in the L1 cache of $\proc$} 
  \\\\
  \mbox{LLC$:\memlocset\to\valset$} & : &\mbox{shared part of $\conf$, defines for each $x\in\memlocset$ its value in the LLC} 
  
%\bnfvar{pgm} &::=&
%\bnfkey{data} \; {\bnfvar{vdecl}}^+
%{\bnfvar{proc}}^+
\\
\end{array}
\]}

\caption{The definition of a configuration $\conf$}
\label{config:fig}
\end{framed}

\end{figure}

\subsection{Configurations}
As illustrated in Figure~\ref{config:fig}, a {\it local configuration} of a process $\proc$ is a triple
$(\lbl,\regval,\loneconf)$, where $\lbl$ is the label of the next statement to execute
in $\proc$, $\regval$ defines the values of the local registers, and
$\loneconf$ defines the state of the L1 cache of $\proc$.
In turn, $\loneconf$ is a triple $(\valid,\lstatus,\lval)$.
Here $\valid\subseteq\memlocset$ defines the set of
shared variables that are currently in the valid state, and
$\lstatus$ is a function from $\valid$ to the set
$\{\dirty, \clean\}$ that defines, for each $\memloc\in\valid$,
whether $x$ is dirty or clean,
and $\lval$ is a function from $\valid$ to $\valset$
that defines for each $\memloc\in\valid$
its current value in the L1 cache of $\proc$.
The {\it shared part} of a configuration
 is given by a function $\llc$ that defines for
each variable $x\in\memlocset$ its value $\llc(x)$ in the LLC.
A configuration $\conf$ then is a pair
$(\localconf,\llc)$ where $\localconf$ is a function
that returns, for each process
$\proc$, the local configuration of $\proc$.
In the formal definition below,
our semantics allows system events to occur
non-deterministically.
\noindent This means that we model not only instructions from the program code
itself, but also events that are caused by unpredictable things
as hardware prefetching, software prefetching, program preemption,
false sharing, multiple threads of the same program being scheduled on
the same core, etc.

A transition $\transition$ is either performed by a given process
when it executes an instruction,
or is a system event. 
In the former case,
 $\transition$ will be of the form
$\lbl$, i.e., $\transition$ models the effect
of a process $\proc$ performing the statement labeled with $\lbl$.
In the latter case, $\transition$ will be equal to
$\evt$ for some system event $\evt$.
For a function $f$,
we use $f\update a b$, to denote the function
$f'$ such that $f'(a)=b$ and $f'(a')=f(a')$ if $a'\neq a$.
We write $f(a)=\bot$ to denote that $f$ is undefined for $a$.

Below, we give an intuitive explanation of each transition. The formal
definition can be found in Figure~\ref{opsem:fig} where we assume
$\conf=\tuple{\localconf,\llc}$, and
$\localconf(\proc)=(\lbl,\regval,\loneconf)$, and
$\loneconf=(\valid,\lstatus,\lval)$, $\procof\lbl=\proc$, and
$\stmtof\lbl=\stmt$.
We leave out the definitions for local instructions, since they have
standard semantics.

\subsection{Semantics}
\label{sec:Semantics}

\subsubsection{Instruction Semantics.}
Let $\proc$ be one of the processes in the program, and let
$\lbl$ be the label of an instruction in $\proc$ whose statement is
$\stmt$.
We will define a {\it transition relation}
$\movesto{\lbl}$, induced by $\lbl$, on the set
of configurations.
The relation is defined in terms of the type
of operation performed by the given statement $\stmt$.
In all the cases only the local state of $\proc$ and
LLC will be changed.
The local states of the rest of the processes will not be affected. This
mirrors the principle in {\sc SiSd} that L1 cache controllers will
communicate with the LLC, but never directly with other L1 caches.

\medskip

\noindent
{\bf Read $(\reg := \xvar)$:}
Process $\proc$ reads the value of $\xvar$
from L1 into the register $\reg$.
The L1 and the LLC will not change.
The transition is only enabled if $\xvar$ is valid in
the L1 cache of $\proc$. This means that if $\xvar$ is not in L1, then a system event $\ifetch$ must occur before $\proc$ is able to execute
the read operation.

\medskip

\noindent
{\bf Write $(\xvar:= e)$:}
An expression $e$ contains only registers and constants.
The value of $\xvar$ in  L1  is updated with the evaluation of
$e$ where registers have values as indicated by $\regval$,
and $\xvar$ becomes dirty.
The write is only enabled if $\xvar$ is valid
for $\proc$.

\medskip

\noindent
{\bf Fence $(\fence)$:}
A full fence transition is only enabled when the L1 of $\proc$
is empty. This means that before the fence can be executed, all
entries in its L1 must be evicted (and written to the LLC if
dirty). So $\proc$ must stall until the necessary system events
($\iwrllc$ and $\ievict$) have occurred. Executing the fence has no
further effect on the caches.

\begin{figure}%[h]
\begin{math}
\begin{array}{@{}c@{}}
\textbf{Instruction Semantics}
\\ \\
\textrm{\footnotesize $\conf=\tuple{\localconf,\llc}$, $\localconf(\proc)=(\lbl,\regval,\loneconf)$,}
\\
\textrm{\footnotesize $\loneconf=(\valid,\lstatus,\lval)$, $\procof\lbl=\proc$, $\stmtof\lbl=\stmt$}
\\ \\
{\displaystyle
\frac 
{\stmt=(\reg := \xvar) \;,\; \xvar\in\valid}
{\conf\movesto{\lbl}
\tuple{
\localconf\update{\proc}{
\tuple{\nextof\lbl,\regval\update\reg{\lval(\xvar)},\loneconf}},
\llc}}}
\\ \\
\frac
    {\stmt=(\xvar := e) \;,\; \xvar\in\valid
      \;}
{\conf\movesto{\lbl}
\tuple{
\localconf\update{\proc}{\tuple{\nextof\lbl,\regval,
\tuple{\valid,\lstatus\update\xvar\dirty,\lval\update\xvar{\regval(e)}}}},
\llc}}
\\ \\

{\displaystyle
\frac
{\stmt=\fence \;,\; \valid=\emptyset}
{\conf\movesto{\lbl}
\tuple{
\localconf\update{\proc}{\tuple{\nextof\lbl,\regval,
\loneconf}},
\llc}}}
\\ \\
{\displaystyle
\frac
{\stmt=\ssfence\;,\;\forall\xvar\in\memlocset.\;
\left(\xvar\in\valid\Rightarrow\lstatus(\xvar)=\clean\right)}
{\conf\movesto{\lbl}
\tuple{
\localconf\update{\proc}{\tuple{\nextof\lbl,\regval,\loneconf}},
\llc}}}
\\ \\
{\displaystyle
\frac
{\stmt=\llfence\;,\;\forall\xvar\in\memlocset.\;
\left(\xvar\in\valid\Rightarrow\lstatus(\xvar)=\dirty\right)}
{\conf\movesto{\lbl}
\tuple{
\localconf\update{\proc}{\tuple{\nextof\lbl,\regval,\loneconf}},
\llc}}}
\\ \\
{\displaystyle
\frac
{\stmt=(\syncwr{\xvar}{e})\;,\;\xvar\not\in\valid}
{\conf\movesto{\lbl}
\tuple{
\localconf\update{\proc}{\tuple{\nextof\lbl,\regval,\loneconf}},
\llc\update\xvar{\regval(e)}}}}
\\ \\
{\displaystyle
\frac
{\stmt=\cas(\xvar,e_0,e_1)\;,\;\xvar\not\in\valid\;,\;\llc(\xvar)=\regval(e_0)}
{\conf\movesto{\lbl}
\tuple{
\localconf\update{\proc}{\tuple{\nextof\lbl,\regval,\loneconf}},
\llc\update\xvar{\regval(e_1)}}}}
\\ \\
{\displaystyle
\frac
{\stmt=(\cbranch(e)\;\lbl'),\;\regval(e)={\tt true}}
{\conf\movesto{\lbl}
\tuple{
\localconf\update{\proc}{\tuple{\lbl',\regval,\loneconf}},
\llc}}}
\\ \\
{\displaystyle
\frac
{\stmt=(\cbranch(e)\;\lbl'),\;\regval(e)={\tt false}}
{\conf\movesto{\lbl}
\tuple{
\localconf\update{\proc}{\tuple{\nextof\lbl,\regval,\loneconf}},
\llc}}}
\\ \\
\textbf{System Event Semantics}
\\ \\
{\displaystyle
\frac
    {\evt=(\fetch{\proc}{\xvar}) \;,\; \xvar\not\in\valid
    \;}%,\; \textnormal{\textsf{S}}'=\lstatus\update\xvar\clean}
{\conf\movesto\evt
\tuple{
\localconf\update{\proc}{\tuple{\lbl,\regval,\tuple{\valid\cup\{\xvar\},\lstatus\update\xvar\clean,\lval\update\xvar{\llc(\xvar)}}}},
\llc}}}
\\ \\
{\displaystyle
\frac
{\evt=(\wrllc{\proc}{\xvar}) \;,\; \xvar\in\valid \;,\; \lstatus(\xvar)=\dirty \;}%,\; \textnormal{\textsf{S}}'=\lstatus\update\xvar\clean}
{\conf\movesto\evt
\tuple{
\localconf\update\proc{\tuple{\lbl,\regval,\tuple{\valid,\lstatus\update\xvar\clean,\lval}}},
\llc\update\xvar{\lval(\xvar)}}}}
\\ \\
{\displaystyle
\frac
{\evt=(\evict{\proc}{\xvar}) \;,\; \xvar\in\valid \;,\; \lstatus(\xvar)=\clean}
{\conf\movesto\evt
\tuple{
\localconf\update\proc{\tuple{\lbl,\regval,\tuple{\valid\setminus\{\xvar\},\lstatus\update\xvar\bot,\lval\update\xvar\bot}}},
\llc}}}

\end{array}
\end{math}
\caption{Semantics of programs running under {\sc SiSd}.}
\label{opsem:fig}

\end{figure}

\medskip

\noindent
{\bf SS-Fence $(\ssfence)$:}
Similarly, an $\ssfence$ transition is only enabled when there are no
dirty entries in the L1 cache of $\proc$. So $\proc$ must stall until
all dirty entries have been written to the LLC by $\iwrllc$ system
events. In contrast to a full fence, an $\ssfence$ permits clean
entries to remain in the L1.

\medskip

\noindent
{\bf LL-Fence $(\llfence)$:}
This is the dual of an SS-Fence.
An $\llfence$ transition is only enabled when there are no clean
entries in the L1 cache of $\proc$. In other words, the read instructions
before and after an $\llfence$ cannot be reordered.

\medskip

\noindent
{\bf Synchronized write $(\syncwr{\xvar}{e})$:}
A synchronized write is like an ordinary write, but acts
directly on the LLC instead of the L1 cache. For a $\plainsyncwr$
transition to be enabled, $\xvar$ may not be in the L1. (I.e., the
cache must invalidate $\xvar$ before executing the $\plainsyncwr$.)
When it is executed, the value of $\xvar$ in the LLC is
updated with the evaluation of the expression $e$ under the register
valuation $\regval$ of $\proc$. The L1 cache is not
changed.

\medskip

\noindent
{\bf CAS $(\cas(\xvar,e_0,e_1))$:}
A compare and swap transition acts directly on the LLC. The $\cas$ is
only enabled when $\xvar$ is not in the L1 cache of $\proc$, and the value
of $\xvar$ in the LLC equals $e_0$ (under
$\regval$). When the instruction is executed, it atomically writes
the value of $e_1$ directly to the LLC in the same way as a
synchronized write would.

\subsubsection{System Event Semantics.}
The system may non-deterministically
(i.e., at any time) perform a {\it system event.}
A system event is not a program instruction, and so will not change
the program counter (label) of a process. %
We will define a {\it transition relation}
$\movesto{\evt}$, induced by the system event $\evt$.
There are three types of system events as follows.

\medskip

\noindent
{\bf Eviction $(\evict{\proc}{\xvar})$:}
An $\evict\proc\xvar$ system event may occur when $\xvar$ is valid and
clean in the  L1  of process $\proc$. When the event occurs,
$\xvar$ is removed from the L1 of $\proc$.

\medskip

\noindent
{\bf Write-LLC $(\wrllc\proc\xvar)$:}
If the entry of $\xvar$ is dirty in  the L1  of $\proc$, then a
$\wrllc\proc\xvar$ event may occur. The value of $\xvar$ in the LLC is
then updated with the value of $\xvar$ in the L1 of $\proc$. The entry
of $\xvar$ in the L1 of $\proc$ becomes clean.

\medskip

\noindent
{\bf Fetch $(\fetch\proc\xvar)$:}
If $\xvar$ does not have an entry in the L1  of $\proc$, then
$\proc$ may fetch the value of $\xvar$ from the LLC, and create a new,
clean entry with that value for $\xvar$ in its L1.

\subsection{Program Semantics under an \textsc{SI} Protocol}

In a self-invalidation protocol without self-downgrade, a writing
process will be downgraded and forced to communicate its dirty data
when another process accesses that location in the LLC.
This behavior can be modelled by a semantics where writes take effect
atomically with respect to the LLC.
Hence, to modify the semantics given in Section~\ref{sec:Semantics}
such that it models a program under an \textsc{Si} protocol, it
suffices to interpret all write instructions as the corresponding
$\plainsyncwr$ instructions.

\subsection{Transition Graph and  the Reachability Algorithm}
Our semantics allows to construct, for a given
program $\prog$, a finite {\it transition graph},  where
each node in the graph is a configuration
in $\prog$, and each edge is a transition.
Figure~\ref{graph:fig} shows four nodes in the transition
graph of the program in Figure~\ref{running:fig}.
The configurations $\conf_2$ and $\conf_3$ are those depicted in
Figure~\ref{confs:fig}, while $\conf_6$ is the configuration
we get from $\conf_2$ by adding a clean copy of {\tt y} with value
$1$ to the L1 of {\tt P1};
and $\conf_7$ is the configuration we get from $\conf_6$
by updating the label of {\tt P1} to {\tt L7}, and the value of {\tt \$r2} to $1$.
A {\it run} is a sequence
$\conf_0\movesto{\transition_1}\conf_1\movesto{\transition_2}\conf_2\cdots\movesto{\transition_n}\conf_n$,
 which is a path in the transition graph, where $\transition_i (0\leq{}i\leq{}n)$ is either a label $\lbl$ or a system event $\evt$. 
Figure~\ref{graph:fig} shows the path of the run $\rrun_3$.

\begin{figure}[h] 
\begin{center}

\begin{tikzpicture}[show background rectangle]
\node[name=dummy]{};

\node[graphnode,name=c2, scale=1.2] at (dummy) {$\conf_2$};
\node[graphnode,name=c6, name=c6, scale=1.2] at ($(c2.east)+(20mm,0mm)$)  {$\conf_6$};
\node[graphnode,name=c6, name=c7, scale=1.2] at ($(c6.east)+(8mm,0mm)$)  {$\conf_7$};
\node[graphnode,name=c3, name=c3, scale=1.2] at ($(c7.east)+(8mm,0mm)$)  {$\conf_3$};

\node[name=n21] at ($(c2)+(-6.5mm,-6.5mm)$){};
\node[name=n22] at ($(c2)+(-6.5mm,6.5mm)$){};
\node[name=n23] at ($(c2)+(6.5mm,-6.5mm)$){};

\node[name=n61] at ($(c6)+(-6.5mm,-6.5mm)$){};
\node[name=n63] at ($(c6)+(6.5mm,-6.5mm)$)  {};

\node[name=n71] at ($(c7)+(0mm,-6.5mm)$){};
\node[name=n73] at ($(c7)+(6mm,-6.5mm)$)  {};

\node[name=n31] at ($(c3)+(0mm,-6.5mm)$){};
\node[name=n32] at ($(c3)+(6.5mm,6.5mm)$)  {};
\node[name=n33] at ($(c3)+(6.5mm,-6.5mm)$)  {};

\draw[dotted,line width=0.7pt,->] (n21) -- (c2);
\draw[dotted,line width=0.7pt,->] (n22) -- (c2);
\draw[dotted,line width=0.7pt,->] (c2) -- (n23);

\draw[line width=0.7pt,->] (c2) to node[textnode,text=black,above=0pt,scale=1,sloped] {{\tt fetch(P1,y)}} (c6);
\draw[dotted,line width=0.7pt,->] (n61) -- (c6);
\draw[dotted,line width=0.7pt,->] (c6) -- (n63);

\draw[line width=0.7pt,->] (c6) to node[textnode,text=black,above=0pt,scale=1,sloped] {{\tt L6}} (c7);
\draw[dotted,line width=0.7pt,->] (n71) -- (c7);
\draw[dotted,line width=0.7pt,->] (c7) -- (n73);

\draw[line width=0.7pt,->] (c7) to node[textnode,text=black,above=0pt,scale=1,sloped] {{\tt L7}} (c3);

\draw[dotted,line width=0.7pt,->] (n31) -- (c3);
\draw[dotted,line width=0.7pt,->] (n32) -- (c3);
\draw[dotted,line width=0.7pt,->] (c3) -- (n33);

\end{tikzpicture}

\caption{Part of the transition graph of the program
in Figure~\ref{running:fig}.}
\label{graph:fig}

\end{center}\end{figure}

Together with the program, the user provides a {\it safety property}
$\phi$ that describes a set $\bad$ of configurations that are
considered to be errors.
Checking $\phi$ for
a program $\prog$ amounts to checking whether there is a run
leading from the initial configuration to a
configuration in $\bad$.
To do that, the input program under {\sc SiSd} is translated to the code recognized by 
the reachability analysis tool chosen by the user. The translated code simulates all the
behaviors which are allowed in the {\sc SiSd} semantics. Also, there is instrumentation added to 
simulate the caches. Verifying the input program amounts
to verifying the translated code which is analyzed under SC.
If a bad configuration is encountered, a witness run is returned by the tool.
Otherwise, the program is declared to be correct.

%%% Local Variables:
%%% mode: latex
%%% TeX-master: "main.tex"
%%% End:

\section{Litmus Tests and Comparison with Other Memory Models} 
In this section, we will first compare the behavior of {\sc SiSd} with other weak memory models.
We do this by presenting a sequence of litmus tests
that differentiate the {\sc SiSd} semantics from the other models.
Each program consists of a number of threads that share a number of variables.
Later, we will describe a number of additional litmus tests to clarify the bahavior of the {\sc SiSd} model.

\subsection{Comparing with Other Memory Models}
\paragraph{Sequential Consistency (SC)}
The {\sc SiSd} model is weaker than SC.
Consider the classic SB (Dekker's) algorithm shown in Figure~\ref{dekker:fig}.
Under the SC semantics, the program does not have any runs satisfying the assertion, 
since the operations performed by the process $P0$ are not reordered.
If $\reg_1=0$, then $\xvar := 1$ must have been executed and
the value 1 of $\xvar$ is updated to the memory.
At this time, $P0$ sees the value of y as 0 in the memory,
which means that $\yvar:=1$ is not executed by $P1$ yet.
When $P1$ executes $\yvar:=1$ and $\reg_2:=\xvar$,
the value of $\xvar$ in the memory is already $1$ and
thus $1$ is assigned to $\reg_2$.

Any run of a program $\prog$ under SC can be simulated by a run of $\prog$
under {\sc SiSd} as the following: 
1) right after each write operation, the assigned value of the variable is updated to LLC
and the variable is invalidated from the local cache immediately;
2) right before each read operation, the value of the variable is fetched from LLC and
after reading the value, the variable is immediately invalidated from the local cache.

\begin{figure} \begin{center}

\begin{tikzpicture}
[background rectangle/.style={rounded corners,fill=green!5,line width=1pt,draw},show background rectangle]

\node[name=dummy]{};

\node[pgmtextnode,name=data,scale=1.1] at (dummy)
{\tt Initially: x=0, y=0};

%%%%%%%%%%%%%%% Proc 0 %%%%%%%%%%%%%%%

\node[pgmtextnode,name=p0,anchor=north west,scale=1.1] at ($(data.south west)+(0pt,-12pt)$)
{\tt process P0};

\node[pgmtextnode,name=r0,anchor=north west,scale=1.1] at ($(p0.south west)+(0pt,-2pt)$)
{\tt registers \$r1};

\node[pgmtextnode,name=b0,anchor=north west,scale=1.1] at ($(r0.south west)+(0pt,-2pt)$)
{\tt begin};

\node[pgmtextnode,name=l1,anchor=north west,scale=1.1] at ($(b0.south west)+(2pt,-2pt)$)
{\tt L1:  \vphantom{y\$}x := 1;};

\node[pgmtextnode,name=l2,anchor=north west,scale=1.1] at ($(l1.south west)+(0pt,-2pt)$)
{\tt L2:  \vphantom{y\$}\$r1 := y;};

\node[pgmtextnode,name=e0,anchor=north west,scale=1.1] at ($(l2.south west)+(0pt,-2pt)$)
{\tt end};

\begin{pgfonlayer}{foreground}
  \node[fitnode,dotted,fill=red!5,fit= (b0) (p0) (r0) (l1) (l2) (e0)]{};
\end{pgfonlayer}

%%%%%%%%%%%%%%% Proc 1 %%%%%%%%%%%%%%%

\node[pgmtextnode,name=p1,anchor=west,scale=1.1] at ($(p0.east)+(31pt,0pt)$)
{\tt process P1};

\node[pgmtextnode,name=r1a,anchor=north west,scale=1.1] at ($(p1.south west)+(0pt,-2pt)$)
{\tt registers \$r2};

\node[pgmtextnode,name=b1,anchor=north west,scale=1.1] at ($(r1a.south west)+(0pt,-2pt)$)
{\tt begin};

\node[pgmtextnode,name=l3,anchor=north west,scale=1.1] at ($(b1.south west)+(2pt,-2pt)$)
{\tt L3:  y := 1;};

\node[pgmtextnode,name=l4,anchor=north west,scale=1.1] at ($(l3.south west)+(0pt,-2pt)$)
{\tt L4:  \$r2 := x;};

\node[pgmtextnode,name=e1,anchor=north west,scale=1.1] at ($(l4.south west)+(-2pt,-2pt)$)
{\tt end};

\begin{pgfonlayer}{foreground}
  \node[fitnode,dotted,fill=blue!5,fit= (b1) (p1) (r1a) (l3) (l4) (e1)]{};
\end{pgfonlayer}

\node[pgmtextnode,name=registers,anchor=north west,scale=1.1] at ($(p0.south west)+(0pt,-55pt)$)
{\tt Assertion: \$r1=0, \$r2=0};

\end{tikzpicture}
\caption{SB (Dekker's) algorithm.}
\label{dekker:fig}
\end{center}\end{figure}

\paragraph{Total Store Order (TSO)}
The {\sc SiSd} model and TSO are not comparable. 
As illustrated in Figure~\ref{TSO:fig},
the TSO model inserts a store buffer between each process
and the shared memory. 
When a process executes a write instruction, the instruction 
is appended to the end of the buffer of the process.
At any point of the execution, the instruction at the head of the buffer may 
nondeterministically be removed and applied to update
the memory.
When a process reads the value of a variable, it fetches
the value from the most recent write operation on the variable
in its buffer. 
If such a write operation is missing, then the value is fetched from 
the memory.

\begin{figure}
  \includegraphics[width=260pt]{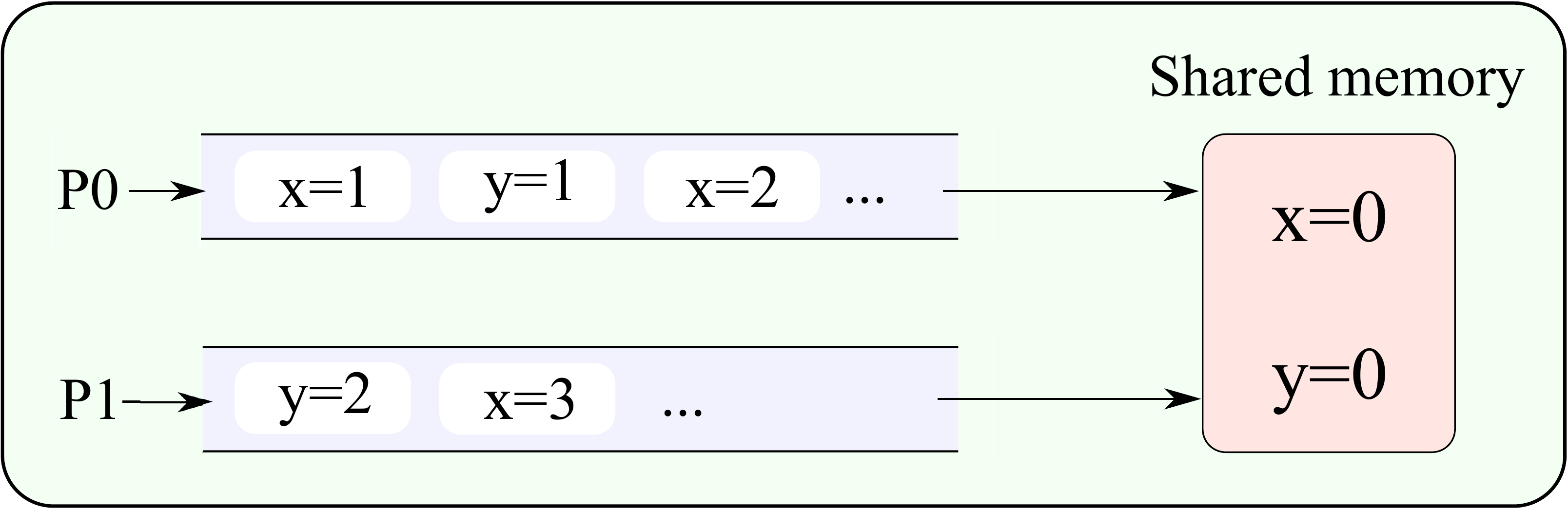}
  \caption{TSO}
  \label{TSO:fig}
\end{figure}

%\begin{figure}
%%
%\begin{tikzpicture}[background rectangle/.style={rounded corners,dotted,line width=1pt,draw},show background rectangle]
%\node[name=n11]  {$\xvar:= 1$};
%\node[name=n12,anchor=west] at ($(n11.west)+(0pt,-15pt)$) {$\yvar:= 1$};
%
%
%\node[name=n21,anchor=west] at ($(n11.east)+(15pt,0pt)$) {$\reg_1 :=\xvar$};
%\node[name=n22,anchor=west] at ($(n21.west)+(0pt,-15pt)$) {$\reg_2:=\yvar$};
%
%\node[name=th1,anchor=south west]at ($(n11.north west)+(0pt,1pt)$) {$P_1:$};
%\node[name=th2,anchor=south west]at ($(n21.north west)+(0pt,1pt)$) {$P_2:$};
%
%\draw[line width= 0.5pt] ($(n21.north west)+(-8.5pt,17pt)$) -- ($(n22.south west)+(-8.5pt,-10pt)$);
%\draw[line width= 0.5pt] ($(n21.north west)+(-6.5pt,17pt)$) -- ($(n22.south west)+(-6.5pt,-10pt)$);
%
%\node[anchor=south] at  ($(n11.north east)+(7.5pt,15pt)$) {Initially: $\xvar=0$ , $\yvar=0$};
%
%\node[anchor=north,draw] at ($(n22.south west)+(-7.5pt,-12pt)$) {$\reg_1=0$ and $\reg_2=1$};
%\end{tikzpicture}
%\caption{Program {\tt MP}.}
%\label{mp:fig}
%\end{figure}
%%

\begin{figure} \begin{center}

\begin{tikzpicture}
[background rectangle/.style={rounded corners,fill=green!5,line width=1pt,draw},show background rectangle]

\node[name=dummy]{};

\node[pgmtextnode,name=data,scale=1.1] at (dummy)
{\tt Initially: x=0, y=0};

%%%%%%%%%%%%%%% Proc 0 %%%%%%%%%%%%%%%

\node[pgmtextnode,name=p0,anchor=north west,scale=1.1] at ($(data.south west)+(0pt,-12pt)$)
{\tt process P0};

\node[pgmtextnode,name=r0,anchor=north west,scale=1.1] at ($(p0.south west)+(0pt,-2pt)$)
{\tt registers};

\node[pgmtextnode,name=b0,anchor=north west,scale=1.1] at ($(r0.south west)+(0pt,-2pt)$)
{\tt begin};

\node[pgmtextnode,name=l1,anchor=north west,scale=1.1] at ($(b0.south west)+(2pt,-2pt)$)
{\tt L1:  \vphantom{y\$}x := 1;};

\node[pgmtextnode,name=l2,anchor=north west,scale=1.1] at ($(l1.south west)+(0pt,-2pt)$)
{\tt L2:  \vphantom{y\$}y := 1;};

\node[pgmtextnode,name=e0,anchor=north west,scale=1.1] at ($(l2.south west)+(0pt,-2pt)$)
{\tt end};

\begin{pgfonlayer}{foreground}
  \node[fitnode,dotted,fill=red!5,fit= (b0) (p0) (r0) (l1) (l2) (e0)]{};
\end{pgfonlayer}

%%%%%%%%%%%%%%% Proc 1 %%%%%%%%%%%%%%%

\node[pgmtextnode,name=p1,anchor=west,scale=1.1] at ($(p0.east)+(23pt,0pt)$)
{\tt process P1};

\node[pgmtextnode,name=r1a,anchor=north west,scale=1.1] at ($(p1.south west)+(0pt,-2pt)$)
{\tt registers \$r1 \$r2};

\node[pgmtextnode,name=b1,anchor=north west,scale=1.1] at ($(r1a.south west)+(0pt,-2pt)$)
{\tt begin};

\node[pgmtextnode,name=l3,anchor=north west,scale=1.1] at ($(b1.south west)+(2pt,-2pt)$)
{\tt L3:  \$r1 := x;};

\node[pgmtextnode,name=l4,anchor=north west,scale=1.1] at ($(l3.south west)+(0pt,-2pt)$)
{\tt L4:  \$r2 := y;};

\node[pgmtextnode,name=e1,anchor=north west,scale=1.1] at ($(l4.south west)+(-2pt,-2pt)$)
{\tt end};

\begin{pgfonlayer}{foreground}
  \node[fitnode,dotted,fill=blue!5,fit= (b1) (p1) (r1a) (l3) (l4) (e1)]{};
\end{pgfonlayer}

\node[pgmtextnode,name=registers,anchor=north west,scale=1.1] at ($(p0.south west)+(0pt,-55pt)$)
{\tt Assertion: \$r1=0, \$r2=1};

\end{tikzpicture}
\caption{Program {\tt MP}.}
\label{mp:fig}
\end{center}\end{figure}

First, we show that TSO is not weaker than {\sc SiSd}. 
Consider the program {\tt MP} in Figure~\ref{mp:fig}.
Under the TSO semantics, the program {\tt MP} does not have any runs that satisfy the assertion.
The reason is that the two write operations 
performed by $P0$ will reach the memory in the same order
as they are performed, i.e, $\xvar:=1$ and then $\yvar:=1$.
Furthermore, the two read operations performed by $P1$ are
not re-ordered according to the TSO semantics.
Therefore, if $\reg_2=1$ then $P1$ has already seen that the value of $\xvar$ is equal
to $1$ when it performs the assignment $\reg_1:=\xvar$.
In contrast, the program {\tt MP} has the following run under {\sc SiSd}
that satisfies the assertion.
First, the process $P0$ assigns $1$ to $\xvar$ and $\yvar$ respectively, but only updates 
the value of $\yvar$ to the LLC.
Next, $P1$ fetches the values of $\xvar$ and $\yvar$ ($0$ and $1$ respectively) from the LLC, and assigns
them to the registers using the instructions $\reg_1:=\xvar$ and $\reg_2:=\yvar$, which means
the assertion will be satisfied.

\begin{figure} \begin{center}

\begin{tikzpicture}
[background rectangle/.style={rounded corners,fill=green!5,line width=1pt,draw},show background rectangle]

\node[name=dummy]{};

\node[pgmtextnode,name=data,scale=1.1] at (dummy)
{\tt Initially: x=0, y=0};

%%%%%%%%%%%%%%% Proc 0 %%%%%%%%%%%%%%%

\node[pgmtextnode,name=p1,anchor=north west,scale=1.1] at ($(data.south west)+(0pt,-12pt)$)
{\tt process P0};

\node[pgmtextnode,name=r1,anchor=north west,scale=1.1] at ($(p1.south west)+(0pt,-2pt)$)
{\tt registers \$r1};

\node[pgmtextnode,name=r234,anchor=north west,scale=1.1] at ($(r1.south west)+(0pt,-2pt)$)
{\tt \$r2 \$r3 \$r4};

\node[pgmtextnode,name=b0,anchor=north west,scale=1.1] at ($(r234.south west)+(0pt,-2pt)$)
{\tt begin};

\node[pgmtextnode,name=l1,anchor=north west,scale=1.1] at ($(b0.south west)+(2pt,-2pt)$)
{\tt L1:  \vphantom{y\$}x := 1;};

\node[pgmtextnode,name=l2,anchor=north west,scale=1.1] at ($(l1.south west)+(0pt,-2pt)$)
{\tt L2:  \vphantom{y\$}x := 2;};

\node[pgmtextnode,name=l3,anchor=north west,scale=1.1] at ($(l2.south west)+(0pt,-2pt)$)
{\tt L3:  \vphantom{y\$}x := 3;};

\node[pgmtextnode,name=l4,anchor=north west,scale=1.1] at ($(l3.south west)+(0pt,-2pt)$)
{\tt L4:  \vphantom{y\$}x := 4;};

\node[pgmtextnode,name=l5,anchor=north west,scale=1.1] at ($(l4.south west)+(0pt,-2pt)$)
{\tt L5:  \vphantom{y\$}\$r1 := y;};

\node[pgmtextnode,name=l6,anchor=north west,scale=1.1] at ($(l5.south west)+(0pt,-2pt)$)
{\tt L6:  \vphantom{y\$}\$r2 := y;};

\node[pgmtextnode,name=l7,anchor=north west,scale=1.1] at ($(l6.south west)+(0pt,-2pt)$)
{\tt L7:  \vphantom{y\$}\$r3 := y;};

\node[pgmtextnode,name=l8,anchor=north west,scale=1.1] at ($(l7.south west)+(0pt,-2pt)$)
{\tt L8:  \vphantom{y\$}\$r4 := y;};

\node[pgmtextnode,name=e0,anchor=north west,scale=1.1] at ($(l8.south west)+(0pt,-2pt)$)
{\tt end};

\begin{pgfonlayer}{foreground}
  \node[fitnode,dotted,fill=red!5,fit= (b0) (p1) (r1) (r234) (l1) (l2) (l3) (l3) (l4) (l5) (l6) (l7) (l8) (e0)]{};
\end{pgfonlayer}

%%%%%%%%%%%%%%% Proc 1 %%%%%%%%%%%%%%%

\node[pgmtextnode,name=p2,anchor=west,scale=1.1] at ($(p1.east)+(31pt,0pt)$)
{\tt process P1};

\node[pgmtextnode,name=r5,anchor=north west,scale=1.1] at ($(p2.south west)+(0pt,-2pt)$)
{\tt registers \$r5};

\node[pgmtextnode,name=r678,anchor=north west,scale=1.1] at ($(r5.south west)+(0pt,-2pt)$)
{\tt \$r6 \$r7 \$r8};

\node[pgmtextnode,name=b1,anchor=north west,scale=1.1] at ($(r678.south west)+(0pt,-2pt)$)
{\tt begin};

\node[pgmtextnode,name=l9,anchor=north west,scale=1.1] at ($(b1.south west)+(2pt,-3pt)$)
{\tt L9:  y := 1;};

\node[pgmtextnode,name=l10,anchor=north west,scale=1.1] at ($(l9.south west)+(0pt,-3pt)$)
{\tt L10:  y := 2;};

\node[pgmtextnode,name=l11,anchor=north west,scale=1.1] at ($(l10.south west)+(0pt,-2pt)$)
{\tt L11:  y := 3;};

\node[pgmtextnode,name=l12,anchor=north west,scale=1.1] at ($(l11.south west)+(0pt,-2pt)$)
{\tt L12:  y := 4;};

\node[pgmtextnode,name=l13,anchor=north west,scale=1.1] at ($(l12.south west)+(0pt,-3pt)$)
{\tt L13:  \$r5 := x;};

\node[pgmtextnode,name=l14,anchor=north west,scale=1.1] at ($(l13.south west)+(0pt,-2pt)$)
{\tt L14:  \$r6 := x;};

\node[pgmtextnode,name=l15,anchor=north west,scale=1.1] at ($(l14.south west)+(0pt,-3pt)$)
{\tt L15:  \$r7 := x;};

\node[pgmtextnode,name=l16,anchor=north west,scale=1.1] at ($(l15.south west)+(0pt,-2pt)$)
{\tt L16:  \$r8 := x;};

\node[pgmtextnode,name=e1,anchor=north west,scale=1.1] at ($(l16.south west)+(0pt,-3pt)$)
{\tt end};

\begin{pgfonlayer}{foreground}
  \node[fitnode,dotted,fill=blue!5,fit= (b1) (p2) (r5) (r678) (l9) (l10) (l11) (l12) (l13) (l14) (l15) (l16) (e1)]{};
\end{pgfonlayer}

\node[pgmtextnode,name=assertion,anchor=north west,scale=1.1] at ($(p0.south west)+(0pt,-128pt)$)
{\tt Assertion:};

\node[pgmtextnode,name=registers1,anchor=north west,scale=1.1] at ($(assertion.south west)+(0pt,-2pt)$)
{\tt \$r1=1, \$r2=2, \$r3=3, \$r4=4,};

\node[pgmtextnode,name=registers2,anchor=north west,scale=1.1] at ($(registers1.south west)+(0pt,-2pt)$)
{\tt \$r5=1, \$r6=2, \$r7=3, \$r8=4};

\end{tikzpicture}
\caption{Program {\tt ReadSeq}.}
\label{read:seq:fig}
\end{center}\end{figure}

The program {\tt ReadSeq} in Figure~\ref{read:seq:fig}
shows that {\sc SiSd} is not weaker than TSO.
Under TSO, the program exhibits a run that satisfies the assertion as follows.
First, $P0$ executes the instructions $x:=1$, $x:=2$, $x:=3$, and $x:=4$,
one by one, and the corresponding operations are put in its buffer.
Then, $P1$ executes the instructions $y:=1$, $y:=2$, $y:=3$, and $y:=4$,
one by one, again putting the corresponding operations in its buffer.
Now, the memory is updated with $x:=1$ after which $P1$ executes
$\reg_5:=\xvar$, thus assigning $1$ to $\reg_5$.
Following this, the memory is updated with $x:=2$ after which $P1$ executes
$\reg_6:=\xvar$, thus assigning $2$ to $\reg_6$.
Finally $3$ and $4$ are assigned to $\reg_7$ and $\reg_8$ respectively
in similar manners.
A similar sequence of operations is performed assigning $1$, $2$, $3$ and $4$
to $\reg_1$, $\reg_2$, $\reg_3$, and $\reg_4$ respectively.

However, under the {\sc SiSd} semantics, the {\tt ReadSeq} program does not have
any runs that can satisfy the assertion.
More specifically, since the processes do not have any store buffers,
at most three different values of a variable can be kept in this example. 
I.e., one in the local cache of the process which executes the write operation of the variable,
one in the local cache of the process which executes the read operation of the variable,
and one in the LLC.
When there are four or more values assigned to a variable in a similar manner as 
in {\tt ReadSeq}, the assertion that each process reads all the values of the
variable in the same order as it is written by the other process cannot be satisfied.

\begin{figure}
  \includegraphics[width=260pt]{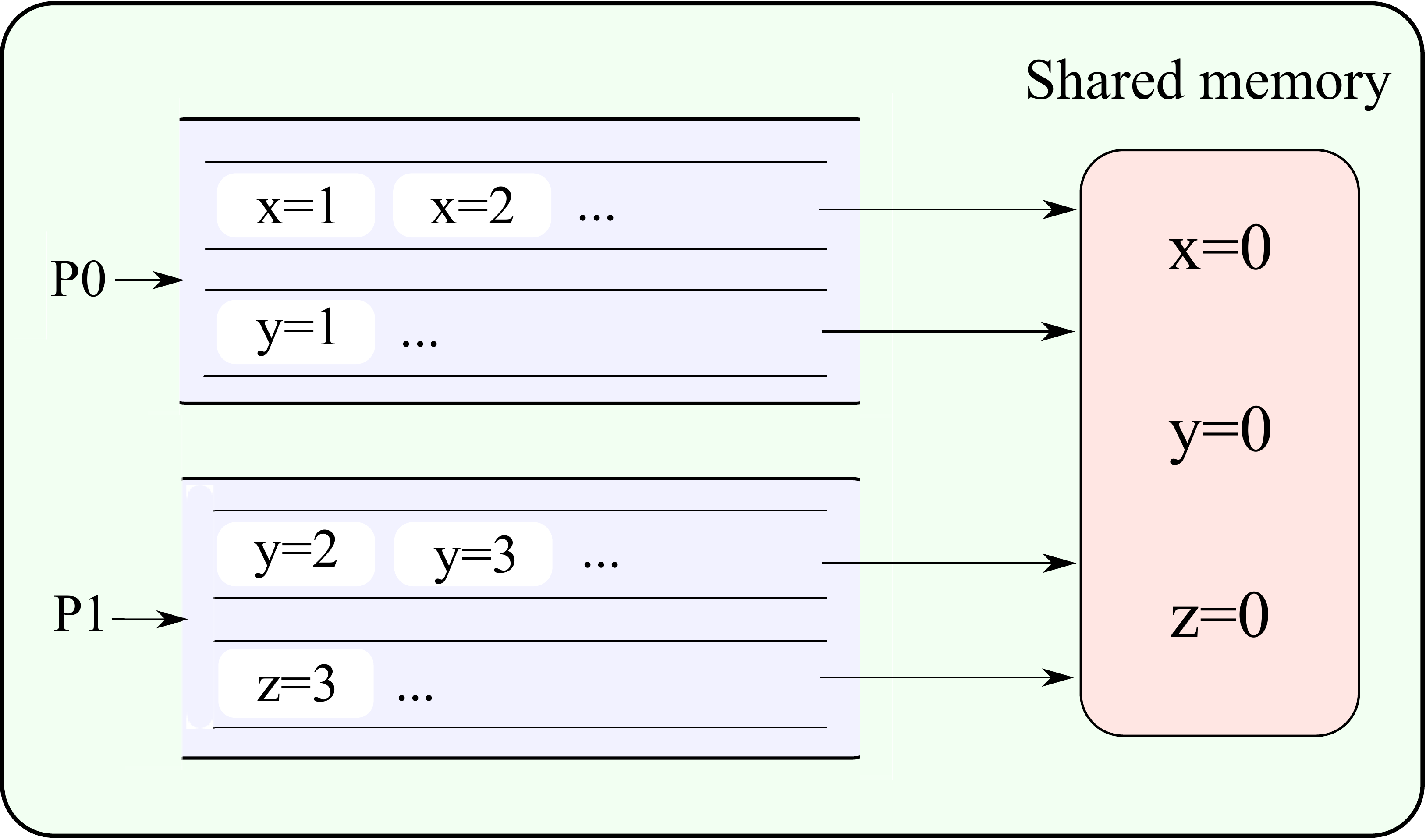}
  \caption{PSO}
  \label{pso:fig}
\end{figure}

\paragraph{Partial Store Order (PSO)}
The {\sc SiSd} model and PSO are not comparable.
In PSO, a store buffer is inserted for each variable, between each process
and the shared memory as illustrated in Figure~\ref{pso:fig} .
%

%\begin{figure}
%\begin{tikzpicture}[background rectangle/.style={rounded corners,dotted,line width=1pt,draw},show background rectangle]
%\node[name=n11]  {$\xvar:= 1$};
%
%
%
%\node[name=n21,anchor=west] at ($(n11.east)+(15pt,7.5pt)$) {$\reg_1 :=\xvar$};
%\node[name=n22,anchor=west] at ($(n11.east)+(15pt,-7.5pt)$) {$\yvar :=1$};
%
%\node[name=n31,anchor=west] at ($(n21.east)+(15pt,0pt)$) {$\reg_2 :=\yvar$};
%\node[name=n32,anchor=west] at ($(n31.west)+(0pt,-15pt)$) {$\reg_3 :=\xvar$};
%
%
%\node[name=th1,anchor=south west]at ($(n11.north west)+(0pt,1pt)$) {$P_1:$};
%\node[name=th2,anchor=south west]at ($(n21.north west)+(0pt,1pt)$) {$P_2:$};
%\node[name=th2,anchor=south west]at ($(n31.north west)+(0pt,1pt)$) {$P_3:$};
%
%\draw[line width= 0.5pt] ($(n21.north west)+(-8.5pt,17pt)$) -- ($(n22.south west)+(-8.5pt,-10pt)$);
%\draw[line width= 0.5pt] ($(n21.north west)+(-6.5pt,17.5pt)$) -- ($(n22.south west)+(-6.5pt,-10pt)$);
%
%\draw[line width= 0.5pt] ($(n31.north west)+(-8.5pt,17pt)$) -- ($(n32.south west)+(-8.5pt,-10pt)$);
%\draw[line width= 0.5pt] ($(n31.north west)+(-6.5pt,17pt)$) -- ($(n32.south west)+(-6.5pt,-10pt)$);
%
%\node[anchor=south] at  ($(n11.north east)+(7.5pt,22.5pt)$) {Initially: $\xvar=0$ , $\yvar=0$};
%
%\node[anchor=north,draw] at ($(n22.south)+(0pt,-19.5pt)$) {$\reg_1=1$ and $\reg_2=1$ and $\reg_3=0$};
%\end{tikzpicture}
%\caption{Program {\tt WRC}.}
%\label{pso:fig}
%\end{figure}

\begin{figure} \begin{center}

\begin{tikzpicture}
[background rectangle/.style={rounded corners,fill=green!5,line width=1pt,draw},show background rectangle]

\node[name=dummy]{};

\node[pgmtextnode,name=data,scale=1.1] at (dummy)
{\tt Initially: x=0, y=0};

%%%%%%%%%%%%%%% Proc 0 %%%%%%%%%%%%%%%

\node[pgmtextnode,name=p0,anchor=north west,scale=1.1] at ($(data.south west)+(0pt,-12pt)$)
{\tt process P0};

\node[pgmtextnode,name=r0,anchor=north west,scale=1.1] at ($(p0.south west)+(0pt,-2pt)$)
{\tt registers};

\node[pgmtextnode,name=b0,anchor=north west,scale=1.1] at ($(r0.south west)+(0pt,-2pt)$)
{\tt begin};

\node[pgmtextnode,name=l1,anchor=north west,scale=1.1] at ($(b0.south west)+(2pt,-2pt)$)
{\tt L1:  \vphantom{y\$}x := 1;};

%\node[pgmtextnode,name=l2,anchor=north west] at ($(l1.south west)+(0pt,-2pt)$)
%{\tt L2:  \vphantom{y\$}\$r1 := y;};

\node[pgmtextnode,name=e0,anchor=north west,scale=1.1] at ($(l1.south west)+(0pt,-2pt)$)
{\tt end};

\begin{pgfonlayer}{foreground}
  \node[fitnode,dotted,fill=red!5,fit= (b0) (p0) (r0) (l1) (e0)]{};
\end{pgfonlayer}

%%%%%%%%%%%%%%% Proc 1 %%%%%%%%%%%%%%%

\node[pgmtextnode,name=p1,anchor=west,scale=1.1] at ($(p0.east)+(23pt,0pt)$)
{\tt process P1};

\node[pgmtextnode,name=r1a,anchor=north west,scale=1.1] at ($(p1.south west)+(0pt,-2pt)$)
{\tt registers \$r1};

\node[pgmtextnode,name=b1,anchor=north west,scale=1.1] at ($(r1a.south west)+(0pt,-2pt)$)
{\tt begin};

\node[pgmtextnode,name=l2,anchor=north west,scale=1.1] at ($(b1.south west)+(2pt,-2pt)$)
{\tt L2:  \$r1 := x;};

\node[pgmtextnode,name=l3,anchor=north west,scale=1.1] at ($(l2.south west)+(0pt,-2pt)$)
{\tt L3:  y := 1;};

\node[pgmtextnode,name=e1,anchor=north west,scale=1.1] at ($(l3.south west)+(-2pt,-2pt)$)
{\tt end};

\begin{pgfonlayer}{foreground}
  \node[fitnode,dotted,fill=blue!5,fit= (b1) (p1) (r1a) (l2) (l3) (e1)]{};
\end{pgfonlayer}

%%%%%%%%%%%%%%% Proc 2 %%%%%%%%%%%%%%%

\node[pgmtextnode,name=p2,anchor=west,scale=1.1] at ($(p1.east)+(32pt,0pt)$)
{\tt process P2};

\node[pgmtextnode,name=r2a,anchor=north west,scale=1.1] at ($(p2.south west)+(0pt,-2pt)$)
{\tt registers \$r2 \$r3};

\node[pgmtextnode,name=b2,anchor=north west,scale=1.1] at ($(r2a.south west)+(0pt,-2pt)$)
{\tt begin};

\node[pgmtextnode,name=l4,anchor=north west,scale=1.1] at ($(b2.south west)+(2pt,-2pt)$)
{\tt L4:  \$r2 := y;};

\node[pgmtextnode,name=l5,anchor=north west,scale=1.1] at ($(l4.south west)+(0pt,-2pt)$)
{\tt L5:  \$r3 := x;};

\node[pgmtextnode,name=e2,anchor=north west,scale=1.1] at ($(l5.south west)+(-2pt,-2pt)$)
{\tt end};

\begin{pgfonlayer}{foreground}
  \node[fitnode,dotted,fill=yellow!5,fit= (b2) (p2) (r2a) (l4) (l5) (e2)]{};
\end{pgfonlayer}

\node[pgmtextnode,name=registers,anchor=north west,scale=1.1] at ($(p0.south west)+(0pt,-58pt)$)
{\tt Assertion: \$r1=1, \$r2=1, \$r3=0};

\end{tikzpicture}
\caption{Program {\tt WRC}.}
\label{wrc:fig}
\end{center}\end{figure}

To illustrate the difference between {\sc SiSd} and PSO,
we consider the program {\tt WRC} in Figure~\ref{wrc:fig}.
Under the PSO semantics, the program {\tt WRC} does not have any runs that satisfy the assertion.
More precisely, if $\reg_1=1$ then the write operation $\xvar:=1$ by $P0$ must have reached the memory
before the write operation $\yvar:=1$ has been performed by $P2$.
Furthermore, if $\reg_2=1$ holds then the write operation $\yvar:=1$
must have reached the memory before the instruction $\reg_3:=\xvar$ has been performed by $P2$.
Since read instructions are not re-ordered in PSO and since $\xvar:=1$ reaches the memory before
$\yvar:=1$ it follows that the value of $\xvar$ is equal to $1$ in the memory when 
$\reg_3:=\xvar$ is performed by $P2$, and hence $\reg_3=1$.

The program {\tt WRC} has the following run under {\sc SiSd}
that satisfies the assertion.
The process $P2$ fetches the initial value $0$ of the variable
$\xvar$.
The process $P0$ assigns $1$ to $\xvar$ and updates the value to the LLC.
The process $P1$ fetches the value $1$ of $\xvar$ from the LLC and then executes
the instruction $\reg_1:=\xvar$ which means that the value of $\reg_1$ is equal to $1$.
Then, $P1$ assigns the value $1$ to the variable $\yvar$ and updates the value to the LLC.
The process $P2$ fetches the value of $\yvar$ from the LLC and then executes
the instruction $\reg_2:=\yvar$ which means that $\reg_2=1$.
Finally, $P2$ executes
the instruction $\reg_3:=\xvar$ and we get $\reg_3=0$.

As PSO is weaker than TSO, we can use the same example in Figure~\ref{read:seq:fig} 
to show that the {\tt ReadSeq} program also has a run under PSO satisfying the assertion,
which means that {\sc SiSd} is not weaker than PSO.

\paragraph{POWER/ARM}
The {\sc SiSd} model and POWER are not comparable.
We show that the POWER memory model is not weaker than {\sc SiSd} with the
program {\tt PwrEg} in Figure~\ref{pwreg:fig}. The program does not
have any runs that satisfy the assertion under POWER, since 
cycles of the form {\tt write-(sync)$\rightarrow$write-(read from)$\rightarrow$ 
read-(address dependency)$\rightarrow$read-(from read)$\rightarrow$} are
not allowed by POWER. The sync instruction maintains the order
between the two write instructions and the address dependency maintains the order
between the two read instructions, which make the cycle impossible.

\begin{figure}  \begin{center}

\begin{tikzpicture}
[background rectangle/.style={rounded corners,fill=green!5,line width=1pt,draw},show background rectangle]

\node[name=dummy]{};

\node[pgmtextnode,name=data,scale=1.1] at (dummy)
{\tt Initially: x=0, y=0};

%%%%%%%%%%%%%%% Proc 0 %%%%%%%%%%%%%%%

\node[pgmtextnode,name=p0,anchor=north west,scale=1.1] at ($(data.south west)+(0pt,-12pt)$)
{\tt process P0};

\node[pgmtextnode,name=r0,anchor=north west,scale=1.1] at ($(p0.south west)+(0pt,-2pt)$)
{\tt registers };

\node[pgmtextnode,name=b0,anchor=north west,scale=1.1] at ($(r0.south west)+(0pt,-2pt)$)
{\tt begin};

\node[pgmtextnode,name=l1,anchor=north west,scale=1.1] at ($(b0.south west)+(2pt,-2pt)$)
{\tt L1:  \vphantom{y\$}x := 1;};

\node[pgmtextnode,name=l2,anchor=north west,scale=1.1] at ($(l1.south west)+(0pt,-2pt)$)
{\tt L2: sync;};

\node[pgmtextnode,name=l3,anchor=north west,scale=1.1] at ($(l2.south west)+(0pt,-2pt)$)
{\tt L3:  \vphantom{y\$}y := 1;};

\node[pgmtextnode,name=e0,anchor=north west,scale=1.1] at ($(l3.south west)+(0pt,-2pt)$)
{\tt end};

\begin{pgfonlayer}{foreground}
  \node[fitnode,dotted,fill=red!5,fit= (b0) (p0) (r0) (l1) (l2) (l3) (e0)]{};
\end{pgfonlayer}

%%%%%%%%%%%%%%% Proc 1 %%%%%%%%%%%%%%%

\node[pgmtextnode,name=p1,anchor=west,scale=1.1] at ($(p0.east)+(25pt,0pt)$)
{\tt process P};

\node[pgmtextnode,name=r1a,anchor=north west,scale=1.1] at ($(p1.south west)+(0pt,-2pt)$)
{\tt registers \$r1 \$r2};

\node[pgmtextnode,name=b1,anchor=north west,scale=1.1] at ($(r1a.south west)+(0pt,-2pt)$)
{\tt begin};

\node[pgmtextnode,name=l4,anchor=north west,scale=1.1] at ($(b1.south west)+(2pt,-2pt)$)
{\tt L4:  \$r1 := y;};

\node[pgmtextnode,name=l5,anchor=north west,scale=1.1] at ($(l4.south west)+(0pt,-2pt)$)
{\tt L5:  \$r2 := (\&x + 0*\$r1);};

\node[pgmtextnode,name=e1,anchor=north west,scale=1.1] at ($(l5.south west)+(-2pt,-2pt)$)
{\tt end};

\begin{pgfonlayer}{foreground}
  \node[fitnode,dotted,fill=blue!5,fit= (b1) (p1) (r1a) (l4) (l5) (e1)]{};
\end{pgfonlayer}

\node[pgmtextnode,name=registers,anchor=north west,scale=1.1] at ($(p0.south west)+(0pt,-66pt)$)
{\tt Assertion: \$r1=1, \$r2=0};

\end{tikzpicture}
\caption{PwrEg}
\label{pwreg:fig}
\end{center}\end{figure}

\begin{figure} \begin{center}

\begin{tikzpicture}
[background rectangle/.style={rounded corners,fill=green!5,line width=1pt,draw},show background rectangle]

\node[name=dummy]{};

\node[pgmtextnode,name=data,scale=1.1] at (dummy)
{\tt Initially: x=0, y=0};

%%%%%%%%%%%%%%% Proc 0 %%%%%%%%%%%%%%%

\node[pgmtextnode,name=p0,anchor=north west,scale=1.1] at ($(data.south west)+(0pt,-12pt)$)
{\tt process P0};

\node[pgmtextnode,name=r0,anchor=north west,scale=1.1] at ($(p0.south west)+(0pt,-2pt)$)
{\tt registers };

\node[pgmtextnode,name=b0,anchor=north west,scale=1.1] at ($(r0.south west)+(0pt,-2pt)$)
{\tt begin};

\node[pgmtextnode,name=l1,anchor=north west,scale=1.1] at ($(b0.south west)+(2pt,-2pt)$)
{\tt L1:  \vphantom{y\$}x := 1;};

\node[pgmtextnode,name=l2,anchor=north west,scale=1.1] at ($(l1.south west)+(0pt,-2pt)$)
{\tt L2: fence;};

\node[pgmtextnode,name=l3,anchor=north west,scale=1.1] at ($(l2.south west)+(0pt,-2pt)$)
{\tt L3:  \vphantom{y\$}y := 1;};

\node[pgmtextnode,name=e0,anchor=north west,scale=1.1] at ($(l3.south west)+(0pt,-2pt)$)
{\tt end};

\begin{pgfonlayer}{foreground}
  \node[fitnode,dotted,fill=red!5,fit= (b0) (p0) (r0) (l1) (l2) (l3) (e0)]{};
\end{pgfonlayer}

%%%%%%%%%%%%%%% Proc 1 %%%%%%%%%%%%%%%

\node[pgmtextnode,name=p1,anchor=west,scale=1.1] at ($(p0.east)+(25pt,0pt)$)
{\tt process P1};

\node[pgmtextnode,name=r1a,anchor=north west,scale=1.1] at ($(p1.south west)+(0pt,-2pt)$)
{\tt registers \$r1 \$r2};

\node[pgmtextnode,name=b1,anchor=north west,scale=1.1] at ($(r1a.south west)+(0pt,-2pt)$)
{\tt begin};

\node[pgmtextnode,name=l4,anchor=north west,scale=1.1] at ($(b1.south west)+(2pt,-2pt)$)
{\tt L4:  \$r1 := y;};

\node[pgmtextnode,name=l5,anchor=north west,scale=1.1] at ($(l4.south west)+(0pt,-2pt)$)
{\tt L5:  \$r2 := x;};

\node[pgmtextnode,name=e1,anchor=north west,scale=1.1] at ($(l5.south west)+(-2pt,-2pt)$)
{\tt end};

\begin{pgfonlayer}{foreground}
  \node[fitnode,dotted,fill=blue!5,fit= (b1) (p1) (r1a) (l4) (l5) (e1)]{};
\end{pgfonlayer}

\node[pgmtextnode,name=registers,anchor=north west,scale=1.1] at ($(p0.south west)+(0pt,-66pt)$)
{\tt Assertion: \$r1=1, \$r2=0};

\end{tikzpicture}
\caption{SisdEg}
\label{sisdeg:fig}
\end{center}\end{figure}

However, under {\sc SiSd} the program {\tt SisdEg} in Figure~\ref{sisdeg:fig} has the run that satisfies
the assertion. The process $P1$ fetches the initial value 0 of the variable $\xvar$.
The process $P0$ assigns 1 to $\xvar$, updates the value to the LLC, assigns 1 to $\yvar$, 
and updates the value to the LLC again. The process $P1$ fetches the value 1 of $\yvar$
from the LLC and then executes the instruction $\reg_1:=\yvar$. Therefore, the value of $\reg_1$
is equal to 1. Finally, $P1$ executes the instruction $\reg_2:=\xvar$ and we get
$\reg_2=0$.

We can show that {\sc SiSd} is not weaker than POWER with the program {\tt ReadSeq}
in Figure~\ref{read:seq:fig}.

We can also show that the {\sc SiSd} model and ARM are not comparable with the
programs in Figure~\ref{read:seq:fig}, Figure~\ref{pwreg:fig} and
Figure~\ref{sisdeg:fig}.

\paragraph{Relaxed Memory Order (RMO)}
We can show that {\sc SiSd} is not weaker than RMO again with the program {\tt ReadSeq}
in Figure~\ref{read:seq:fig}, since RMO allows the reorder between both 
1) write and write operations and 2) read and read/write operations.

We keep the other direction open since we have not found any examples as the proof.

\subsection{Further Litmus Tests}
We describe the behavior of the {\sc SiSd} model for three more litmus tests.

\paragraph{Load-Buffering (LB)} The {\tt LB} program in Figure~\ref{lb:fig} does not have any
runs that can satisfy the assertion under {\sc SiSd}. If $\reg_1=1$, then 
process $P1$ must have executed the write instruction $\xvar:=1$ and updated 
the value 1 of $\xvar$ to the LLC before $P0$ has executed the instruction
$\reg_1:=\xvar$. This means when $P1$ executes the read instruction
$\reg_2:=\yvar$, the value of $\yvar$ has not been updated by $P0$,
and thus $\reg_2=0$.

\begin{figure}[h] \begin{center}

\begin{tikzpicture}
[background rectangle/.style={rounded corners,fill=green!5,line width=1pt,draw},show background rectangle]

\node[name=dummy]{};

\node[pgmtextnode,name=data,scale=1.1] at (dummy)
{\tt Initially: x=0, y=0};

%%%%%%%%%%%%%%% Proc 0 %%%%%%%%%%%%%%%

\node[pgmtextnode,name=p0,anchor=north west,scale=1.1] at ($(data.south west)+(0pt,-10pt)$)
{\tt process P0};

\node[pgmtextnode,name=r0,anchor=north west,scale=1.1] at ($(p0.south west)+(0pt,-2pt)$)
{\tt registers \$r1};

\node[pgmtextnode,name=b0,anchor=north west,scale=1.1] at ($(r0.south west)+(0pt,-2pt)$)
{\tt begin};

\node[pgmtextnode,name=l1,anchor=north west,scale=1.1] at ($(b0.south west)+(2pt,-2pt)$)
{\tt L1:  \vphantom{y\$}\$r1 := x;};

\node[pgmtextnode,name=l2,anchor=north west,scale=1.1] at ($(l1.south west)+(0pt,-2pt)$)
{\tt L2:  \vphantom{y\$}y := 1;};

\node[pgmtextnode,name=e0,anchor=north west,scale=1.1] at ($(l2.south west)+(0pt,-2pt)$)
{\tt end};

\begin{pgfonlayer}{foreground}
  \node[fitnode,dotted,fill=red!5,fit= (b0) (p0) (r0) (l1) (l2) (e0)]{};
\end{pgfonlayer}

%%%%%%%%%%%%%%% Proc 1 %%%%%%%%%%%%%%%

\node[pgmtextnode,name=p1,anchor=west,scale=1.1] at ($(p0.east)+(31pt,0pt)$)
{\tt process P1};

\node[pgmtextnode,name=r1a,anchor=north west,scale=1.1] at ($(p1.south west)+(0pt,-2pt)$)
{\tt registers \$r2};

\node[pgmtextnode,name=b1,anchor=north west,scale=1.1] at ($(r1a.south west)+(0pt,-2pt)$)
{\tt begin};

\node[pgmtextnode,name=l3,anchor=north west,scale=1.1] at ($(b1.south west)+(2pt,-2pt)$)
{\tt L3:  \$r2 := y;};

\node[pgmtextnode,name=l4,anchor=north west,scale=1.1] at ($(l3.south west)+(0pt,-2pt)$)
{\tt L4:  x := 1;};

\node[pgmtextnode,name=e1,anchor=north west,scale=1.1] at ($(l4.south west)+(-2pt,-2pt)$)
{\tt end};

\begin{pgfonlayer}{foreground}
  \node[fitnode,dotted,fill=blue!5,fit= (b1) (p1) (r1a) (l3) (l4) (e1)]{};
\end{pgfonlayer}

\node[pgmtextnode,name=registers,anchor=north west,scale=1.1] at ($(p0.south west)+(0pt,-58pt)$)
{\tt Assertion: \$r1=1, \$r2=1};

\end{tikzpicture}
\caption{Program {\tt LB}.}
\label{lb:fig}
\end{center}\end{figure}

\paragraph{ISA2} 
The program {\tt ISA2} in Figure~\ref{isa2:fig} has the following run
under {\sc SiSd} that satisfies the assertion.
The process $P2$ fetches the initial value $0$ of the variable
$\xvar$.
The process $P0$ assigns $1$ to $\xvar$ and $\yvar$, and updates the value to the LLC.
The process $P1$ fetches the value $1$ of $\yvar$ from the LLC and then executes
the instruction $\reg_1:=\yvar$, which means that the value of $\reg_1$ is equal to $1$.
Then, $P1$ assigns the value $1$ to the variable $\zvar$ and updates the value to the LLC.
The process $P2$ fetches the value of $\zvar$ from the LLC and then executes
the instruction $\reg_2:=\zvar$, which means that $\reg_2=1$.
Finally, $P2$ executes the instruction $\reg_3:=\xvar$ and we get $\reg_3=0$.

\begin{figure} \begin{center}

\begin{tikzpicture}
[background rectangle/.style={rounded corners,fill=green!5,line width=1pt,draw},show background rectangle]

\node[name=dummy]{};

\node[pgmtextnode,name=data,scale=1.1] at (dummy)
{\tt Initially: x=0, y=0, z=0};

%%%%%%%%%%%%%%% Proc 0 %%%%%%%%%%%%%%%

\node[pgmtextnode,name=p0,anchor=north west,scale=1.1] at ($(data.south west)+(0pt,-12pt)$)
{\tt process P0};

\node[pgmtextnode,name=r0,anchor=north west,scale=1.1] at ($(p0.south west)+(0pt,-2pt)$)
{\tt registers};

\node[pgmtextnode,name=b0,anchor=north west,scale=1.1] at ($(r0.south west)+(0pt,-2pt)$)
{\tt begin};

\node[pgmtextnode,name=l1,anchor=north west,scale=1.1] at ($(b0.south west)+(2pt,-2pt)$)
{\tt L1:  \vphantom{y\$}x := 1;};

\node[pgmtextnode,name=l2,anchor=north west,scale=1.1] at ($(l1.south west)+(0pt,-2pt)$)
{\tt L2:  \vphantom{y\$}y := 1;};

\node[pgmtextnode,name=e0,anchor=north west,scale=1.1] at ($(l2.south west)+(0pt,-2pt)$)
{\tt end};

\begin{pgfonlayer}{foreground}
  \node[fitnode,dotted,fill=red!5,fit= (b0) (p0) (r0) (l1) (l2) (e0)]{};
\end{pgfonlayer}

%%%%%%%%%%%%%%% Proc 1 %%%%%%%%%%%%%%%

\node[pgmtextnode,name=p1,anchor=west,scale=1.1] at ($(p0.east)+(23pt,0pt)$)
{\tt process P1};

\node[pgmtextnode,name=r1a,anchor=north west,scale=1.1] at ($(p1.south west)+(0pt,-2pt)$)
{\tt registers \$r1};

\node[pgmtextnode,name=b1,anchor=north west,scale=1.1] at ($(r1a.south west)+(0pt,-2pt)$)
{\tt begin};

\node[pgmtextnode,name=l3,anchor=north west,scale=1.1] at ($(b1.south west)+(2pt,-2pt)$)
{\tt L3:  \$r1 := y;};

\node[pgmtextnode,name=l4,anchor=north west,scale=1.1] at ($(l3.south west)+(0pt,-2pt)$)
{\tt L4:  z := 1;};

\node[pgmtextnode,name=e1,anchor=north west,scale=1.1] at ($(l4.south west)+(-2pt,-2pt)$)
{\tt end};

\begin{pgfonlayer}{foreground}
  \node[fitnode,dotted,fill=blue!5,fit= (b1) (p1) (r1a) (l3) (l4) (e1)]{};
\end{pgfonlayer}

%%%%%%%%%%%%%%% Proc 2 %%%%%%%%%%%%%%%

\node[pgmtextnode,name=p2,anchor=west,scale=1.1] at ($(p1.east)+(31pt,0pt)$)
{\tt process P2};

\node[pgmtextnode,name=r2a,anchor=north west,scale=1.1] at ($(p2.south west)+(0pt,-2pt)$)
{\tt registers \$r2 \$r3};

\node[pgmtextnode,name=b2,anchor=north west,scale=1.1] at ($(r2a.south west)+(0pt,-2pt)$)
{\tt begin};

\node[pgmtextnode,name=l5,anchor=north west,scale=1.1] at ($(b2.south west)+(2pt,-2pt)$)
{\tt L5:  \$r2 := z;};

\node[pgmtextnode,name=l6,anchor=north west,scale=1.1] at ($(l5.south west)+(0pt,-2pt)$)
{\tt L6:  \$r3 := x;};

\node[pgmtextnode,name=e2,anchor=north west,scale=1.1] at ($(l6.south west)+(-2pt,-2pt)$)
{\tt end};

\begin{pgfonlayer}{foreground}
  \node[fitnode,dotted,fill={yellow!5},fit= (b2) (p2) (r2a) (l5) (l6) (e2)]{};
\end{pgfonlayer}

\node[pgmtextnode,name=registers,anchor=north west,scale=1.1] at ($(p0.south west)+(0pt,-58pt)$)
{\tt Assertion: \$r1=1, \$r2=1, \$r3=0};

\end{tikzpicture}
\caption{Program {\tt ISA2}.}
\label{isa2:fig}
\end{center}\end{figure}

\paragraph{IRIW}
The program {\tt IRIW} in Figure~\ref{iriw:fig} has the following run
under {\sc SiSd} that satisfies the assertion.
The processes $P1$ and $P3$ fetch the initial values $0$ of the variables 
$\yvar$ and $\xvar$ respectively.
The process $P0$ assigns $1$ to $\xvar$ and updates the values to the LLC.
The process $P1$ fetches the value $1$ of $\xvar$ from the LLC and executes
the instruction $\reg_1:=\xvar$, which means that the value of $\reg_1$ is equal to $1$.
Then $P1$ executes the instruction $\reg_2:=\yvar$, which means that $\reg_2=0$.
After that, the process $P2$ assigns the value $1$ to the variable $\yvar$ and 
updates the value to the LLC.
The process $P3$ fetches the value of $\yvar$ from the LLC and then executes
the instruction $\reg_3:=\yvar$, which means that $\reg_3=1$.
Finally, $P3$ executes the instruction $\reg_4:=\xvar$ and we get $\reg_4=0$.

\begin{figure} \begin{center}

\begin{tikzpicture}
[background rectangle/.style={rounded corners,fill=green!5,line width=1pt,draw},show background rectangle]

\node[name=dummy]{};

\node[pgmtextnode,name=data,scale=1.1] at (dummy)
{\tt Initially: x=0, y=0};

%%%%%%%%%%%%%%% Proc 0 %%%%%%%%%%%%%%%

\node[pgmtextnode,name=p0,anchor=north west,scale=1.1] at ($(data.south west)+(0pt,-12pt)$)
{\tt process P0};

\node[pgmtextnode,name=r0,anchor=north west,scale=1.1] at ($(p0.south west)+(0pt,-2pt)$)
{\tt registers};

\node[pgmtextnode,name=b0,anchor=north west,scale=1.1] at ($(r0.south west)+(0pt,-2pt)$)
{\tt begin};

\node[pgmtextnode,name=l1,anchor=north west,scale=1.1] at ($(b0.south west)+(2pt,-2pt)$)
{\tt L1:  \vphantom{y\$}x := 1;};

%\node[pgmtextnode,name=l2,anchor=north west,scale=1.1] at ($(l1.south west)+(0pt,-2pt)$)
%{\tt L2:  \vphantom{y\$}y := 1;};

\node[pgmtextnode,name=e0,anchor=north west,scale=1.1] at ($(l1.south west)+(0pt,-2pt)$)
{\tt end};

\begin{pgfonlayer}{foreground}
  \node[fitnode,dotted,fill=red!5,fit= (b0) (p0) (r0) (l1) (e0)]{};
\end{pgfonlayer}

%%%%%%%%%%%%%%% Proc 1 %%%%%%%%%%%%%%%

\node[pgmtextnode,name=p1,anchor=west,scale=1.1] at ($(p0.east)+(23pt,0pt)$)
{\tt process P1};

\node[pgmtextnode,name=r1a,anchor=north west,scale=1.1] at ($(p1.south west)+(0pt,-2pt)$)
{\tt registers \$r1 \$r2};

\node[pgmtextnode,name=b1,anchor=north west,scale=1.1] at ($(r1a.south west)+(0pt,-2pt)$)
{\tt begin};

\node[pgmtextnode,name=l2,anchor=north west,scale=1.1] at ($(b1.south west)+(2pt,-2pt)$)
{\tt L2:  \$r1 := x;};

\node[pgmtextnode,name=l3,anchor=north west,scale=1.1] at ($(l2.south west)+(0pt,-2pt)$)
{\tt L3:  \$r2 := y;};

\node[pgmtextnode,name=e1,anchor=north west,scale=1.1] at ($(l3.south west)+(-2pt,-2pt)$)
{\tt end};

\begin{pgfonlayer}{foreground}
  \node[fitnode,dotted,fill=blue!5,fit= (b1) (p1) (r1a) (l2) (l3) (e1)]{};
\end{pgfonlayer}

%%%%%%%%%%%%%%% Proc 2 %%%%%%%%%%%%%%%

\node[pgmtextnode,name=p2,anchor=west,scale=1.1] at ($(p1.east)+(43pt,0pt)$)
{\tt process P2};

\node[pgmtextnode,name=r2a,anchor=north west,scale=1.1] at ($(p2.south west)+(0pt,-2pt)$)
{\tt registers};

\node[pgmtextnode,name=b2,anchor=north west,scale=1.1] at ($(r2a.south west)+(0pt,-2pt)$)
{\tt begin};

\node[pgmtextnode,name=l4,anchor=north west,scale=1.1] at ($(b2.south west)+(2pt,-2pt)$)
{\tt L4:  y := 1;};

%\node[pgmtextnode,name=l6,anchor=north west,scale=1.1] at ($(l5.south west)+(0pt,-2pt)$)
%{\tt L5:  \$r3 := x;};

\node[pgmtextnode,name=e2,anchor=north west,scale=1.1] at ($(l4.south west)+(-2pt,-2pt)$)
{\tt end};

\begin{pgfonlayer}{foreground}
  \node[fitnode,dotted,fill=yellow!5,fit= (b2) (p2) (r2a) (l4) (e2)]{};
\end{pgfonlayer}

%%%%%%%%%%%%%%% Proc 3 %%%%%%%%%%%%%%%

\node[pgmtextnode,name=p3,anchor=west,scale=1.1] at ($(p2.east)+(23pt,0pt)$)
{\tt process P3};

\node[pgmtextnode,name=r3a,anchor=north west,scale=1.1] at ($(p3.south west)+(0pt,-2pt)$)
{\tt registers \$r3 \$r4};

\node[pgmtextnode,name=b3,anchor=north west,scale=1.1] at ($(r3a.south west)+(0pt,-2pt)$)
{\tt begin};

\node[pgmtextnode,name=l5,anchor=north west,scale=1.1] at ($(b3.south west)+(2pt,-2pt)$)
{\tt L5:  \$r3 := y;};

\node[pgmtextnode,name=l6,anchor=north west,scale=1.1] at ($(l5.south west)+(0pt,-2pt)$)
{\tt L6:  \$r4 := x;};

\node[pgmtextnode,name=e3,anchor=north west,scale=1.1] at ($(l6.south west)+(-2pt,-2pt)$)
{\tt end};

\begin{pgfonlayer}{foreground}
  \node[fitnode,dotted,fill={red!50!blue!5!white},fit= (b3) (p3) (r3a) (l5) (l6) (e3)]{};
\end{pgfonlayer}

\node[pgmtextnode,name=registers,anchor=north west,scale=1.1] at ($(p0.south west)+(0pt,-58pt)$)
{\tt Assertion: \$r1=1, \$r2=0, \$r3=1, \$r4=0};

\end{tikzpicture}
\caption{Program {\tt IRIW}.}
\label{iriw:fig}
\end{center}\end{figure} 

%%% Local Variables:
%%% mode: latex
%%% TeX-master: "main.tex"
%%% End:

\newcommand{\insfcset}[2]{\textrm{${#1}\oplus{#2}$}}
\newcommand{\optimal}{\textrm{\texttt{opt}}}
\newcommand{\constraints}{\textrm{\texttt{req}}}
\newcommand{\hittingsets}{\textrm{\textbf{hits}}}
\newcommand{\reachable}{\textrm{\textbf{reachable}}}
\newcommand{\witanalyze}{\textrm{\textbf{analyze\_witness}}}

\section{Fence Insertion} 
\label{fence:section}

In this section we describe our fence insertion procedure, which is
closely related to the algorithm described in \cite{LiuNPVY12}.
Given a program $\prog$, a cost function $\kappa$ and a safety
property $\phi$, the procedure finds \textit{all} the sets of fences
that are optimal for $\prog$ w.r.t.\ $\phi$ and $\kappa$.

In this section we take \emph{fence constraint} (or \emph{fence} for
short) to mean a pair $(\lbl,f)$ where $\lbl$ is a statement label and
$f$ is a fence instruction. A fence constraint $(\lbl,f)$ should be
interpreted as the notion of inserting the fence instruction $f$ into
a program, between the statement labeled $\lbl$ and the next statement
(labeled by $\nextof\lbl$)\footnote{This definition can be
  generalized. Our prototype tool does indeed support a more general
  definition of fence positions, which is left out of the article for
  simplicity.}. For a program $\prog$ and a set $F$ of fence
constraints, we define $\insfcset{\prog}{F}$ to mean the program
$\prog$ where all fence constraints in $F$ have been inserted. To
avoid ambiguities in the case when $F$ contains multiple fence
constraints with the same statement label (e.g $(\lbl,\llfence)$ and
$(\lbl,\ssfence)$), we assume that fences are always inserted in some
fixed order.

\begin{defi}[Soundness of Fence Sets]
  For a program $\prog$, safety property $\phi$, and set $F$ of fence
  constraints, the set $F$ is sound for $\prog$ w.r.t.\ $\phi$ if
  $\insfcset{\prog}{F}$ satisfies $\phi$ under \textsc{SiSd}.
\end{defi}

A \emph{cost function} $\kappa$ is a function from fence constraints
to positive integer costs. We extend the notion of a cost function to sets of fence constraints
in the natural way: For a cost function $\kappa$ and a set $F$ of
fence constraints, we define $\kappa(F)=\sum_{c\in{}F}\kappa(c)$.

\begin{defi}[Optimality of Fence Sets]
  For a program $\prog$, safety property $\phi$, cost function
  $\kappa$, and set $F$ of fence constraints, $F$ is
  optimal for $\prog$ w.r.t.\ $\phi$ and $\kappa$ if $F$ is sound for
  $\prog$ w.r.t.\ $\phi$, and there is no sound fence set $G$
  for $\prog$ w.r.t.\ $\phi$ where $\kappa(G)<\kappa(F)$.
\end{defi}

Observe that the optimality is evaluated with the number of occurrences
of fences in the source program and the costs of different fences.

In order to introduce our algorithm, we define the notion of a
\emph{hitting set}.

\begin{defi}[Hitting Set]
  For a set $S = \{S_0,\cdots,S_n\}$ of sets $S_0,\cdots,S_n$, and a
  set $T$, we say that $T$ is a hitting set of $S$ if
  $T\cap{}S_i\neq\emptyset$ for all $0\leq{}i\leq{}n$.
\end{defi}

For example $\{a,d\}$ is a hitting set of $\{\{a,b,c\},\{d\},\{a,e\}\}$.
For a set $S$ of sets, hitting sets of $S$ can be computed using
various search techniques, such as constraint programming. We
will assume that we are given a function $\hittingsets(S,\kappa)$
which computes all hitting sets for $S$ which are cheapest
w.r.t. $\kappa$. I.e., for a set $S$ of finite sets, and a cost
function $\kappa$, the call $\hittingsets(S,\kappa)$ returns the set
of all sets $T$ with $T\subseteq\bigcup_{S_i\in{}S}S_i$ such
that 
\begin{itemize}
\item $T$ is a hitting set of $S$, and %\item 
\item there is no hitting set $T'$ of $S$ such that $\kappa(T')<\kappa(T)$.
\end{itemize}

\begin{figure}[h]

\begin{center}
  \begin{algo1}{Fencins($\prog$,$\phi$,$\kappa$)}{algofencinslnctr}
    \lin{\optimal{} := $\emptyset$; \icmt{Optimal fence sets}}
    \lin{\constraints{} := $\emptyset$; \icmt{Known requirements}}
    \while{$\exists\textrm{\texttt{F}}\in\hittingsets(\constraints,\kappa)\setminus\optimal$\label{ln:fencins:loop:begin}\label{ln:fencins:comp:F}}{
      \lin{$\run$ := $\reachable(\insfcset{\prog}{\textrm{\texttt{F}}},\phi)$;\label{ln:fencins:call:reach}}
      \ifunterm[]{$\run = \perp$}{
      \comment{The fence set F is sound }
            \comment{ (and optimal)!}

        \lin{\optimal{} := $\optimal\cup\{\textrm{\texttt{F}}\}$;\label{ln:fencins:add:opt}}
      }\elseterm[$\run$ is a witness run.]{
        \lin{C := $\witanalyze(\insfcset{\prog}{\textrm{\texttt{F}}},\run)$;}
        \comment{C is the set of fences }
                \comment{ that can prevent $\run$.}

        \ifterm[error under SC!]{$\textrm{\texttt{C}} = \emptyset$\label{ln:fencins:sc:error:begin}}{
          \lin{return $\emptyset$;\label{ln:fencins:fail:return}}
        }\lnlabel{ln:fencins:sc:error:end}\lin{\constraints{} := $\constraints\cup\{\textrm{\texttt{C}}\}$;\label{ln:fencins:add:constraint}}
      }
    }\lnlabel{ln:fencins:loop:end}\lin{return \optimal{};} 
  \end{algo1}
  \caption{The fence insertion algorithm.}\label{fig:fencins:algo1}
\end{center}

\end{figure}
We present our fence insertion algorithm in
Figure~\ref{fig:fencins:algo1}. The algorithm keeps two variables
$\optimal$ and $\constraints$. Both are sets of fence constraint sets,
but are intuitively interpreted in different ways. The set $\optimal$ 
contains all the optimal fence constraint sets for $\prog$
w.r.t.\ $\phi$ and $\kappa$ that have been found thus far. The set
$\constraints$ is used to keep track of the requirements that have
been discovered for which fences are necessary for soundness of
$\prog$. We maintain the following invariant for
$\constraints$: Any fence constraint set $F$ which is sound for
$\prog$ w.r.t.\ $\phi$ is a hitting set of $\constraints$. As the
algorithm learns more about $\prog$, the requirements in
$\constraints$ will grow, and hence give more information about what a
sound fence set may look like.
Notice that the invariant holds trivially in the beginning, when
$\constraints=\emptyset$.

In the loop from
lines~\ref{ln:fencins:loop:begin}-\ref{ln:fencins:loop:end} we
repeatedly compute a candidate fence set \texttt{F}
(line~\ref{ln:fencins:comp:F}), insert it into $\prog$, and call the
reachability analysis to check if \texttt{F} is sound
(line~\ref{ln:fencins:call:reach}). We assume that the call
$\reachable(\insfcset{\prog}{\textrm{\texttt{F}}},\phi)$ returns
$\perp$ if $\phi$ is unreachable in
$\insfcset{\prog}{\textrm{\texttt{F}}}$, and a witness run
otherwise.
If $\insfcset{\prog}{\textrm{\texttt{F}}}$ satisfies the safety
property $\phi$, then \texttt{F} is sound. Furthermore, since
\texttt{F} is chosen as one of the cheapest (w.r.t.\ $\kappa$) hitting
sets of $\constraints$, and all sound fence sets are hitting sets of
$\constraints$, it must also be the case that \texttt{F} is
optimal. Therefore, we add \texttt{F} to $\optimal$ on
line~\ref{ln:fencins:add:opt}.

If $\insfcset{\prog}{\textrm{\texttt{F}}}$ does not satisfy the safety
property $\phi$, then we proceed to analyze the witness run $\run$.
The witness analysis procedure is outlined in
Section~\ref{sec:witness:analysis}.
The analysis will return a set \texttt{C} of fence constraints
such that any fence set which is restrictive enough to prevent the
erroneous run $\run$ must contain at least one fence constraint from
\texttt{C}. Since every sound fence set must prevent $\run$, this
means that every sound fence set must have a non-empty intersection
with \texttt{C}. Therefore we add \texttt{C} to $\constraints$ on
line~\ref{ln:fencins:add:constraint}, so that $\constraints$ will
better guide our choice of fence set candidates in the future.

Note that in the beginning, $\hittingsets(\constraints,\kappa)$ will return a 
singleton set of the empty set, namely $\{\emptyset\}$. Then $\texttt{F}$ is chosen as
the empty set $\emptyset$ and the algorithm continues. A special case occurs when the run $\run$ contains no memory access
reorderings. This means that $\prog$ can reach the bad states even
under the SC memory model. Hence it is impossible to correct $\prog$
by only inserting fences. The call
$\witanalyze(\insfcset{\prog}{\textrm{\texttt{F}}},\run)$ will in this
case return the empty set. The main algorithm then terminates, also
returning the empty set, indicating that there are no optimal fence
sets for the given problem.

\subsection{Witness Analysis}\label{sec:witness:analysis}

The $\witanalyze$ function takes as input a program $\prog$ (which may
already contain some fences inserted by the fence insertion
algorithm), and a counter-example run $\run$ generated by the
reachability analysis.
The goal is to find a set $G$ of fences such that
\begin{itemize}
\item all sound fence sets have at least one fence in common with $G$ and
\item $G$ contains no fence which is already in $\prog$.
\end{itemize}
It is desirable to keep $G$ as small as possible, in order to quickly
converge on sound fence sets.

There are several ways to implement $\witanalyze$ to satisfy the above
requirements. One simple way builds on the following insight:
Any sound fence set must prevent the current witness run. The only way
to do that, is to have fences preventing some access reordering that
occurs in the witness. So a set $G$ which contains all fences
preventing some reordering in the current witness satisfies both
requirements listed above.

As an example, consider Figure~\ref{fig:witanalyze:example}. On the
left, we show part of a program $\prog$ where the thread \texttt{P0}
performs three memory accesses \texttt{L0}, \texttt{L1} and
\texttt{L2}. On the right, we show the corresponding part of a
counter-example run $\run$. We see that the store \texttt{L0} becomes
globally visible at line 7, while the loads \texttt{L1} and
\texttt{L2} access the LLC at respectively lines 3 and 5. Hence the
order between the instructions \texttt{L0} and \texttt{L1} and the
order between \texttt{L0} and \texttt{L2} in the program code, is
opposite to the order in which they take effect w.r.t. the LLC in
$\run$. We say that \texttt{L0} is \emph{reordered} with \texttt{L1}
and \texttt{L2}. The loads are not reordered with each other. Let us
assume that $\run$ does not contain any other memory access
reordering.

\begin{figure}
  \centering
  { \begin{tabular}{l@{\rule{10pt}{0pt}}|@{\rule{10pt}{0pt}}l}
    Program fragment & Witness run\\
    \hline
    {\texttt{
      \begin{tabular}{@{}l@{}}
        process P0\\
        ...\\
        L0: \xvar{} := 1;\\
        L1: $\reg_0$ := \yvar;\\
        L2: $\reg_1$ := \zvar;\\
        ...\\
      \end{tabular}
    }}
    &
    {\texttt{
      \begin{tabular}{@{}l@{}l@{}}
           & ...\\
        1. & fetch(P0,\xvar)\\
        2. & L0: \xvar{} := 1\\
        3. & fetch(P0,\yvar)\\
        4. & L1: $\reg_0$ := \yvar\\
        5. & fetch(P0,\zvar)\\
        6. & L2: $\reg_1$ := \zvar\\
           & ... \\
        7. & wrllc(P0,\xvar)\\
           & ...\\
      \end{tabular}
    }}\\
  \end{tabular}}
  \caption{ Left: Part of a program $\prog$, containing three
    instructions of the thread \texttt{P0}. Right:  A
    part of a counter-example run $\run$ of $\prog$.
  }\label{fig:witanalyze:example}

\end{figure}

\noindent  The reordering is caused by the late $\iwrllc$ on line
7. Hence, this particular error run can be prevented by the following
four fence constraints: $c_0=(\texttt{L0},\ssfence)$,
$c_1=(\texttt{L1},\ssfence)$, $c_2=(\texttt{L0},\fence)$, and
$c_3=(\texttt{L1},\fence)$. The fence set returned by
$\witanalyze(\prog,\run)$ is $G=\{c_0,c_1,c_2,c_3\}$. Notice that $G$
satisfies both of the requirements for $\witanalyze$.

%%% Local Variables:
%%% mode: latex
%%% TeX-master: "main.tex"
%%% End:

\section{Experimental Results}
\label{sec:results}
We have implemented our fence insertion algorithm together with a
reachability analysis for {\sc SiSd} in the tool \textsc{Memorax}. It
is publicly available at \url{https://github.com/memorax/memorax}.
We apply the tool
 to a number of benchmarks (Section~\ref{sec:exp:fencins}).
Using simulation, we show the positive impact of
using different types of fences, compared to using only the full
fence, on performance and network traffic
(Section~\ref{sec:exp:simulation}).

\subsection{Fence Insertion Results}\label{sec:exp:fencins}

We evaluate the automatic fence insertion procedure by running our
tool on a number of different benchmarks containing racy code.
For each example, the tool gives us all optimal sets of fences.
We run our tool on the same benchmarks both for \textsc{SiSd} and for
the \textsc{Si} protocol.%
\footnote{Our methods could also run under a plain
  \textsc{Sd} protocol. However, to our knowledge, no cache
  coherence protocol employs only \textsc{Sd} without \textsc{Si}.}
The results for \textsc{SiSd} are given in
Table~\ref{tab:fencins:results}. We give the benchmark sizes in lines
of code. All benchmarks have 2 or 3 processes.
The fence insertion procedure was run single-threadedly on a 3.07 GHz
Intel i7 CPU with 6 GB RAM.

\begin{table}[ht]
  \centering
  \footnotesize{
    \newcommand{\ssfnc}{{\texttt{ss}}}
    \newcommand{\llfnc}{\texttt{ll}}
    \newcommand{\sw}{\texttt{sw}}
    \begin{tabular}{@{}l@{}r|r@{$\;$}c@{$\;$}c|r@{$\;$}c@{}c@{}}
      \hline
      \multicolumn{2}{@{}l|}{}& \multicolumn{3}{@{}c|}{Only full fence} & \multicolumn{3}{@{}c@{}}{Mixed fences}\\
      \hline
      Benchmark & Size (LOC) & Time & \#{}solutions & \#{}fences & Time & \#{}solutions & Fences / proc\\
      \hline
      bakery      &  45 &  17.3 s &  4 &     5 & 108.1 s &  16 & 2 $\sw$, 4 $\llfnc$, 1 $\ssfnc$ \\
      cas         &  32 &  $<$0.1 s &  1 &     2 &  $<$0.1 s &   1 & 1 $\llfnc$, 1 $\ssfnc$ \\
      clh         &  37 &   4.4 s &  4 &     4 &   3.7 s &   1 & 3 $\sw$, 2 $\llfnc$, 1 $\ssfnc$ \\
      dekker      &  48 &   2.0 s & 16 &     3 &   2.9 s &     16 & 1 $\sw$,2 $\llfnc$, 1 $\ssfnc$ \\
      mcslock     &  67 &  15.6 s &  4 &     2 &  33.0 s &   4 & 1 $\llfnc$, 1 $\ssfnc$ \\
      testtas     &  38 &  $<$0.1 s &  1 &     2 &  $<$0.1 s &   1 & 1 $\llfnc$, 1 $\ssfnc$ \\
      srbarrier   &  60 &   0.3 s &  9 &     3 &   0.4 s &   4 & 2 $\llfnc$, 1 $\ssfnc$ \\
      treebarrier &  56 &  33.2 s & 12 & 1 / 2 & 769.9 s & 132 & 1 $\llfnc$, 1 $\ssfnc$ \\
      dclocking   &  44 &   0.8 s & 16 &     4 &   0.9 s &  16 & 1 $\sw$, 2 $\llfnc$, 1 $\ssfnc$ \\
      postgresql  &  32 &  $<$0.1 s &  4 &     2 &   0.1 s &   4 & 1 $\llfnc$, 1 $\ssfnc$ \\
      \hline
      barnes 1    &  30 &   0.2 s &  1 &     1 &   0.5 s &   1 & 1 $\llfnc$ / 1 $\ssfnc$ \\
      barnes 2    &  96 &  16.3 s & 16 &     1 &  16.1 s &  16 & 1 $\ssfnc$ \\
      cholesky    &  98 &   1.6 s &  1 &     0 &   1.6 s &   1 & 0           \\
      radiosity   & 196 &  25.1 s &  1 &     0 &  24.6 s &   1 & 0           \\
      raytrace    & 101 &  69.3 s &  1 &     0 &  70.1 s &   1 & 0           \\
      volrend     &  87 & 376.2 s &  1 &     0 & 376.9 s &   1 & 0           \\
      \hline
    \end{tabular}
  }
  \caption{Automatic fence insertion for \textsc{SiSd}.}
  \label{tab:fencins:results}
\end{table}

The first set of benchmarks are classical examples from the context of lock-free synchronization. They contain mutual exclusion algorithms:
a simple CAS lock --{\em cas}--,
a test \& TAS lock \mbox{--\emph{tatas}--}~\cite{scott:shmemsync},
Lamport's bakery algorithm --{\em bakery}--~\cite{Lamport:Bakery},
the MCS queue lock --{\em mcsqueue}--~\cite{MCS91},
the CLH queue lock --{clh}--~\cite{magnusson1994queue},
and Dekker's
algorithm --{\em dekker}--~\cite{Dijkstra:cooperating}.
They also contain a work scheduling algorithm
--{\em postgresql}--%
\footnote{\href{http://archives.postgresql.org/pgsql-hackers/2011-08/msg00330.php}{http://archives.postgresql.org/pgsql-hackers/2011-08/msg00330.php}},
and an idiom for double-checked locking
--{\em dclocking}--~\cite{Schmidt97double-checkedlocking},
as well as two process barriers
--{\em srbarrier}--~\cite{scott:shmemsync}
and
--{\em treebarrier}--~\cite{MCS91}.
The second set of benchmarks are based on the \mbox{Splash-2}
benchmark suite~\cite{splash95}. We use the race detection tool
Fast\&Furious~\cite{aros-parmaditam15} to detect racy parts in the
Splash-2 code. We then manually extract models capturing the core of
those parts.

In four cases the tool detects bugs
in the original Splash-2 code.
The \emph{barnes} benchmark is an n-body simulation, where the bodies
are kept in a shared tree structure.
We detect two bugs under \textsc{SiSd}: When bodies are inserted
(\emph{barnes 2}), some bodies may be lost. When the center of mass is
computed for each node (\emph{barnes 1}), some nodes may neglect
entirely the weight of some of their children. Our tool inserts fences
that prevent these bugs.
The \emph{radiosity} model describes a work-stealing queue that appears
in the Splash-2 \emph{radiosity} benchmark. Our tool detects data race from 
it. After careful code inspection we find that it is
possible for all workers but one to terminate prematurely, leaving one
worker to do all remaining work, which is caused by data race.~\cite{sakalis16}
The \emph{volrend} model is also a work-stealing queue. Our tool
detects that it is possible for some tasks to be performed twice.
The bugs in \emph{radiosity} and \emph{volrend} can occur even under
SC. Hence the code cannot be fixed only by adding fences. Instead
we manually correct it.

For each benchmark, we apply the fence insertion procedure in two
different modes. In the first one (``Only full fence''), we use only
full fences.
In the table, we give the total time for computing all optimal sets,
the number of such sets, and the number of fences to insert into each
process.
For \emph{treebarrier}, one process (the root process) requires only
one fence, while the others require two.
Notice also that if a benchmark has one solution with zero fence, that
means that the benchmark is correct without the need to insert any
fences.

In the second set of experiments (``Mixed fences''),
we allow all four types of fences,
using a cost function  assigning a cost of
ten units for a full fence, five units for an $\ssfence$ or an $\llfence$,
and one unit for a synchronized write.
These cost assignments are reasonable in light of our empirical
evaluation of synchronization cost in Section~\ref{sec:exp:simulation}.
We list the number of inserted fences of each kind.
In \emph{barnes 1}, the processes in the model run
different codes. One process requires an $\llfence$, the other an
$\ssfence$.

In addition to running our tool for \textsc{SiSd}, we have also run
the same benchmarks for \textsc{Si}.
As expected, $\ssfence$ and $\plainsyncwr$ are no longer necessary,
and $\fence$ may be downgraded to $\llfence$. Otherwise, the inferred
fence sets are the same as for \textsc{SiSd}.
Since \textsc{Si} allows fewer behaviors than \textsc{SiSd}, the
inference for \textsc{Si} is mostly faster. Each benchmark is fenced
under \textsc{Si} within 71 seconds.

\subsection{Simulation Results}\label{sec:exp:simulation}

Here we show the impact of different choices of fences when executing
programs. In particular we show that an optimal fence set using the
``Mixed fences'' cost function yields a better performance and network
traffic compared to an optimal fence set using the ``Only full fence''
cost function. 
Here network traffic refers to the traffic in both the on-chip
interconnection network and the memory bus. We account for all
the traffic due to coherence messages.
We evaluate the micro-benchmarks analyzed in the previous section and
the whole Splash-2 benchmark suite~\cite{splash95}, running the
applications from beginnig to end, but presenting results only for the
parallel phase of the applications.
All programs are fenced according to the optimal fence
sets produced by our tool as described above.

\paragraph{Simulation Environment:}
We use the Wisconsin GEMS simulator~\cite{gems05}. We model an
in-order processor that with the Ruby cycle-accurate memory simulator
(provided by GEMS) offers a detailed timing model. The simulated
system is a 64-core chip multiprocessor implementing the \textsc{SiSd}
protocol described in Section \ref{sec:sisd} and 32KB, 4-way, private
L1 caches and a logically shared but physically distributed L2 with 64
banks of 256KB, 16-way each.

\paragraph{Cost of Fences:}

Our automatic fence insertion tool employs different weights in order to insert the optimal amount of fences given the cost of each fence. Here, we
calculate the weights based on an approximate cost of fences obtained
by our simulations.

The effect of fences on performance is twofold. First, there is a cost to 
execute the fence instructions (fence latency); the more fences
and the more dirty blocks to self-downgrade, the higher the
penalty. Second, fences affect cache miss ratio
(due to self-invalidation) and network traffic (due to extra fetches
caused by self-invalidations and write-throughs caused by
self-downgrades). The combined effect on cache misses and network
traffic also affects performance.

\begin{figure*}[h]
\centering
\subfloat[Cycles]{
\label{fig:cost_time}
\includegraphics[width=7.5cm]{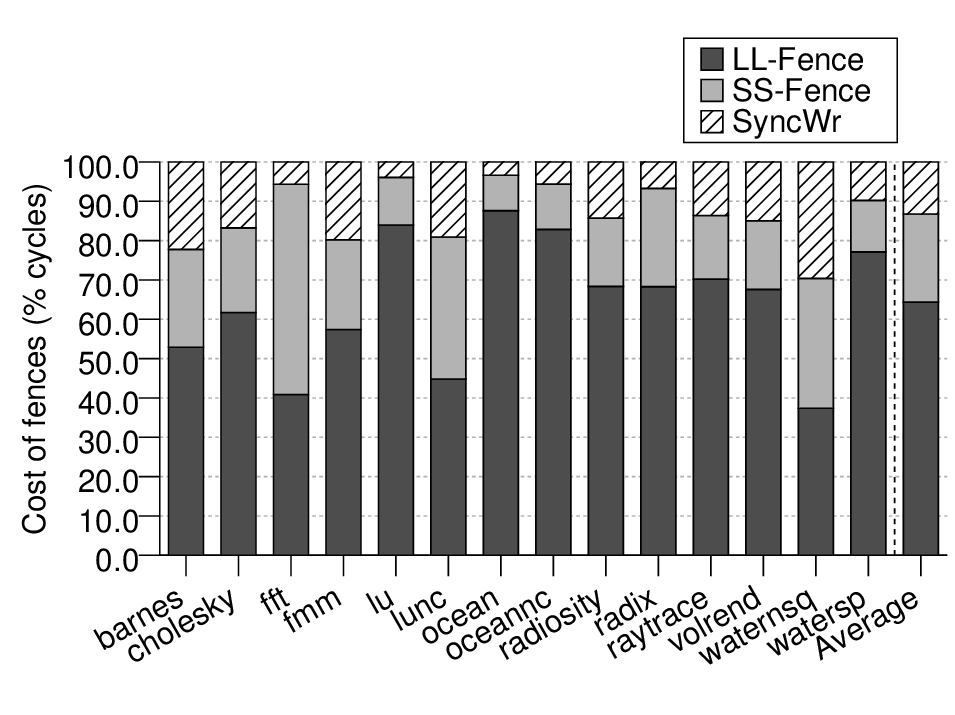}
}
\subfloat[Traffic]{
\label{fig:cost_traffic}
\includegraphics[width=7.5cm]{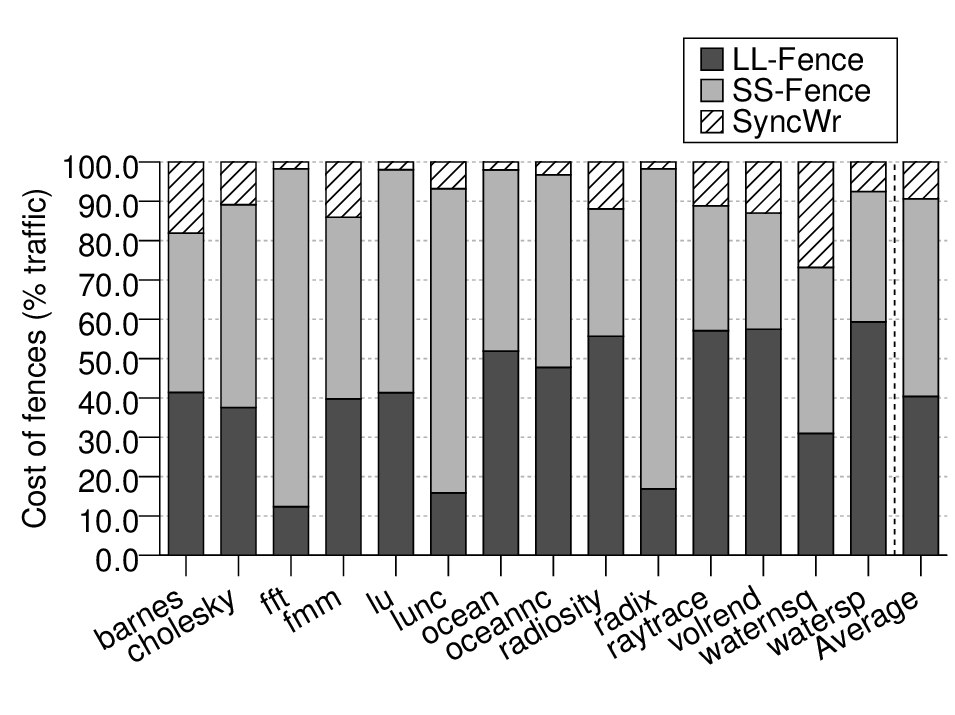}
}
\caption{Percentage of cycles and traffic that each type of fence cost}
\label{fig:energy}
\end{figure*}

We calculate the cost of fences in terms of execution as indicated in
equation \ref{eq:time_cost}, where $\mathit{latency}_{\mathit{fence}}$
is the time in cycles required by the fence,
$\mathit{misses}_{\mathit{si}}$ is the number of misses caused by
self-invalidation, and $\mathit{latency}_{\mathit{miss}}$ is the
average latency of such misses. According to this equation, and
considering a protocol implementing the DoI state described in Section
\ref{sec:sisd}, the average percentage of cycles (execution time)
employed by each type of fence when running the Splash2 benchmarks is
the following: the cost of an \llfence{} is 64.4\%, the cost of an
\ssfence{} is 22.4\%, and the cost of a \plainsyncwr{} is 13.2\%, as
shown in Figure \ref{fig:cost_time}.

\begin{equation}\label{eq:time_cost}
\mathit{time}_{\mathit{fence}} = \mathit{latency}_{\mathit{fence}} + \mathit{misses}_{\mathit{si}} \times \mathit{latency}_{\mathit{miss}}
\end{equation}

The cost of the fences in traffic is calculated as indicated in
equation \ref{eq:traffic_cost}, where $\mathit{sd}$ is the number of
self-downgrades, $\mathit{traffic}_{\mathit{wt}}$ is the traffic
caused by a write-through, and $\mathit{traffic}_{\mathit{miss}}$ is
the traffic caused by a cache miss. In percentage, the cost of the
each type of fence on average in terms of traffic is 40.4\% for an
\llfence{}, 50.3\% for an \ssfence{}, and 9.3\% for a \plainsyncwr{},
as shown in Figure \ref{fig:cost_traffic}. Thus, the weights assigned
to fences in our tool seem reasonable.

\begin{equation}\label{eq:traffic_cost}
\mathit{traffic}_{\mathit{fence}} = \mathit{sd} * \mathit{traffic}_ {\mathit{wt}} +  \mathit{misses}_ {\mathit{si}} \times \mathit{traffic}_ {\mathit{miss}}
\end{equation}

\paragraph{Cache Misses:} 
As mentioned, the fences affect the cache miss rate. Figure
\ref{fig:misses} shows clearly the effect of self-invalidation and
self-downgrade on misses. First we show the misses due to cold
capacity and conflict misses (\emph{Cold-cap-conf}), which, in
general, are not affected by the type of fences. However, in some
cases reducing the self-invalidation can give the appearance of extra
capacity misses because of having a more occupied cache. The graph
does not plot 

\begin{figure*}[ht]
\centering
\subfloat[Micro-benchmarks]{
\label{fig:misses-micro}
\includegraphics[width=12cm]{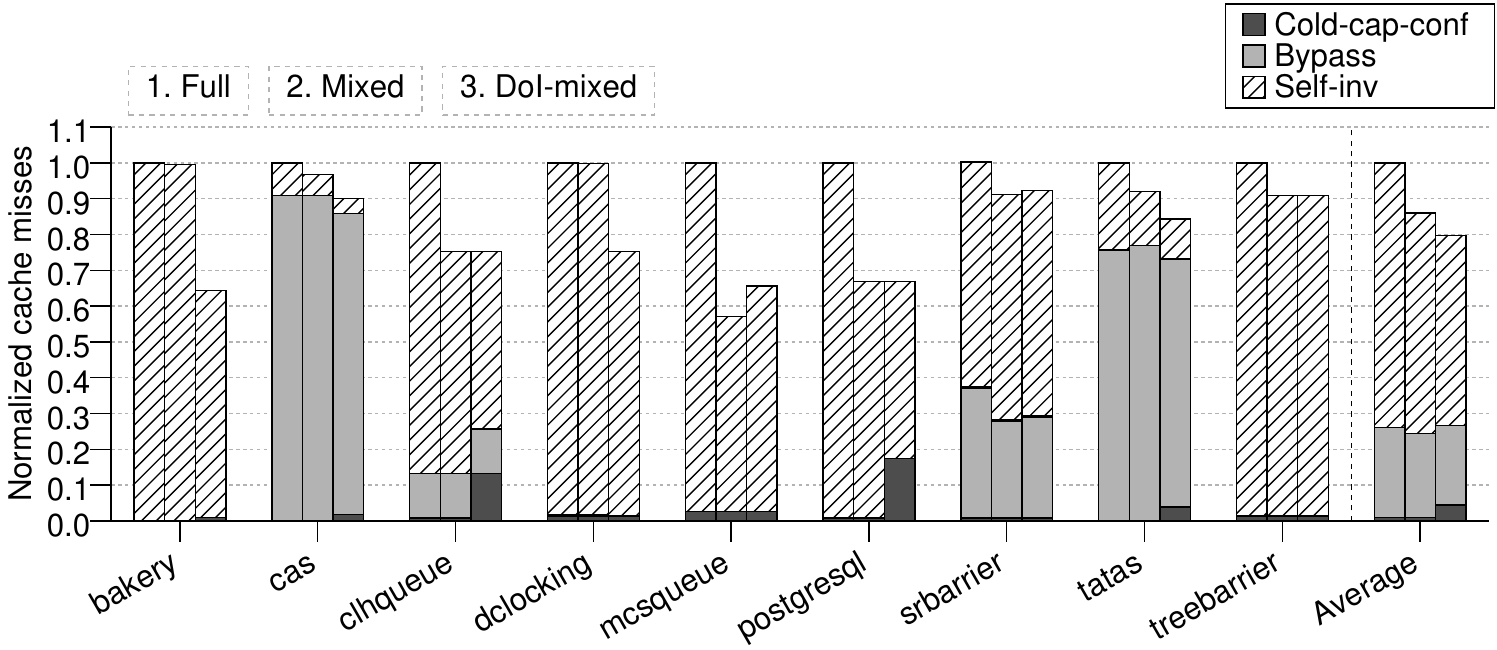}
}\\
\subfloat[Splash2 benchmarks]{
\label{fig:misses-splash2}
\includegraphics[width=12cm]{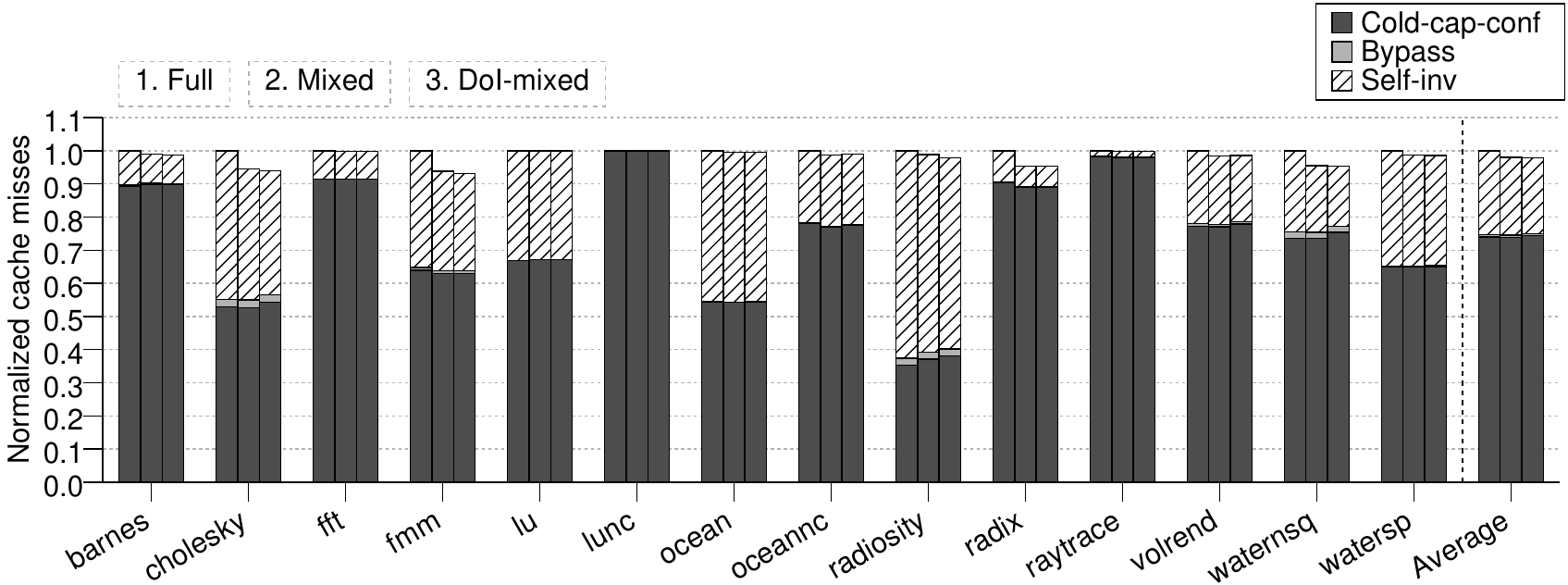}
}
\caption{Normalized cache misses under different fence sets and protocol states}
\label{fig:misses}
\end{figure*}

\noindent coherence misses since fenced programs on {\sc Sisd}
coherence do not induce cache misses due to coherence
transactions. The second kind of miss is named as \emph{Bypass}. These
misses are due to atomic operations which cannot use the data in the
private cache, but need to access it from the shared cache. They are
very frequent in the micro-benchmarks (Figure \ref{fig:misses-micro}), 
which are synchronization
intensive, but almost unnoticeable for the Splash2
benchmarks (Figure \ref{fig:misses-splash2}). Finally, we show the misses caused by self-invalidation
\emph{Self-inv}. These are the ones which number is reduced, when
applying the mixed fences, but also when employing the DoI state,
since dirty words are not invalidated.

\paragraph{Traffic:}
As already mentioned, traffic is also affected by the type of fences employed. 
Figure \ref{fig:traffic} shows the traffic in the on-chip network
generated by these applications. The use of \llfence{}, \ssfence{},
\plainsyncwr{} is able to reduce the traffic requirements by 11.1\%
for the micro-benchmarks and 1.6\% for the Splash2 applications, on
average, compared to using full fences. Additionally, when employing
the DoI state, this reduction reaches 21.3\% and 1.9\%, on average,
for the micro-benchmarks and the Splash2, respectively. Again, the
more synchronization is required by the applications, the more traffic
can be saved by employing mixed fences.

\begin{figure*}[ht]
\centering
\subfloat[Micro-benchmarks]{
\includegraphics[width=12cm]{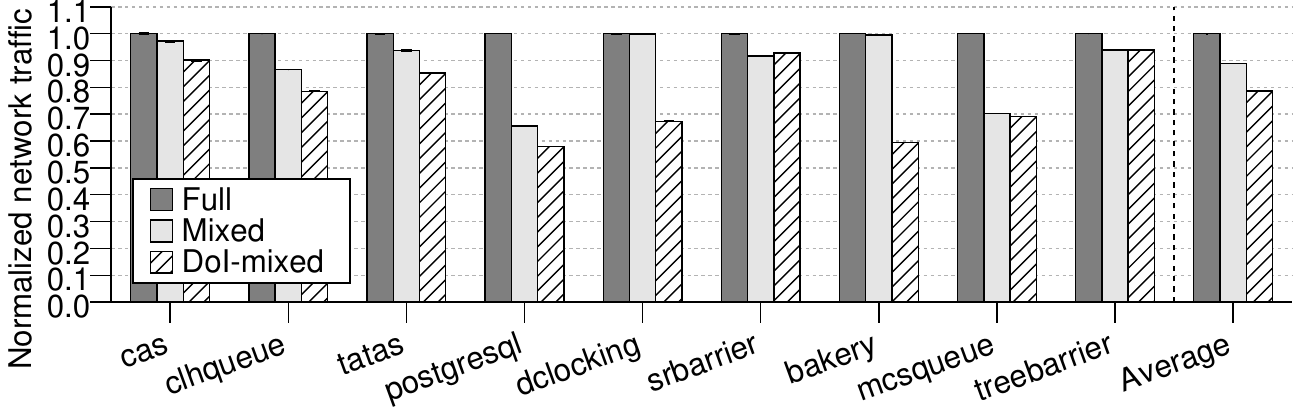}
}\\
\subfloat[Splash2 benchmarks]{
\includegraphics[width=12cm]{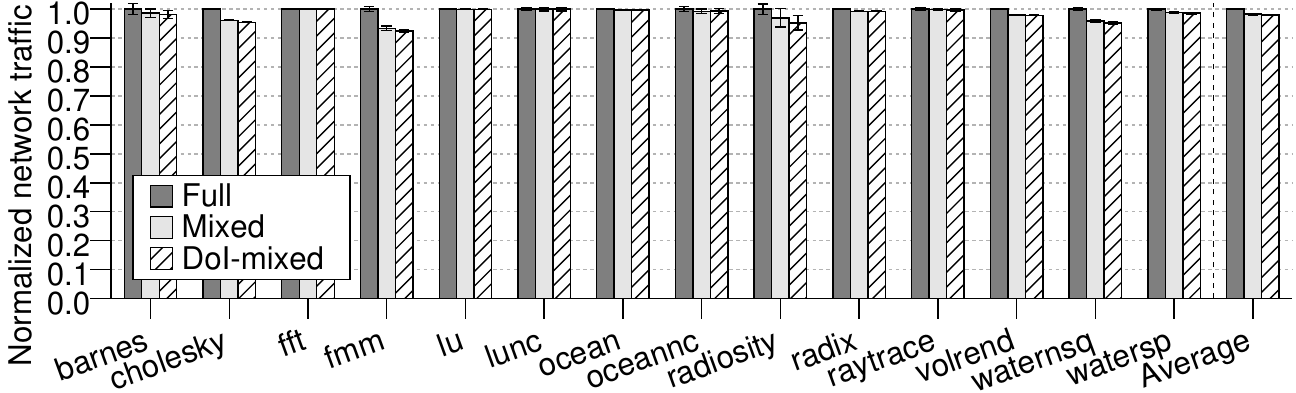}
}
\caption{Normalized network traffic under different fence sets and protocol states}
\label{fig:traffic}
\end{figure*}

\paragraph{Execution Time:}
Finally, we show the impact on execution time, which is affected by
the reductions in cache misses and traffic.  Figure
\ref{fig:exec_time} shows simulated execution time for both the
micro-benchmarks (Figure \ref{fig:exec_time-micro}) and the Splash2
benchmarks (Figure \ref{fig:exec_time-splash2}). The use of mixed
fences improves the execution time compared to using full fences by
10.4\% for the micro-benchmarks and by 1.0\% for the Splash2
benchmarks.
The DoI-mixed column shows the execution time results for the same
mixed fence sets as the mixed column. But in DoI case, \llfence{}s are
implemented in GEMS using an extra L1 cache line state (the
Dirty-or-Invalid state). This feature is an architectural optimization
of the {\sc SiSd} protocol.
Implementing the DoI state further improves the performance of the
mixed fences, by 20.0\% for the micro-benchmarks and 2.1\% for the
Splash2, on average, compared to using of full fences. Mixed fences
are useful for applications with more synchronization. Applications
using more synchronization would benefit to a large extent from the
use of mixed fences.

\begin{figure*}[t]
\centering
\subfloat[Micro-benchmarks]{
\label{fig:exec_time-micro}
\includegraphics[width=12cm]{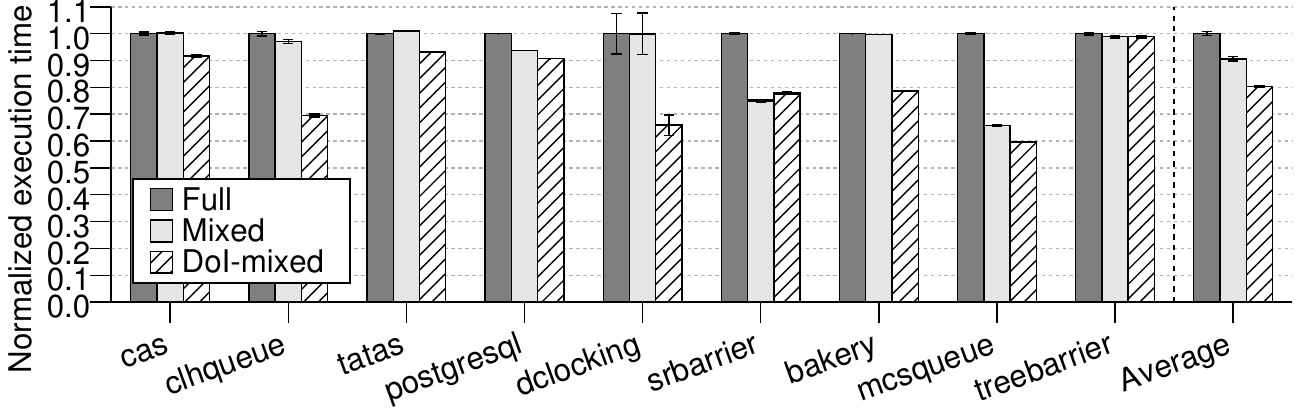}
}\\
\subfloat[Splash2 benchmarks]{
\label{fig:exec_time-splash2}
\includegraphics[width=12cm]{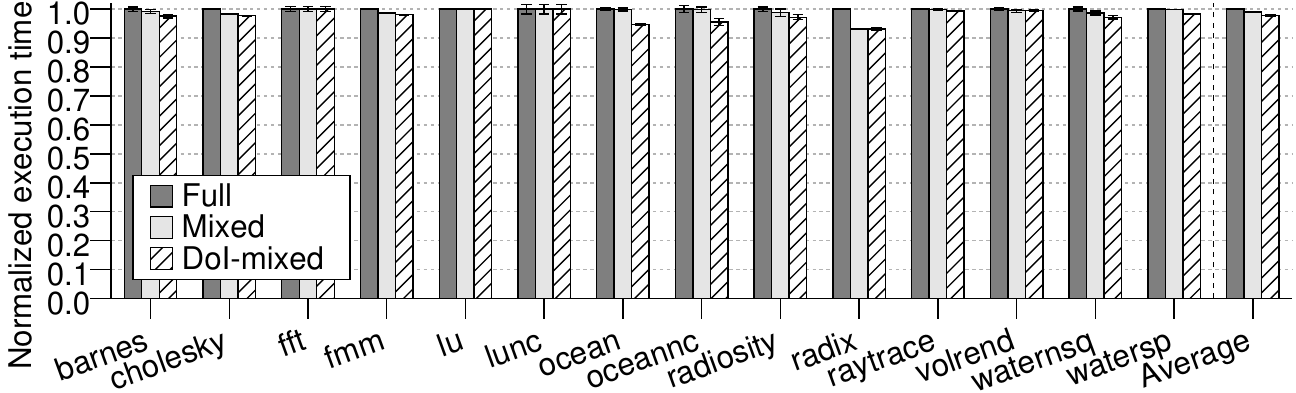}
}
\caption{Execution time under different fence sets and protocol states}
\label{fig:exec_time}
\end{figure*}

%%% Local Variables:
%%% mode: latex
%%% TeX-master: "main.tex"
%%% End:

\section{Conclusions and Future Work}
\label{conclusions:section}
We have presented a uniform framework for automatic fence
insertion in programs that run on architectures that provide
self-invalidation and self-downgrade. We have implemented a tool
and applied it on a wide range of benchmarks. There are several 
interesting directions for future work. One is to instantiate our 
framework in the context of abstract interpretation and stateless 
model checking. While this will compromise the optimality criterion, 
it will allow more scalability and application to real program code.  
Another direction is to consider {\it robustness} properties 
\cite{DBLP:conf/icalp/BouajjaniMM11}. In our framework, this would 
mean that we consider program traces (in the sense of Shasha and 
Snir~\cite{shasha1988efficient}), and show that the program will 
not exhibit more behaviors under {\sc SiSd} than under SC. While 
this may cause over-fencing, it frees the user from providing correctness 
specifications such as safety properties. Also, the optimality of fence 
insertion can be evaluated with the number of the times that each fence 
is executed. This measurement will provide more accuracy when, for instance, 
fences with different weights are inserted in a loop computation in a
branching program. 

\paragraph{\bf Acknowledgment}
This work was supported by the Uppsala Programming for 
Multicore Architectures Research Center (UPMARC), the 
Swedish Board of Science project, ``Rethinking the Memory System'',  
the ``Fundaci\'{o}n Seneca-Agencia de Ciencia y Tecnolog\'{i}a de la Regi\'{o}n de Murcia'' 
under the project ``J\'{o}venes L\'{i}deres en Investigaci\'{o}n'' and 
European Commission FEDER funds.

%%% Local Variables:
%%% mode: latex
%%% TeX-master: "main.tex"
%%% End:

\bibliographystyle{alpha}
\bibliography{main}

\end{document}